\documentclass[a4paper,11pt]{article}
\usepackage{jheppub} 
\usepackage[dvipsnames]{xcolor}

\newcommand{\matr}[4]{\left(\!\begin{array}{cc}
    #1 \, & \, #2 \\
    #3 \, & \, #4
\end{array}\!\right)}

\usepackage{lineno}
\usepackage[table, dvipsnames]{xcolor}
\usepackage{hyperref}
\usepackage{graphicx}
\usepackage{subfig}
\usepackage{amsmath,amssymb,amsthm,amsfonts}
\usepackage{dsfont}
\usepackage{mathtools}
\usepackage[all]{xy}
\usepackage{tikz}
\usepackage{float}
\usetikzlibrary{calc}
\usepackage[toc,page]{appendix}
\usepackage{enumerate}
\usepackage{booktabs}
\usepackage{cancel}
\usepackage{setspace}
\usepackage{bbm}
\usepackage{youngtab}
\usepackage{slashed}
\usepackage{fancyhdr}
\usepackage{empheq}
\usepackage{multirow}
\usepackage{tabularx}
\usepackage{makecell}
\usepackage{float}
\usepackage{comment}
\usepackage{tensor}
\usepackage{subcaption}
\usepackage{verbatim}
 \usepackage[framemethod=tikz]{mdframed}
\usepackage{soul}
\usepackage{cancel}
\usepackage[normalem]{ulem}
\usepackage{mathtools}
\usepackage{bbm}
\usepackage{physics}
\usepackage{braket}
\usepackage{scalerel}
\usepackage{array}
\usepackage{enumitem}
\usepackage{tcolorbox}
\usepackage[backgroundcolor=white!95!red, linecolor=purple, bordercolor=purple, textsize=footnotesize]{todonotes}
\usepackage{tikz-3dplot}
\usetikzlibrary{calc}

\usepackage[T1]{fontenc}
\usepackage{lmodern}

\newcommand{\ab}{|}

\renewcommand{\arraystretch}{1.7}
\definecolor{Gray}{gray}{0.9}
\definecolor{mycolor}{RGB}{240,240,240}
\newcommand{\nsns}{\text{NSNS}}
\newcommand{\rr}{\text{RR}}
\newcommand{\adj}{\text{Adj}}
\newcommand{\usp}{\text{USp}}
\newcommand{\So}{\text{SO}}

\newcommand{\mpl}{M_{\text{Pl}}}
\newcommand{\ms}{M_{s}}
\newcommand{\tree}{\text{tree}}

\newcommand{\np}{\text{np}}
\newcommand{\gs}{g_{s}}
\newcolumntype{C}{>{\centering\arraybackslash}X}

\begin{document}

\begin{titlepage}

\setcounter{page}{1} \baselineskip=15.5pt \thispagestyle{empty}

\bigskip\

\vspace{1cm}
\begin{center}

{\fontsize{18}{22}\selectfont  \sffamily \bfseries {The Cosmological Constant and Dark Dimensions \\ from Non-Supersymmetric Strings
 \\}}

\end{center}

\vspace{0.2cm}

\begin{center}
{\fontsize{13}{30}\selectfont 
Emilian Dudas$^{a,}$\footnote{\texttt{Emilian.Dudas@polytechnique.edu}}, 
Susha Parameswaran$^{b,}$\footnote{\texttt{Susha.Parameswaran@liverpool.ac.uk}}, 
Marco Serra$^{b,}$\footnote{\texttt{Marco.Serra@liverpool.ac.uk}}} 
\end{center}

\begin{center}

\vskip 8pt
\textsl{$^a$ CPHT, Ecole Polytechnique, 91128 Palaiseau cedex, France} \\
\textsl{$^b$ Department of Mathematical Sciences, University of Liverpool, Liverpool, L69 7ZL, United Kingdom}
\vskip 6pt

\end{center}

\vspace{1.2cm}
\hrule \vspace{0.3cm}
\noindent 
  We present a string theory construction in which the particle physics contributions to the one-loop vacuum energy exactly cancel, whilst the gravitational contributions are suppressed in the size of one or two large extra dimensions. This provides an ultraviolet realisation of the Dark Dimension and Supersymmetric Large Extra Dimensions scenarios, with, moreover, an explanation as to why the Standard Model contributions to the vacuum energy cancel without the need of eV mass-splittings.  Gravity propagates in micron sized dark dimension(s), whilst the visible and hidden sectors are supported on D-branes.  Supersymmetry is broken in the dark dimension(s) \`a la Scherk-Schwarz, whereas supersymmetry is broken at the string scale, \`a la Brane Supersymmetry Breaking, in the D-branes sector, without inducing tadpoles, similarly to a different construction proposed a long time ago by Angelantonj and Antoniadis. Vacuum energy from the visible sector is cancelled by the vacuum energy of the hidden sector branes.  We also discuss moduli stabilization in this set-up, finding that the interplay between the Scherk-Schwarz one-loop contribution and non-perturbative effects can fix the size of the dark dimension(s) to be exponentially large in the inverse string-coupling, leading to an exponentially small total vacuum energy, with all moduli stabilised in a dS saddle.

\vskip 10pt
\hrule

\vspace{0.4cm}
 \end{titlepage}

 \tableofcontents

\section{Introduction}

The Cosmological Constant Problem \cite{Weinberg:1988cp} lies at the heart of fundamental physics, arising when trying to bring together our theories of particle physics and cosmology.  On the one hand, vacuum energy is one of the most basic consequences of quantum physics, with differences in vacuum energy experimentally verified via, for example, the Casimir effect.  On the other hand, in the presence of gravity the absolute value of the vacuum energy takes on physical significance, yet the value inferred from the observed Dark Energy is some 120 orders of magnitude smaller than the value expected from particle physics, which is naturally driven toward the quantum gravity scale, $\lesssim \mathcal{O}(\mpl^4)$.  In this paper, we present an explicit string construction whose gauge and matter sectors from open-strings contribute precisely zero to the one-loop cosmological constant, $\Lambda_{\text{open}}=0$, without the need for new light fields in the putative visible sector.  Moreover, we propose a moduli stabilisation scenario that ensures the contributions from the closed-string gravitational sector are exponentially suppressed in the inverse string-coupling, approaching $\Lambda_{\text{closed}}=\Lambda_{\text{observed}}$.

Supersymmetry is well-known to ameliorate the Cosmological Constant Problem, but only down to the scale set by the supersymmetry-breaking mass splittings, at least $\mathcal{O}(\text{TeV})$ for the visible sector.  String theory offers new possibilities.  These include the proposal of non-supersymmetric string models with vanishing one-loop vacuum energy, thanks to a matching between the number of fermion and boson states at every mass level, similar to supersymmetry \cite{Kachru:1998hd, Harvey:1998rc}.  In open-string descendants of such constructions, D-branes could help hide this degeneracy from a putatitive visible sector.   Although in the first such models \cite{Blumenhagen:1998uf, Angelantonj:1999gm}, supersymmetry turned out to be unbroken on the D-branes \cite{Angelantonj:1999gm}, subsequent extensions \cite{Angelantonj:2003hr} of non-supersymmetric orientifolds managed, remarkably, to identify genuinely non-supersymmetric D-brane spectra featuring Bose-Fermi degeneracy at the massless level, mass splittings of order $M_{\text{s}}$, and a one-loop cosmological constant induced by a Scherk-Schwarz supersymmetry breaking in the closed-string sector that scales as $\sim 1/R^4$.  The construction of Angelantonj and Antoniadis \cite{Angelantonj:2003hr} was also the first to combine Scherk-Schwarz and Brane Supersymmetry Breaking mechanisms, in the presence of a discrete antisymmetric tensor background $B_{ab} = \alpha'/2$.  The coexistence of TeV-scale mass-splittings in a putative visible sector and a suppressed one-loop cosmological constant in these constructions seems particularly compelling.  Further significant progress was made in \cite{carlo1}, which constructed a non-supersymmetric orientifold yielding D-brane spectra with Bose–Fermi degeneracy at all mass levels, potentially allowing for cancellations beyond one-loop, albeit with the D-branes in an unstable configuration.\footnote{See also \cite{Abel:2017rch, Abel:2018zyt, Abel:2020ldo} for further constructions with Bose-Fermi degeneracy at massless level and thus an exponentially suppressed one-loop vacuum energy.}

Our construction builds on these results by employing a non-supersymmetric orientifold that combines Brane Supersymmetry Breaking with Scherk–Schwarz supersymmetry breaking.   Starting from the non-supersymmetric orientifold in\footnote{The heterotic dual of this model was proposed in \cite{Fraiman:2025yrx}.} \cite{Coudarchet:2021qwc}, our chosen configuration of D-branes and O-planes is stable and exhibits an exact Bose–Fermi degeneracy at all mass levels in the open-string sector, mass splittings of order $M_{\text{s}}$, and a one-loop cosmological constant scaling as $\sim 1/R^{4}$. By balancing this one-loop contribution\footnote{See \cite{Parameswaran:2024mrc} and \cite{ValeixoBento:2025yhz} for recent moduli stabilisation scenarios that balance Casimir contributions against flux and curvature.} against non-perturbative effects, such as D(–1) instantons and ED3-branes, the radius $R$ can be stabilized at a value exponentially large in the inverse string coupling. Including further -- tadpole-free -- fluxes, the remaining closed-string moduli can be fixed as well, in a de Sitter saddle with exponentially small vacuum energy.

Depending on the number and size of the Scherk–Schwarz supersymmetry-breaking directions, this set-up provides a string theory realisation of either the\footnote{For previous realizations of large extra dimensions, see \cite{add,ddg,aadd}. For an early discussion on how brane supersymmetry breaking and large extra dimensions could help with the cosmological constant problem, see \cite{Antoniadis:1999xk}.} Supersymmetric Large Extra Dimensions scenario \cite{Aghababaie:2003wz, Burgess:2004ib} or the Dark Dimension scenario \cite{Montero:dd}, together with an explanation as to why the Standard Model contributions to the vacuum energy cancel. In our framework, gravity propagates in the bulk of the dark dimension(s), while the Standard Model is localised on a stack of D-branes orthogonal to them, with additional stacks of hidden-sector branes located elsewhere in the bulk. Supersymmetry is broken at the micron scale along the dark dimension(s) via the Scherk–Schwarz mechanism, whereas it is broken at the string scale on both the Standard Model and hidden-sector branes by Brane Supersymmetry Breaking. The vacuum energy cancels between the Standard Model and hidden-sector brane stacks. As a result, the total vacuum energy is governed by the closed-string gravitational sector and is tied to the size of the dark dimension(s), as in the original proposals.  Dark dimension(s) sit tantalizingly at the current observational bounds, with order one parameters already being important for Supersymmetric Large Extra Dimensions to be consistent with all the constraints (see \cite{ParticleDataGroup:2024cfk}, \cite{antoniadis2DD} and also \cite{Burgess:2023pnk} for an optimistic view on the viability of models with two large extra dimensions).  As such, they are testable in the near-future table-top gravity experiments, as well as across cosmology, astrophysics and accelerators.

The paper is organized as follows.  Section \ref{S:SBintro}, which can be skipped by experts, provides a pedagogical introduction to supersymmetry-breaking mechanisms and one-loop vacuum amplitudes in type II string theory with D-branes.  In Section \ref{S:CCeq0}, we present our explicit string construction that, via a non-supersymmetric orientifolding, achieves $\Lambda_{\text{open}}=0$ and $\Lambda_{\text{closed}} \sim 1/R^4$.  Section \ref{S:modstab} develops a moduli stabilisation scenario that realises large extra dimensions and an exponentially small vacuum energy at weak string coupling.  After confronting our construction against current observations, we close in Section \ref{S:concl} with a brief summary and a discussion of the open questions.  We further include a number of technical appendices referred to in the main text. 

\section{Supersymmetry-breaking in type II string theory}
\label{S:SBintro}

In this section, we briefly review the mechanisms of supersymmetry-breaking in type II models with D-branes and discuss their implications for the spectra and vacuum energy.  We are particularly interested in mechanisms that can be implemented directly in string theory and not just at the level of effective field theory, together with the interplay between bulk and brane supersymmetry.  Experts can skip this section and for readers wanting more details we point to the reviews \cite{Angelantonj:2002ct}, \cite{Leone:2025mwo} and \cite{dms-review}.

\subsection{Type IIB, toroidal compactifications and supersymmetric orientifolds} \label{S:susytoriorients}
Before entering our discussion of supersymmetry-breaking, it will be useful to have in mind the one-loop amplitudes of the supersymmetric closed-string type IIB string theory, supersymmetric toroidal compactifications, and the supersymmetric orientifold to the open-string type I theory.  The one-loop amplitudes both encode the full perturbative string spectrum and correspond to the one-loop vacuum energy. 

Starting with the type IIB theory, the one-loop vacuum amplitude is given by the unique oriented genus-zero worlsheet, the torus:
\begin{equation}
    \mathcal{T}_{\text{IIB}} = \int_{\mathcal{F}} \frac{d^2\tau}{\tau_2^2} \frac{1}{\tau_2^4}\text{Str}\left(q^{L_0-\frac12}\bar{q}^{\bar{L}_0-\frac12}\right) = \int_{\mathcal{F}} \frac{d^2\tau}{\tau_2^2} \frac{1}{\tau_2^4}\frac{\left(V_8-S_8\right)\left(\overline{V_8}-\overline{S_8}\right)}{\left\lvert \eta\right\rvert^{16}}[\tau]\,, \label{eq:torusIIB}
\end{equation}
where $\tau=\tau_1+i\tau_2$ is the modular parameter of the worldsheet torus and $\mathcal{F}$ its fundamental domain under the $\text{SL}(2,\mathbb{Z})$ modular symmetry.   Moreover, $\eta(\tau)$ is the modular-covariant  Dedekind $\eta$-function, comprising the contributions of the eight transverse bosons, whereas the modular-covariant SO(8) characters, $V_8$ and $S_8$, comprise the contributions of the eight transverse fermions, organized by their SO(8) representations -- respectively, vector ${\bf 8_v}$ and spinor ${\bf 8_s}$ -- and excitation towers.  Later, we will also need the other two SO(8) characters, $O_8$ and $C_8$, corresponding to the tachyonic singlet ${\bf 1}$ and conjugate spinor ${\bf 8_c}$ representations, along with excitation towers.  The $q$-expansions of the SO(8) characters are given by:
\begin{equation}\label{qexp}
       \begin{split}
       O_8 &= \frac{\vartheta_3^4(0|\tau)+\vartheta_4^4(0|\tau)}{2\,\eta^4(\tau)}
= q^{-1/2}\!\left(1 + 28\,q + 134\,q^2 + 568\,q^3 + \cdots\right),\\[4pt]
V_8 &= \frac{\vartheta_3^4(0|\tau)-\vartheta_4^4(0|\tau)}{2\,\eta^4(\tau)}
= 8 + 56\,q + 224\,q^2 + 720\,q^3 + \cdots,\\[4pt]
S_8 &= \frac{\vartheta_2^4(0|\tau)+\vartheta_1^4(0|\tau)}{2\,\eta^4(\tau)}
= 8 + 56\,q + 224\,q^2 + 720\,q^3 + \cdots,\\[4pt]
C_8 &= \frac{\vartheta_2^4(0|\tau)-\vartheta_1^4(0|\tau)}{2\,\eta^4(\tau)}
= 8 + 56\,q + 224\,q^2 + 720\,q^3 + \cdots,
      \end{split}
   \end{equation}
with $q=e^{2\pi i \tau}$. Note that $\theta_1^4(0|\tau)$ vanishes numerically, yet the relative sign in $S_8$ and $C_8$ distinguishes the two possible spinor chiralities. The supersymmetric cancellation of the torus amplitude takes place via the Jacobi abstruse identity $V_8=S_8(=C_8)$.

Next, we present the one-loop vacuum amplitude for a supersymmetric toroidal compactification of type IIB.  Compactifying on a $d$-dimensional torus $\mathbb{T}^d$, with metric $G_{ij}$ and its inverse $G^{-1}_{ij}=G^{ij}$, the associated internal momenta are discrete and given in the left- and right-moving sectors by:
\begin{equation}\label{leftrightmomenta}   
p_{i}^L=m_i+G_{ij}\,n^j\,,\quad p_{i}^R=m_i-G_{ij}\,n^j\, ,
\end{equation}
where $m_i,n^i\in \mathbb{Z}$ are, respectively, the quantised Kaluza-Klein (KK) momentum and winding numbers along the compact direction $i=D,\dots,9$, with $D=10-d$. 
The $d=10-D$ compact bosons then contribute to the torus partition function as a lattice sum and one readily finds
\begin{equation}\label{toruscomp}
\mathcal{T}_{\mathrm{IIB}_{\mathbb{T}^d}} = \int_\mathcal{F}\frac{d^2\tau}{\tau_2^2}\frac{1}{\tau_2^{\frac{D-2}{2}}}\left|\frac{V_8-S_8}{\eta^8}\right|^2 [\tau]\sum_{\vec{m},\vec{n}} \Lambda_{\vec{m},\vec{n}}[\tau]\, ,
\end{equation}
with
\begin{equation}
    \sum_{\vec{m},\vec{n}}\Lambda_{\vec{m},\vec{n}}[\tau]:=q^{\frac{1}{4}P^L_iG^{ij}P^L_j}\bar{q}^{\frac{1}{4}P^R_iG^{ij}P^R_j}\ \,.
\end{equation}
Note that acting with a T-duality transformation $G_{ij} \leftrightarrow G^{ij}$ and $m_i \leftrightarrow n_i$ takes us to the type IIA theory compactified on the dual torus $T(G_{ij}^{-1})$, and vice versa; an even number of T-dualities then
 gives back the same type II theory that one started with, compactified on the dual torus.

The supersymmetric open-string type I string theory is obtained via an orientifold of type IIB, modding out by the worldsheet parity operator, $\Omega$, which exchanges the left- and right-moving sectors.  Consequently, the torus amplitude \eqref{eq:torusIIB} is halved, and (half of) the Klein-bottle is introduced, together with the annulus and M\"obius strip\footnote{The M\"obius strip involves ``hatted'' characters, which differ from the ordinary ones by a phase, to ensure that the states contribute with integer degeneracies even though the modular parameter of the covering torus has a real part \cite{Angelantonj:2002ct}.} open-string contributions that ensure modular invariance and cancellation of the RR tadpole:
\begin{align}
    &\mathcal{T}_{\text{I}} = \frac12 \mathcal{T}_{\text{IIB} }\nonumber \\
    &\mathcal{K}_{\text{I}} =\frac12 \int_0^\infty  \frac{d\tau_2}{\tau_2^2}\frac{1}{\tau_2^4} \text{Str}(\Omega\, q^{L_0 - \frac12} \bar{q}^{\bar{L}_0-\frac12}) = \frac12  \int_{0}^{\infty} \frac{d\tau_2}{\tau_2^2}\frac{1}{\tau_2^4}\frac{V_8-S_8}{\eta^8}[2i\tau_2] \nonumber\\
    &\mathcal{A}_{\text{I}} = \frac12 \int_{0}^{\infty} \frac{d\tau_2}{\tau_2^2}  \frac{1}{\tau_2^4}\text{Str}\left(q^{\frac12\left(L_0-\frac12\right)}\right)=\frac{N^2}{2} \int_{0}^{\infty} \frac{d\tau_2}{\tau_2^2}\frac{1}{\tau_2^4}\frac{V_8-S_8}{\eta^8}\left[\frac{i\tau_2}{2}\right]\nonumber \\
     &\mathcal{M}_{\text{I}} = \frac12 \int_{0}^{\infty} \frac{d\tau_2}{\tau_2^2}  \frac{1}{\tau_2^4}\text{Str}\left(\Omega \,q^{\frac12\left(L_0-\frac12\right)}\right) =  \epsilon\frac{N}{2} \int_{0}^{\infty} \frac{d\tau_2}{\tau_2^2}\frac{1}{\tau_2^4}\frac{\hat{V}_8-\hat{S}_8}{\hat{\eta}^8}\left[\frac{i\tau_2}{2}+\frac12\right]\,, \label{eq:typeI}
\end{align}
with $N$ the Chan-Paton factors and $\epsilon$ so far unfixed.  For $\epsilon=\mp 1$, there are then $N(N\mp1)/2$ massless gauge bosons and fermions from the open-string sector, leading to gauge groups SO($N$) and USp($N$) respectively.

In order to identify the types of O-planes and D-branes that are now present in the spacetime geometry, we consider a modular transformation of the above amplitudes, which takes one from the direct open-channel loop amplitudes to the transverse closed-channel tree amplitudes:
\begin{align}
    &\tilde{\mathcal{T}}_{\text{I}} = {\mathcal{T}}_{\text{I}} \nonumber \\
    &\tilde{\mathcal{K}}_{\text{I}} = \frac{2^5}{2} \int_{0}^{\infty} d\ell \frac{V_8-S_8}{\eta^8}[il] \nonumber\\
    &\tilde{\mathcal{A}}_{\text{I}} = \frac{2^{-5}N^2}{2} \int_{0}^{\infty} d\ell\frac{V_8-S_8}{\eta^8}\left[il\right]\nonumber \\
     &\tilde{\mathcal{M}}_{\text{I}} =  \frac{2 \epsilon N}{2} \int_{0}^{\infty} d\ell\frac{\hat{V}_8-\hat{S}_8}{\hat{\eta}^8}\left[il+\frac12\right]\,. \label{eq:typeItilde}
\end{align}
This transforms UV divergences in $\mathcal{K}, \mathcal{A}, \mathcal{M}$ (at $\tau_2 \rightarrow \infty$) into IR divergences in $\tilde{\mathcal{K}}, \tilde{\mathcal{A}}, \tilde{\mathcal{M}}$ (at $l\rightarrow \infty$), which correspond to the exchange of zero-momentum massless modes in the NSNS and RR closed-string sectors.  The tadpole cancellation condition then ensures that the coefficients of the woud-be divergence cancel:
\begin{equation}
    \frac{2^5}{2} + {2^{-5}N^2}{2} + \frac{2\epsilon N}{2} = \frac{2^{-5}}{2}(N+32\epsilon)^2 = 0\,,
\end{equation}
which has the unique solution $N=32$ and $\epsilon=-1$, corresponding to an SO(32) gauge group.  The spacetime interpretation is then of an O9$^-$-plane, with negative tension and negative RR charge, and 32 D9-branes, with positive tension and positive RR charge.  Indeed, the charge and tension assignments of the D9-branes compared to the O9$^-$-plane, can be read from the M\"obius amplitude describing the D9-O9 interactions: because of the overall minus sign $\epsilon=-1$ in front of both $V_8$ and $-S_8$, the D9-branes have opposite signs to the O9$^{-}$-plane, in both charge and tension.  

It is straightforward to combine the toroidal compactifications and orientifolding to obtain the one-loop vacuum amplitudes of supersymmetric toroidal orientifolds.   In this context,  T-duality is of particular relevance to us, as a way to obtain orientifolds with lower-dimensional D$p$-branes and O$p$-planes.  Indeed, in the open-string sector T-duality exchanges Neumann with Dirichlet boundary conditions. Therefore, if a T-duality is performed in a direction longitudinal to a D$p$-brane worldvolume, the
duality gives a D$(p-1)$-brane; if it is transverse to the D$p$-brane, the duality gives a D$(p+1)$-brane.  Wilson line and brane-position moduli are also appropriately interchanged.  Consider, for example, the T-dual of type I theory compactified on the circle $S^1_R$, which corresponds to the type I$'$ theory defined as the $\Omega \Pi_9$ orientifold\footnote{We remind the reader that type IIA does not possess a 10d orientifold because $\Omega$ is not a symmetry of the theory, being the theory not left-right symmetric. Instead, the 9d orbifold $\Omega\Pi_9$, with $\Pi_9$ a parity along ${S}^{1}_{R'}$, is a symmetry of the theory that can then be gauged.} of type IIA compactified on the dual circle, $S^1_{R'}$, with $R'=\alpha'/R$, where the parity operator $\Pi_9: X^9 \rightarrow -X^9$, turns the circle  $S^1_{R'}$ into the interval $S^1_{R'}/\mathbb{Z}_2 \simeq [0,\pi R']$. The type I O9$^-$-plane is then dualised into two O8$^-$-planes localised at the two endpoints of the interval and the 32 type I D9-branes\footnote{A better terminology would be 32 half-branes  on the O-plane, which would become 16 whole branes plus mirror images when moving into the bulk (and this is possible only for even numbers of half-branes).  When this terminology becomes too cumbersome, we will drop it.} are dualised into a stack of 32 D8-branes, localised on the O8$^-$-plane at the origin.

\subsection{Brane supersymmetry breaking}

In brane supersymmetry breaking \cite{Antoniadis:1999xk, Mourad:2017rrl}, an orientifold projection leaves the bulk, closed-string sector exactly supersymmetric to lowest order, but breaks supersymmetry at the string scale in the open-string sector by introducing mutually non-BPS combinations of BPS D-branes and O-planes.  The breaking of supersymmetry in these constructions is explicit; there is no order parameter to restore it and, indeed, the spectrum of open-string states has no underlying supersymmetric pairing, but rather a misaligned supersymmetry, with a mismatch between boson and fermion degeneracy at each mass-level that grows exponentially as one moves up the excitation tower  \cite{Kutasov:1990sv, Dienes:1994np, Angelantonj:2010ic, Angelantonj:2023egh, Cribiori:2020sct}.  At the same time, supersymmetry is non-linearly realised in the low-energy effective field theory \cite{Dudas:2000nv, Antoniadis:1999xk}, with a gravitino present in the low-lying closed-string spectrum, along with a massless singlet fermion amongst the open-strings playing the role of the goldstino. Typically, in brane supersymmetry breaking tachyons can be avoided thanks to the non-dynamical nature of the orientifold planes.  However, an instability still arises from a tree-level NSNS tadpole, which leads to a runaway dilaton potential in the low energy effective field theory, consistently with the non-linearly realised supergravity.  Recently a novel realisation of brane supersymmetry breaking has been developed in which the disk NSNS tadpole cancels \cite{Coudarchet:2021qwc}; we will discuss this construction in detail in the following section.

The simplest example of brane supersymmetry breaking is the ten-dimensional USp(32) Sugimoto model \cite{Sugimoto:1999tx}.  This is obtained by modding out type IIB string theory by the orientifold $\Omega (-1)^F$, with $F$ the spacetime fermion number.  The resulting amplitudes are:
\begin{align}
    &\mathcal{T}_{\text{BSB}}=\frac12 \mathcal{T}_{\text{IIB}} =  \mathcal{T}_{\text{I}}= \frac12 \int_{\mathcal F} \frac{d^2\tau}{\tau_2^2}\frac{1}{\tau_2^4}\left\lvert \frac{V_8-S_8}{\eta^8}\right\rvert \nonumber \\
    &\mathcal{K}_{\text{BSB}}  = \mathcal{K}_{\text{I}}=  \frac12 \int_{0}^{\infty} \frac{d\tau_2}{\tau_2^2}\frac{1}{\tau_2^4}\frac{V_8-S_8}{\eta^8}[2i\tau_2] \nonumber\\
    &\mathcal{A}_{\text{BSB}} = \mathcal{A}_{\text{I}} =  \frac{N^2}{2} \int_{0}^{\infty} \frac{d\tau_2}{\tau_2^2}\frac{1}{\tau_2^4}\frac{V_8-S_8}{\eta^8}\left[\frac{i\tau_2}{2}\right]\nonumber \\
     &\mathcal{M}_{\text{BSB}} =  \epsilon\frac{N}{2} \int_{0}^{\infty} \frac{d\tau_2}{\tau_2^2}\frac{1}{\tau_2^4}\frac{\hat{V}_8+\hat{S}_8}{\hat{\eta}^8}\left[\frac{i\tau_2}{2}+\frac12\right]\,.
\end{align}
Notice that all the amplitudes are the same as for the type I theory, obtained via the supersymmetric orientifolding of type IIB by worldsheet parity $\Omega$, apart from the M\"obius strip, where the fermion parity reverses the sign of the $S_8$ term. 

Similarly to the type I theory, translating to the transverse-channel reveals the tadpole cancellation conditions, now:
\begin{align}
    \text{RR:}& \quad 32 - \epsilon N = 0 \nonumber \\
    \text{NSNS:}& \quad 32 + \epsilon N = 0\,.
\end{align}
We see that, because of the sign flip of the $S_8$ term in the M\"obius strip amplitude, the RR and NSNS tadpole cancellation conditions are distinct, signalling the breaking of supersymmetry. RR tadpoles are related to inconsistencies in the field equations for the RR forms and the presence of gauge and gravitational anomalies, and hence must be cancelled.  In contrast, in the presence of NSNS tadpoles, the field equations are not inconsistent but do signal an instability away from the assumed flat vacuum.  Ensuring the mandatory RR tadpole cancellation then imposes $\epsilon=1$ and $N=32$, which corresponds to the presence of an O9$^+$-plane, with positive tension and positive charge, and the cancellation of the net RR charge with $N=32$ $\overline{\text{D}9}$-branes, with positive tension and negative charge.  Correspondingly, the NSNS tadpole is not cancelled and the one-loop vacuum energy, whose non-vanishing contributions come from the M\"obius, diverges. 

The non-vanishing NSNS tadpole leads to a disk-level contribution to the 10d string frame action (genus-1/2 dilaton 1-point function):
\begin{equation}
    S_{\text{disk}} = - \int d^{10}x \sqrt{g_{10}} T e^{-\phi} \label{eq:disktad}
\end{equation}
where $T = 64 \frac{2\pi}{(4\pi^2\alpha')^5}\equiv 64 T_{\overline{\text{D}9}}$, with $T_{\overline{\text{D}9}}$ the tension of an $\overline{\text{D}9}$-brane.  As already mentioned, the disk tadpole contribution in \eqref{eq:disktad} is actually necessary for the non-linear realisation of supersymmetry; it corresponds to the leading term in the Volkov-Akulov action.  The massless content of the theory consists of the gauge bosons in the adjoint of a USp(32) gauge group and massless fermions in the rank-2 antisymmetric representation; the latter $(496)$ decomposes into $(496)=(495) \oplus (1)$, with the singlet playing the role of the goldstino \cite{Dudas:2000nv, pr-nonlinear}.

 It will be useful to have in mind a generalisation of the annulus and M\"obius strip amplitudes for the type I and Sugimoto models given above, corresponding to $n_+$ D9-branes and $n_-$ $\overline{\text{D}9}$-branes on top of an O9$^{\mp}$-plane (see {\it{e.g.}} \cite{Angelantonj:2002ct}).  Starting from the transverse-channel:
\begin{equation} \label{eq:O9sD9s}
\begin{split}
    \tilde{\mathcal{{A}}}&=\frac{2^{-5}}{2}\int_0^\infty d\ell\,\frac{(n_++n_-)^2\,V_8-(n_+-n_-)^2\,S_8}{\eta^8}\, ,\\
   \tilde{ \mathcal{M}}&=\frac{2}{2}\int_0^\infty d\ell\,\frac{\epsilon_{\text{NS}}\,(n_++n_-)\,\hat{V}_8-\epsilon_{\text{R}}(n_+-n_-)\,\hat{S}_8}{\hat{\eta}^8}\, ,\\
 \end{split}   
\end{equation}
we allow for different signs $\epsilon_{\text{NS}}$ and $\epsilon_{\text{R}}$, which, respectively, correspond to the relative signs between the tensions and charges of the D/$\overline{\text{D}}$-branes and the O$^\mp$-plane. The tadpole cancellation conditions become:
\begin{equation}\label{tadnsr}
\begin{split}
    S_8&:\qquad 32+\epsilon_\text{R}\,(n_+-n_-)=0\\
    V_8&:\qquad 32+\epsilon_\text{NS}\,(n_++n_-)=0\,.
 \end{split}   
\end{equation}
Of course, $\mathrm{\overline{D9}}$-branes contribute negatively to the RR tadpole and positively to the NSNS tadpole. We learn more by switching back to the one-loop direct-channel
\begin{equation} 
\begin{split}
    {\mathcal{{A}}}&=\frac{1}{2}\int_0^\infty \frac{d\tau_2}{\tau_2^6}\,\frac{(n_+^2+n_-^2)\,(V_8-S_8)+2\,n_+n_-\,(O_8-C_8)}{\eta^8}\, ,\\
   { \mathcal{M}}&=\frac{\epsilon }{2}\int_0^\infty \frac{d\tau_2}{\tau_2^6}\,\frac{\epsilon_{\text{NS}}\,(n_++n_-)\,V_8-\epsilon_{\text{R}}(n_+-n_-)\,S_8}{\eta^8}\, .\\
 \end{split}   
\end{equation}
Notice that, whenever both $n_+$ and $n_-$ are non-vanishing, $O_8$ and $C_8$ characters also appear; this signals the presence of tachyons in D$-\overline{\text{D}}$ systems. Type I SO(32) superstring corresponds to the tachyon-free solution to the RR tadpole constraint in (\ref{tadnsr}) with $n_-=0$, $\epsilon_{\text{NS}}=\epsilon_{\text{R}}=-1$ and $n_+=32$.  The brane supersymmetry breaking Sugimoto USp(32) model corresponds to the tachyon-free solution $n_{+}=0$, $n_{-}=32$, $\epsilon_{\text{NS}}=\epsilon_{R}=1$, with an $\mathrm{O9}^+$-plane and 32 $\mathrm{\overline{D9}}$-branes.  We anticipate that O$p^+$-planes and O$p^-$-planes appear together in toroidal orientifold compactifications with non-trivial NSNS $B$-field turned on \cite{Witten:1997bs,Angelantonj:2002ct}, whilst $\overline{\text{O}p^\pm}$-planes appear together in orientifolds of Scherk-Schwarz compactifications.

\subsection{The Scherk-Schwarz mechanism}  The Scherk-Schwarz mechanism breaks supersymmetry with a compactification that is twisted by a symmetry which acts differently on bosons and fermions; typically an $R$-symmetry or the spacetime fermion number $(-1)^F$.    The supersymmetry-breaking, being induced by different boundary conditions for bosons and fermions, is again explicit.  However, at the level of the effective\footnote{Strictly speaking, because $M_{3/2} \sim M_{\text{KK}}$, there is no lower-dimensional effective supergravity theory.  However, as we will illustrate explicitly in Section \ref{S:modstab}, integrating in the gravitino allows an effective description of the spontaneous supersymmetry breaking.} supergravity theory, it appears as a spontaneous breaking, with the order parameter corresponding to the compactification radius, mass-splittings of would-be superpartners of order the KK scale, and supersymmetry being restored in the decompactification limit.  We now outline the implications of Scherk-Schwarz supersymmetry-breaking, first for closed-string theories and then for theories including open-strings.

\subsubsection{Closed-strings}
For example, consider string theory compactified on a circle with radius $R_9$ in the $X^9$ direction:
\begin{equation}
    X^9 \sim X^9  + 2\pi R_9\,.
\end{equation}
The torus amplitude is then given by:
\begin{equation}
    \mathcal{T}_{\text{IIB}_{S^1}} = \int_{\mathcal{F}} \frac{d^2\tau}{\tau_2^{\,\frac{11}{2}}} \left\lvert\frac{V_8-S_8}{\eta^8}\right\rvert^2 \sum_{m_9,n_9} \Lambda_{m_9,n_9}\,,
\end{equation}
where the lattices describing the closed-string along the circle take the form:
\begin{equation}
    \Lambda_{m_9,n_9}=q^{\frac{\alpha'}{4}\left(\frac{m_9}{R_9}+\frac{n_9 R_9}{\alpha'}\right)^2} \bar{q}^{{\frac{\alpha'}{4}\left(\frac{m_9}{R_9}-\frac{n_9 R_9}{\alpha'}\right)^2}}\,, \quad q=e^{2\pi i \tau}
\end{equation}
with $m_9$ and $n_9$, respectively, the KK momentum and the winding numbers.

To implement a Scherk-Schwarz compactification, one orbifolds the circle compactification by the $\mathbb{Z}_2$ symmetry $g'=(-1)^F \delta_{p_9}$, with $\delta_{p_9}$ the freely-acting momentum shift:
\begin{equation}
    \delta_{p_9}: X_L^9 \rightarrow X_L^9 + \frac{\pi}{2} R_9 \quad \text{and} \quad X_R^9 \rightarrow X_R^9 + \frac{\pi}{2} R_9   \,.
\end{equation}
Inserting the generator $\frac12(1+g')$ into the amplitudes, changes the sign of the spinorial character $S_8$ and produces a sign $(-1)^{m_9}$ in the circle lattice.  Moreover, further twisted sectors appear to restore modular invariance of the amplitude, leading to:
\begin{align}
\mathcal{T}_{\text{IIB}_{S^1/g}} &= \frac12 \int_{\mathcal{F}} \frac{d^2\tau}{\tau_2^{\,\frac{11}{2}}}\left(\left(\left\lvert \frac{V_8-S_8}{\eta^8}\right\rvert^2 + (-1)^{m_9} \left\lvert \frac{V_8+S_8}{\eta^8}\right\rvert^2 \right)\sum_{m_9,n_9} \Lambda_{m_9,n_9} \right. \nonumber \\ 
&\quad\quad \left. + \left(\left\lvert \frac{O_8-C_8}{\eta^8}\right\rvert^2 + (-1)^{m_9} \left\lvert \frac{O_8+C_8}{\eta^8}\right\rvert^2 \right)\sum_{m_9,n_9} \Lambda_{m_9,n_9+\frac12}\right)\,.
\end{align}
Finally, rescaling $R_9 \rightarrow 2 R_9$ and splitting the spacetime bosons and fermions gives:
 \begin{align}
\mathcal{T}_{\text{IIB}_{S^1/g}} &=  \int_{\mathcal{F}} \frac{d^2\tau}{\tau_2^{\,\frac{11}{2}}}\left(\left(\left\lvert V_8\right\rvert^2 + \left\lvert S_8 \right\rvert^2\right) \sum_{m_9,n_9} \Lambda_{m_9,2n_9} - \left( V_8 \bar{S}_8 + S_8 \bar{V_8} \right) \sum_{m_9,n_9}\Lambda_{m_9+\frac12,2n_9} \right.\nonumber \\ 
&\quad\quad \left. + (\left(\left\lvert O_8\right\rvert^2 + \left\lvert C_8 \right\rvert^2\right) \sum_{m_9,n_9}\Lambda_{m_9,2n_9+1} - \left( O_8 \bar{C}_8 + C_8 \bar{O_8} \right) \sum_{m_9,n_9}\Lambda_{m_9+\frac12,2n_9+1} \right) \frac{1}{\left\lvert \eta^8 \right\rvert^2}\,. \label{eq:torusSS}
\end{align}
From here we see that, as in field theory Scherk-Schwarz compactifications, all the spacetime fermions now have half-integer KK number and hence their zero-modes acquire mass, whilst spacetime bosons keep their massless modes ($m_9=0=n_9$); supersymmetry is evidently broken. The gravitino mass, in the untwisted NSR sector, sets the scale of supersymmetry-breaking, which is of order the  KK scale:
\begin{equation}
    M_{3/2} = \frac{1}{2 R_9} = M_{\text{KK}}\,.
\end{equation}
Note further that the lowest-lying state from the $O_8$ character receives a winding contribution to its mass:
\begin{equation}
    M_{\text{tachyon}}^2 = -\frac{2}{\alpha'} + \frac{R_9^2}{\alpha'^2}\,,
\end{equation}
and so the would-be tachyon is lifted so long as $R_9 > \sqrt{2\alpha'}$.  

Whilst the Scherk-Schwarz orbifolding thus far described has a natural field theory limit, string theory offers more possibilities.  As we will use below, one can also orbifold using freely-acting winding shifts, which are T-dual to the momentum shifts:
\begin{equation}
    \delta_{w_i}: X_L^i \rightarrow X_L^i + \frac{\pi\,\alpha'}{2R_i} \quad \text{and} \quad X_R^i \rightarrow X_R^i - \frac{\pi\,\alpha'}{2R_i}   \,.
\end{equation}
or, indeed, using combinations of momentum and winding shifts.

\subsubsection{Open-strings}
Scherk-Schwarz supersymmetry-breaking can also be implemented in the presence of O-planes and D-branes \cite{Blum:1997cs,Blum:1997gw,Antoniadis:1998ki,Antoniadis:1998ep}.  The pattern of supersymmetry-breaking is distinct depending on whether the branes are longitudinal to the direction of Scherk-Schwarz breaking \cite{Blum:1997cs,Blum:1997gw} or perpendicular to it \cite{Antoniadis:1998ki,Antoniadis:1998ep}.  In the longitudinal case -- called again ``Scherk-Schwarz breaking'' -- the massless D-brane spectrum manifests supersymmetry-breaking already at tree-level.  In the perpendicular case -- called ``brane supersymmetry'' -- the massless D-brane spectrum is supersymmetric at tree-level; at the same time, $\overline{\text{D}}$-branes are introduced into the background, which interact with the D-branes, and supersymmetry-breaking is eventually transmitted  to the massless modes by radiative corrections from the massive open-strings or from the closed-string sector.   Note that, in contrast to brane supersymmetry breaking set-ups, all RR and NSNS tadpoles can be vanishing in Scherk-Schwarz breaking with D-branes.  Moreover, in these models the closed sector has a softly broken supersymmetry.

\paragraph{Scherk-Schwarz breaking:}
Let us first illustrate Scherk-Schwarz breaking by considering the Scherk-Schwarz orbifold $g=(-1)^F \delta_{p_9}$ of type I theory on $S^1$ (equivalent to modding the torus amplitude \eqref{eq:torusSS} by worldsheet parity $\Omega$).  The amplitudes are found to be: 
\begin{align}
    &\mathcal{K}_{\text{IIB}_{S^1_g/\Omega}} = \frac12 \int_0^{\infty}\frac{d\tau_2}{\tau_2^{11/2}}\frac{V_8-S_8}{\eta^8}\sum_mP_m\,,\nonumber \\
     &\mathcal{A}_{\text{IIB}_{S^1_g/\Omega}} = \frac12\int_0^{\infty}\frac{d\tau_2}{\tau_2^{11/2}}  \sum_{\alpha, \beta=1}^{16} \sum_m \biggl(\frac{V_8}{\eta^8}\left( \,P_{m+a_\alpha-a_\beta}+P_{m-a_\alpha+a_\beta}+P_{m+a_\alpha+a_\beta}+P_{m-a_\alpha-a_\beta}\right)\nonumber\\
     &\qquad\qquad\qquad\qquad\qquad \qquad\qquad\quad- \frac{S_8}{\eta^8}\left ( \,P_{m+\frac{1}{2}+a_\alpha-a_\beta}+P_{m+\frac{1}{2}-a_\alpha+a_\beta}+P_{m+\frac{1}{2}+a_\alpha+a_\beta}+P_{m+\frac{1}{2}-a_\alpha-a_\beta}\right)\biggl)\,, \nonumber\\
    &\mathcal{M}_{\text{IIB}_{S^1_g/\Omega}} = -\frac12 \int_0^{\infty}\frac{d\tau_2}{\tau_2^{11/2}}\sum_{\alpha=1}^{16} \sum_m\left(\frac{\hat{V}_8}{\hat{\eta}^8} \,\left(P_{m+2a_\alpha}+P_{m-2a_\alpha}\right) - \frac{\hat{S}_8}{\hat{\eta}^8}\, \left(P_{m+\frac12+2a_\alpha}+P_{m+\frac12-2a_\alpha}\right)\right)\,.
\end{align}
where only the momentum numbers are present through the lattice $P_m \equiv \Lambda_{m,0}$ and we have allowed also for the possibility of Wilson lines on the $S^1$, which break the SO(32) with $\mathcal{W} = \text{diag}(e^{2\pi i a_\alpha}, e^{-2\pi i a_\alpha}\, ,\alpha=1, \dots, 16)$.  The spacetime interpretation of these amplitudes is an O9$^-$-plane and 32 D9-branes, giving an U(1)$^{16}$ gauge group when  generic Wilson lines are turned on\footnote{This is equivalent, in the T-dual picture, to all the D8-branes being displaced in the bulk.} $(a_\alpha\neq a_\beta\neq \{0,1/2\})$ and cancellation of both RR and NSNS tadpoles. 
Notice that, whilst the Klein-bottle sector is still supersymmetric, supersymmetry is broken in the annulus and M\"obius amplitudes due to the Scherk-Schwarz shift, $m \rightarrow m+\frac12$, in the momentum tower of the fermion $S_8$ character.  This is of course consistent with the fact that the D-branes wrap the Scherk-Schwarz direction, so the fermions on the branes are affected by the supersymmetry-breaking.

 The orientifold symmetry, acting as $a_\alpha = -a_\alpha \text{ mod }1$, ensures that the Wilson line moduli have extrema at $a_{\alpha}=0$ or 1/2. Notice that for vanishing Wilson lines all the brane fermions become massive, and -- recalling that all the bulk fermions have also been lifted by the Scherk-Schwarz twist -- the total one-loop effective potential reaches its most negative value.  Wilson lines can compensate the Scherk-Schwarz shift to leave some fermions massless and uplift the effective potential.  Interestingly, compactifying to lower dimensions, there exists some stable configurations of Wilson lines for which the effective potential can be vanishing or even positive \cite{Abel:2018zyt}.

\paragraph{Brane supersymmetry:}  A simple illustration of brane supersymmetry in Scherk-Schwarz compactifications can be made by T-dualising the type I string theory compactified on a circle to type I$'$ theory and performing a Scherk-Schwarz twist along the circle, orthogonal to the D8-branes.  The amplitudes can be obtained by starting from \eqref{eq:torusSS} and modding out by the orientifold projection $\Omega \Pi_9$.  Going immediately to the transverse amplitudes, the Klein-bottle result can be written as:
\begin{equation}
    \tilde{\mathcal{K}}_{\text{IIB}_{{S^1_g/\Omega\Pi_9}}} = \frac{2^5\sqrt{\alpha'}}{2\,R_9}\int d\ell\sum_m\left(\frac{V_8}{\eta^8}\left(\frac{1+(-1)^m}{2}\right)-\frac{S_8}{\eta^8}\left(\frac{1-(-1)^m}{2}\right)\right)P_m\,,
\end{equation}
which reveals an O8$^-$-plane at the position $X^9=0$ and an $\overline{\text{O}8}^-$-plane at position $X^9=\pi R_9'$ in the internal $S^1_{R_9'}/\mathbb{Z}_2$.  This makes sense, as -- whilst $X^9=0$ is a fixed-point of the orientifold $\Omega\Pi_9$ -- the point $X^9=\pi R_9'$ is fixed only after a $2\pi R_9'$ shift, which is dressed by the Scherk-Schwarz orbifold generator, $g$.  The $(-1)^F$ action inside $\Omega\,\Pi_9\,g$ then reverses the charge of the type I$'$ O-plane.

Notice that the total RR charge already vanishes, but the NSNS charge can also be cancelled by adding 16 D8-branes and 16 $\overline{\text{D}8}$-branes.   Choosing a locally BPS configuration, with the 16 D8-branes on top of the O8$^-$-plane and the 16 $\overline{\text{D}8}$-branes on top of the $\overline{\text{O}8}^-$-plane ensures, moreover, local RR and NSNS tadpole cancellations.  The transverse annulus describing this brane set-up is
\begin{equation}
    \tilde{\mathcal{A}}_{\text{IIB}_{{S^1_g}/\Omega\Pi_9} }= \frac{2^{-5}\sqrt{\alpha'}}{2R_9}\int d\ell\sum_m\left(\left(N_\text{D} +(-1)^m N_{\overline{\text{D}}}\right)^2 \frac{V_8}{\eta^8} - \left(N_\text{D} -(-1)^m N_{\overline{\text{D}}}\right)^2 \frac{S_8}{\eta^8} \right)P_m\,,
\end{equation}
with $N_\text{D} = 16 = N_{\overline{\text{D}}}$.  To see explicitly the brane supersymmetry, we write the open-string amplitudes in the direct-channel:
\begin{align}
     &\mathcal{A}_{\text{IIB}_{{S^1_g}/\Omega\Pi_9}} = \frac12 \int \frac{d\tau_2}{\tau_2^{11/2}}\sum_n\left(\left(N_D^2 +N_{\overline{D}}^2\right) \left(\frac{V_8 -S_8}{\eta^8}\right)W_n + N_D N_{\overline{D}}\left(\frac{O_8-C_8}{\eta^8} \right)W_{n+\frac12}\right)\,,\nonumber\\
     &\mathcal{M}_{\text{IIB}_{{S^1_g}/\Omega\Pi_9}} = -\frac12 \int \frac{d\tau_2}{\tau_2^{11/2}}\sum_n\left(N_D +N_{\overline{D}}\right) \left(\frac{\hat{V}_8}{\hat{\eta}^8} -(-1)^n\frac{\hat{S}_8}{\hat{\eta}^8}\right)W_n\,,
\end{align}
and observe that, whilst supersymmetry is broken at the massive level, the massless sector remains supersymmetric, comprising of a vector supermultiplet for the gauge group $\text{SO}(16)\times \text{SO}(16)$.  This is expected, as the D8- and $\overline{\text{D}8}$-branes are perpendicular to the Scherk-Schwarz direction, and moreover, in locally BPS -- but globally mutually non-BPS -  configurations.

\subsection{\texorpdfstring{General O$^\mp$ and D/$\overline{\text{D}}$ set-ups}{General O minus/plus and D/Dbar set-ups}}
We close this introductory discussion by presenting the one-loop vacuum amplitudes for general set-ups, containing some combinations of $\text{O}^\mp/\overline{\text{O}}^\mp$-planes and D/$\overline{\text{D}}$-branes.

While the O$p$-planes are forced to be stuck at the fixed points of the corresponding orientifold involution, D$p$-branes can be given a dynamical position in the internal torus by turning on non-vanishing expectation values for the scalars $a^i\in [0,\frac{1}{2}]$ giving the brane positions along the transverse directions $X^i=2\pi \,a^i\,\sqrt{G_{ii}}$. Because branes have to move in brane/mirror-brane pairs, only the position moduli of 16 branes along each direction are actually independent degrees of freedom and the full brane positions are encoded in the vectors $(a^i_\alpha,-a^i_\alpha),\, \alpha=1,\dots,16$. In the following we assume $d=9-p$, such that the worldvolume of the localised sources is completely orthogonal to the compact directions. When such deformations are taken into account in the lower-dimensional orientifold theory, the open-string amplitudes in the transverse-channel, which describe D$p$-branes/O$p$-planes interactions, can be written down generally as \cite{Coudarchet:2021qwc}
\begin{equation}\label{transverseamplpos}
    \begin{split}
        \mathcal{\tilde{K}}&\propto\int_0^\infty d\ell\, \sum_{\vec{m}}\,\frac{1}{\eta^8}\,\left(V_8\,\Pi^\mathcal{K}_{\text{NSNS}}-S_8\Pi^\mathcal{K}_{\text{RR}}\right)P_{\vec{m}}[\ell/2]\,,\\
         \mathcal{\tilde{A}}&\propto\int_0^\infty d\ell\, \sum_{\vec{m}}\,\frac{1}{\eta^8}\,\left(V_8\,\Pi^\mathcal{A}_{\text{NSNS}}-S_8\Pi^\mathcal{A}_{\text{RR}}\right)P_{\vec{m}}[\ell/2]\,,\\
          \mathcal{\tilde{M}}&\propto\int_0^\infty d\ell\, \sum_{\vec{m}}\,\frac{1}{\hat\eta^8}\, \left(\hat V_8\,\Pi^\mathcal{M}_{\text{NSNS}}-\hat S_8\Pi^\mathcal{M}_{\text{RR}}\right)P_{\vec{m}}[\ell/2]\,,
    \end{split}
\end{equation}
up to prefactors that can be determined by requiring that the direct-channel amplitudes have the correct interpretation of one-loop vacuum amplitudes. The projectors are defined as
\begin{equation}\label{projdef}
\begin{split}
    \Pi_{\text{NSNS}}&=\sum_{A,B}T_A\,T_B\,e^{2\pi i\vec{m}\cdot(\vec{a}_A-\vec{a_B})}\,,\\\ \Pi_{\text{RR}}&=\sum_{A,B}Q_A\,Q_B\,e^{2\pi i\vec{m}\cdot(\vec{a}_A-\vec{a_B})}\,,
\end{split}    
\end{equation}
with the sums in the Klein, annulus and M\"obius amplitudes respectively on all the possible O-plane/O-plane, D-brane/D-brane and O-plane/D-brane pairs, mirrors included, with $Q$ and $T$  their charge and tensions.  It then follows  that the interaction between two mutually BPS objects is mediated by the supersymmetric combination $V_8-S_8$, whilst the interactions of mutually non-BPS objects are described by the non-vanishing combination $V_8+S_8$.  Note that the general expression for lower-dimensional O$p^{\mp}$/$\overline{\text{O}p}^{\mp}$-planes and D$p/\overline{\text{D}p}$-branes in \eqref{transverseamplpos} extends the one given for O9$^\mp$-planes and D9/$\overline{\text{D}9}$-branes given in \eqref{eq:O9sD9s}.
 
\section{Towards a vanishing cosmological constant and dark energy} \label{S:CCeq0}
In this section, we present a non-supersymmetric string construction that is a combination of brane supersymmetry breaking and Scherk-Schwarz compactification.  The open-string sector arising from D-branes is non-supersymmetric, with supersymmetry broken already at the string scale, {\it \`a la} brane supersymmetry breaking, and can be considered a toy model for the Standard Model.  The closed-string sector has supersymmetry broken {\it \`a la} Scherk-Schwarz boundary conditions, with supersymmetry-breaking mass-splittings tied to the compactification scale.  Remarkably, in our construction, the non-supersymmetric open-string ``Standard Model'' contributions to the one-loop vacuum energy exactly cancel with those from distant ``mirror'' sectors that are ensured by the orientifold symmetry\footnote{See \cite{Angelantonj:2003hr} and \cite{carlo1} for precursor papers.}.  Moreover,  the closed-string contributions are suppressed in the Scherk-Schwarz compactification scale and can be at the scale of the observed Dark Energy for one or two large extra dimensions.  In the following section, we will present a moduli stabilisation scheme that fixes the large extra dimensions at weak string coupling in a de Sitter saddle-point.

\subsection{Supersymmetry-breaking and the cosmological constant}
It is typically argued that supersymmetry can only go so far in helping with the cosmological constant problem: we know that supersymmetry-breaking in the Standard Model sector has be at least $M_{\text{susy}} \sim \mathcal{O}(\text{TeV})$ scale, and therefore, the one-loop vacuum energy can  be protected only down to $\Lambda_{\text{open}} \sim M_{\text{susy}}^4 \sim \mathcal{O}(\text{TeV}^4)$, which is around sixty orders of magnitude too large.  However, in string theory, the interplay between brane and bulk supersymmetry leads to a more interesting discussion.

In particular, so far we have seen the following three scenarios for supersymmetry-breaking and the cosmological constant in the presence of O-planes and D-branes:
\begin{itemize}
\item Scherk-Schwarz breaking: when the D-branes sourcing the Standard Model wrap a Scherk-Schwarz supersymmetry-breaking direction, both closed and open sectors feel the supersymmetry-breaking at the Scherk-Schwarz KK scale.  The scales for the one-loop vacuum energy and the gaugini masses are tied to each other:
\begin{align}
    &\Lambda = \Lambda_{\text{closed}} + \Lambda_{\text{open}} \;\text{  with  } \;\Lambda_{\text{closed}}\sim \Lambda_{\text{open}}\sim \frac{\ms^4}{R_{\text{SS}}^4} \nonumber \\
    &M_{\text{3/2}} \sim \frac{\ms}{R_{\text{SS}}} \quad \text{and} \quad M_{\text{1/2}} \sim \frac{\ms}{R_{\text{SS}}}\,,
\end{align}
where we momentarily express the Scherk-Schwarz radius $R_{\text{SS}}$ in units of $\ms$.  In this scenario, it is impossible to obtain simultaneously a small cosmological constant and heavy gaugini masses.

\item Brane supersymmetry: when the D-branes sourcing the Standard Model are perpendicular to a Scherk-Schwarz supersymmetry-breaking direction, the massless open-string sector is supersymmetric to leading order.  The one-loop vacuum energy from open-strings therefore originates exclusively from massive modes with non-vanishing windings in the Scherk-Schwarz direction and hence is exponentially suppressed in the large radius limit.  At the same time, the supersymmetry-breaking in the bulk will be transmitted to the massless modes via radiative corrections in a gravity mediation, resulting in a suppression by inverse powers of $\mpl$ \cite{Antoniadis:mediation}.  Despite this suppression:
\begin{align}
    &\Lambda = \Lambda_{\text{closed}} + \Lambda_{\text{open}} \;\text{  with  } \;\Lambda_{\text{closed}}\sim \frac{\ms^4}{R_{\text{SS}}^4} \; \text{ and }\; \Lambda_{\text{open}}\sim \ms^4 e^{-R_{\text{SS}}/\ms}\nonumber \\
    &M_{3/2} \sim \frac{\ms}{R_{\text{SS}}}  \quad \text{and} \quad M_{\text{1/2}} \sim \frac{M_{3/2}^3}{\mpl^2}\,,
\end{align}
a sufficient hierarchy between a small cosmological constant and heavy gaugini masses remains impossible.

\item Brane supersymmetry breaking: here the bulk is supersymmetric at leading order, whilst the D-brane sector sourcing the Standard Model is non-supersymmetric at all scales.  The open-string contribution to the vacuum energy begins already at genus-1/2, from the tensions of the D-brane and O-plane sources and the corresponding NSNS dilaton tadpole.  We have then for the one-loop vacuum energy and mass of the heavy open-string states:
\begin{align}
    &\Lambda = \Lambda_{\text{closed}} + \Lambda_{\text{open}} \;\text{  with  } \;\Lambda_{\text{closed}}\sim 0 \; \text{ and }\; \Lambda_{\text{open}}\sim \gs^{-1}\ms^4 \nonumber \\
    &M_{3/2} \sim 0  \quad \text{and} \quad M_{\text{heavy}} \sim \ms\,.
\end{align}
Here, the open-string contributions to the vacuum energy are too large, already at the disk-level.
\end{itemize}

We see that Scherk-Schwarz supersymmetry-breaking leads to a one-loop vacuum energy contribution from the closed-string sector that can correspond to the observed Dark Energy scale for one or two large extra dimensions.  On the other hand, brane supersymmetry breaking leads to a non-supersymmetric open-string sector, with a high mass gap corresponding to the string scale.  If the two mechanisms can be combined, in such a way that the open-string contributions to the vacuum energy exactly cancel, we would obtain simultaneously the observed Dark Energy and no supersymmetric light states that would violate experimental bounds.

We now proceed to develop such a construction. The first step in our programme, namely the construction of brane supersymmetry breaking  open sectors with no disk tadpoles, has recently been achieved in a model by Coudarchet-Dudas-Partouche (CDP) \cite{Coudarchet:2021qwc}. Despite this tree-level cancellation, the string-scale contributions to the open-string vacuum energy are only postponed to  one-loop. Our construction will be a deformation of the CDP model, in which it is possible to achieve an exact cancellation in both massless and massive open-string sectors at one-loop\footnote{A precursor model with similar features was proposed in \cite{carlo1}.  The present construction builds upon this by achieving stability and a clear spacetime interpretation.}.

\subsection{Brane supersymmetry breaking without disk tadpoles}

Let us first review in some detail the simplest 8d model put forward in \cite{Coudarchet:2021qwc}.  The CDP model is a non-supersymmetric 8d model obtained by  modding type IIB compactified on a 2-torus ${\mathbb T}^2=S^1(R_8)\times S^1(R_9)$ by the Scherk-Schwarz like orbifold $g=(-1)^{F}\delta_{w_8}\delta_{p_9}$ and then orientifolding with the non-supersymmetric O7-projection $\Omega''=\Omega\,\Pi_8\Pi_9 (-1)^{F_L}(-\delta_{w_9})^{F}$ introduced first in \cite{carlo1}.   

The closed-string spectrum is described by the non-supersymmetric torus amplitude 
\begin{equation}\label{toruscdp}
\begin{aligned}
\mathcal{T}=\int \frac{\mathrm{d}^2 \tau}{\tau_2^5}\sum_{\vec{m},\vec{n}}\biggl\{ & \left(\Lambda_{m_8, 2 n_8} \Lambda_{m_9, 2 n_9}+\Lambda_{m_8, 2 n_8+1} \Lambda_{m_9+\frac{1}{2}, 2 n_9}\right)\left(\left|V_8\right|^2+\left|S_8\right|^2\right) \\
& -\left(\Lambda_{m_8, 2 n_8+1} \Lambda_{m_9, 2 n_9}+\Lambda_{m_8, 2 n_8} \Lambda_{m_9+\frac{1}{2}, 2 n_9}\right)\left(V_8 \bar{S}_8+\bar{V}_8 S_8\right) \\
& +\left(\Lambda_{m_8+\frac{1}{2}, 2 n_8} \Lambda_{m_9, 2 n_9+1}+\Lambda_{m_8+\frac{1}{2}, 2 n_8+1} \Lambda_{m_9+\frac{1}{2}, 2 n_9+1}\right)\left(\left|O_8\right|^2+\left|C_8\right|^2\right)
\\
&-\left(\Lambda_{m_8+\frac{1}{2}, 2 n_8+1} \Lambda_{m_9, 2 n_9+1}+\Lambda_{m_8+\frac{1}{2}, 2 n_8} \Lambda_{m_9+\frac{1}{2}, 2 n_9+1}\right)\left(O_8 \bar{C}_8+\bar{O}_8 C_8\right)\biggr\} \frac{1}{\left|\eta^8\right|^2}
\end{aligned}
\end{equation}
and shows the expected features of a Scherk-Schwarz mechanism when a KK shift in one direction is accompanied by a winding shift along a second direction: the gravitini acquire masses
\begin{equation} \label{eq:SSm32}
    M_{3/2}=\frac{1}{2R_9}\,\quad \text{or}\,\quad M_{3/2}=\frac{R_8}{\alpha'}\,, 
\end{equation}
which vanish in the two supersymmetric limits $R_8\rightarrow0$ and/or $R_9\rightarrow \infty$, and the scalar from the $O_8 \overline{O}_8$ sector
becomes tachyonic when the radii satisfy
\begin{equation}
    \frac{1}{4R_8^2}+\frac{R^2_9}{\alpha'^2}<\frac{2}{\alpha'}\, .
\end{equation}
To understand the open-string sector that results from the orbifold $g=(-1)^{F}\delta_{w_8} \delta_{p_9}$ and orientifold  $\Omega''=\Omega\,\Pi_8\Pi_9 (-1)^{F_L}(-\delta_{w_9})^{F}$ projections it is helpful to  make the following recollections. It was shown by Pradisi \cite{Pradisi:1995pp} that type IIB compactified on $\mathbb{T}^2$ and modded by the freely-acting orbifold $\delta_{w_8}\delta_{p_9}$ is equivalent to IIB compactified on $\mathbb{T}^2$ with a quantised background of the NSNS $B$-field, $B_{89}=\frac{\alpha'}{2}\mathbb{Z}$ \cite{Pradisi:Bfield}. At same time it was shown by Witten \cite{Witten:1997bs} that,
in the presence of a $B$-field background, the $\Omega\Pi_8\Pi_9(-1)^{F_L}$ orientifold of IIB yields one $\mathrm{O7^+}$ plane at the origin and three $\mathrm{O7^{-}}$ planes localised at the other fixed points\footnote{In the T-dual language, this can be seen by acting with a T-duality transformation $(R_8\leftrightarrow \frac{\alpha'}{2R_8}, R_9\leftrightarrow\frac{\alpha'}{2 R_9})$ and $(m_8\leftrightarrow n_8,m_9\leftrightarrow n_9)$ on the Klein amplitude of type I theory compactified on $\mathbb{T}^2$ with the $B$-field background and then going to the transverse-channel to read off the O-planes charge and tension assignments.}, rather than the four $\mathrm{O7}^-$ planes T-dual to the $\mathrm{O9}^-$ plane of the type I SO(32) theory. A clear consequence is that the RR tadpole is now halved and only 16 (half) D7-branes need to be introduced into the background in order to cancel the tadpole. Hence, summarising, the $\Omega \Pi_8\Pi_9(-1)^{F_L}$ orientifold of IIB on $\mathbb{T}^2$ modded by the freely acting orbifold $\delta_{\omega_8}\delta_{p_9}$ corresponds to IIB on $\mathbb{T}^2$ with 16 (half) D7-branes, one $\mathrm{O7}^+$ plane and three $\mathrm{O7}^+$ planes. When the branes are placed on top of the $\mathrm{O7}^+$ plane at the origin, towards which they are indeed attracted, the open-string spectrum describes, at the massless level,  
 a supersymmetric USp(16) gauge theory. Note that, due to supersymmetry, NSNS tadpoles are also cancelled. 
\begin{figure}
\centering
\begin{tikzpicture}[scale=1.3]

\def\L{3}

\draw[->] (\L+0.3,0) -- (\L+0.5,0) node[right] {$X^8$};
\draw[->] (0,\L+0.3) -- (0,\L+0.5) node[above] {$X^9$};


\tikzset{
  ominus/.style={fill=teal, circle, minimum size=4pt, inner sep=0pt, font=\tiny},
  obarplus/.style={fill=purple, circle, minimum size=4pt, inner sep=0pt,font=\tiny},   
  obarminus/.style={fill=cyan, circle, minimum size=4pt, inner sep=0pt,font=\tiny}  
}

\node[obarplus] (Opp) at (0,0) {$\overline{\mathrm{O7}}^{+}$};
\node[ominus] (Omp) at (0,\L) {$\mathrm{O7}^{-}$};
\node[obarminus] (Opm) at (\L,0) {$\overline{\mathrm{O7}}^{-}$};
\node[ominus] (Omm) at (\L,\L) {$\mathrm{O7}^{-}$};

\draw (Opp.east) -- (Opm.west);
\draw (Opm.north) -- (Omm.south);
\draw (Omm.west) -- (Omp.east);
\draw (Omp.south) -- (Opp.north);

\def\nBranes{8}
\def\braneStart{0.38}
\def\braneStep{0.05}



\def\wDB{0.4}        
\def\hDB{1}        
\def\dOffsetDB{0.02} 
\def\nStackDB{8}     
\def\phiDB{0}       

\begin{scope}[rotate around={\phiDB:(0,0)}]
  \foreach \i in {0,...,15} {
      \coordinate (Pos) at ($ (0,0) + (\i/30, \i*\dOffsetDB) $);

      
     \path[fill=gray!20!white, draw=black, thick, opacity=0.1]
             ($ (Pos) - (0.3,0.5)$) rectangle ++(\wDB,\hDB);}
\end{scope}
\node[below=18pt] at (0,0) {$\underbrace{}_{\mathrm{USp}(16)}$};

\end{tikzpicture}
\caption{The O-plane/D-brane configuration corresponding to the non-supersymmetric USp(16) model by CDP \cite{Coudarchet:2021qwc}.  The tension and charge combinations result in a cancellation of the disk tadpoles.  The branes are dynamically attracted to the $\overline{\text{O}7}^+$-plane, giving rise to a USp(16) gauge group.  Tadpoles arise, however, at one-loop, corresponding to a one-loop vacuum energy $\Lambda_{\text{open}}\sim M_s^4$.}
\label{fig:USp16}
\end{figure}
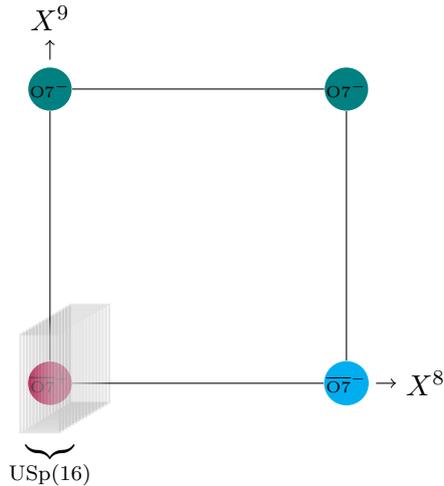

In the CDP model, the combined action of the  orbifold $g$ and orientifold $\Omega''$ projections implements a non-supersymmetric deformation of this USp(16) theory. In the closed-string sector, the deformation is Scherk-Schwarz like, with fermions acquiring tree-level masses, as already described. In the open-string sector,  the O-plane configuration is replaced according to $(\mathrm{O7^+},\mathrm{O7^-},\mathrm{O7}^-,\mathrm{O7}^-)\rightarrow (\mathrm{\overline{O7}^+},\mathrm{\mathrm{\overline{O7}}^-},\mathrm{O7}^-,\mathrm{O7}^-)$. Since the pairs $(\mathrm{O7^+},\mathrm{O7^-})$ and $(\mathrm{\overline{O7}^+},\mathrm{\overline{O7}^-})$ share the same quantum numbers, no additional charge or tension is introduced into the background, hence NSNS and RR tadpoles continue to vanish.  Supersymmetry, however, is now completely broken at the string scale, {\it{à la}} brane supersymmetry breaking, by the presence of the mutually non-BPS objects in the background. Thus, when all the $N=8$ branes are on top of the $\mathrm{\overline{{O7}}^+}$, which is the stable configuration for the brane position moduli (Wilson lines in the T-dual language) \cite{Coudarchet:2021qwc}, the open-string sector describes a Sugimoto-like non-supersymmetric USp(16) gauge theory but, remarkably, the NSNS disk tadpole is cancelled.  We illustrate the O-plane/D-brane configuration just described in Figure \ref{fig:USp16}.

For the CDP choice of brane position/Wilson line moduli, the amplitudes describing the open string sector are, in the direct-channel \cite{Coudarchet:2021qwc},
\begin{equation}\label{directCDP}
\begin{aligned}
\mathcal{K} & =\frac{1}{2} \int_0^{\infty} \frac{\mathrm{d} \tau_2}{\tau_2^5} \sum_{\vec{{n}}} W_{2 n_9} W_{2 n_8} \frac{V_8-S_8}{\eta^8}\left(2 i \tau_2\right) \,,\\
\mathcal{A} & =\frac{8^2}{2} \int_0^{\infty} \frac{\mathrm{d} \tau_2}{\tau_2^5}  \sum_{\vec{{n}}} W_{n_9} W_{n_8} \frac{V_8-S_8}{\eta^8}\left(\frac{i \tau_2}{2}\right)\,, \\
\mathcal{M} & =\frac{8}{2} \int_0^{\infty} \frac{\mathrm{d} \tau_2}{\tau_2^5}  \sum_{\vec{{n}}} W_{n_9}\left[(-1)^{n_9} W_{2 n_8}-W_{2 n_8+1}\right] \frac{\hat{V}_8+(-1)^{n_9} \hat{S}_8}{\hat{\eta}^8}\left(\frac{i \tau_2}{2}+\frac{1}{2}\right)\,.
\end{aligned}
\end{equation}
Unsurprisingly, the supersymmetry-breaking is not visible in the annulus amplitude, since it describes D7-D7 interactions, which are clearly still supersymmetric. Interestingly, the Klein-bottle amplitude also remains supersymmetric, because of a non-trivial overall cancellation of the supersymmetry-breaking contributions resulting from the particular orientifold configuration.
 The only amplitude affected by the supersymmetry-breaking is thus the M\"obius amplitude, which describes D-brane/O-plane interactions that are clearly non-BPS.

The outcome of the CDP model is thus a spontaneous supersymmetry-breaking in the bulk and  supersymmetry completely broken at the string-scale -- yet non-linearly realised -- on the worldvolume of the D7 branes, without RR and NSNS disk tadpoles.
Going to 4d, we  find the closed-string contribution to the one-loop cosmological constant from the torus amplitude (computed in Appendix \ref{A:ssvacuumenergy})  to  behave in the large $R_9$ limit as $\Lambda_{\text{closed}}\sim M_{\text{KK},9}^4\sim R_9^{-4}$, which could match the Dark Energy scale for large extra dimensions. However, the one-loop vacuum energy from the open-string sector, which is finite thanks to the cancellation of both NSNS and RR (disk) tadpoles and comes entirely from the M\"obius amplitude, is of order the string scale $\Lambda_{\text{open}}\sim \ms^{d}$ in $d$ spacetime dimensions, and thus far too large to address the Cosmological Constant Problem and allow a matching of the Dark Energy scale.

We now present a deformation of the CDP model such that the open-string sector exhibits an exact Bose-Fermi degeneracy at any mass level. This is a sufficient condition for the vanishing of  the one-loop vacuum energy  contribution from the open-sector. To describe how the model works, we first present its realisation in 8d, and then go down to 4d.

\subsection{Cancelling the open-sector one-loop vacuum energy}
\subsubsection{\texorpdfstring{The $\mathrm{USp(8)\times SO(8)}$ 8d model}{The USp(8) x SO(8) 8d model}}
\begin{figure}
\centering
\begin{tikzpicture}[scale=1.3]
\def\L{3}

\draw[->] (\L+0.3,0) -- (\L+0.5,0) node[right] {$X^8$};
\draw[->] (0,\L+0.3) -- (0,\L+0.5) node[above] {$X^9$};


\tikzset{
  ominus/.style={fill=teal, circle, minimum size=4pt, inner sep=0pt, font=\tiny},
  obarplus/.style={fill=purple, circle, minimum size=4pt, inner sep=0pt,font=\tiny},   
  obarminus/.style={fill=cyan, circle, minimum size=4pt, inner sep=0pt,font=\tiny}  
}

\node[obarplus] (Opp) at (0,0) {$\overline{\mathrm{O7}}^{+}$};
\node[ominus] (Omp) at (0,\L) {$\mathrm{O7}^{-}$};
\node[obarminus] (Opm) at (\L,0) {$\overline{\mathrm{O7}}^{-}$};
\node[ominus] (Omm) at (\L,\L) {$\mathrm{O7}^{-}$};

\draw (Opp.east) -- (Opm.west);
\draw (Opm.north) -- (Omm.south);
\draw (Omm.west) -- (Omp.east);
\draw (Omp.south) -- (Opp.north);

\def\nBranes{8}
\def\braneStart{0.38}
\def\braneStep{0.05}



\def\wDB{0.4}        
\def\hDB{1}        
\def\dOffsetDB{0.02} 
\def\nStackDB{8}     
\def\phiDB{0}       

\begin{scope}[rotate around={\phiDB:(0,0)}]
  \foreach \i in {0,...,7} {
      \coordinate (Pos) at ($ (0,0) + (\i/30, \i*\dOffsetDB) $);

      
     \path[fill=gray!20!white, draw=black,thick, opacity=0.1]
             ($ (Pos) - (0.3,0.5)$) rectangle ++(\wDB,\hDB);}
\end{scope}

\begin{scope}[rotate around={360:(\L,0)}]
  \foreach \i in {0,...,7} {
      \coordinate (Pos) at ($ (\L,0) + (-\i/30, \i*\dOffsetDB) $);

       \path[fill=gray!20!white, draw=black,thick, opacity=0.1]
             ($ (Pos) -(0.1,0.5)$) rectangle ++(\wDB,\hDB);
             }
\end{scope}

\node[below=18pt] at (0,0) {$\underbrace{}_{\mathrm{USp}(8)}$};
\node[below=18pt] at (\L,0) {$\underbrace{}_{\mathrm{SO}(8)}$};
\end{tikzpicture}
\caption{The O-plane/D-brane configuration for the USp(8)$\times$SO(8) deformation of the CDP model.  Thanks to an exact matching between the numbers of bosons and fermions at every mass level, when counting contributions from both the USp(8) and SO(8) stacks, the one-loop vacuum energy from the open-string sectors cancels exactly, $\Lambda_{\text{open}}=0$.  However, the SO(8) branes are dynamically attracted to the $\overline{\text{O}7}^+$-plane, so the configuration is unstable.  The endpoint of the instability is the CDP model illustrated in Figure \ref{fig:USp16}, which has $\Lambda_{\text{open}}\sim M_s^4$.}
\label{fig:usp8so8}
\end{figure}
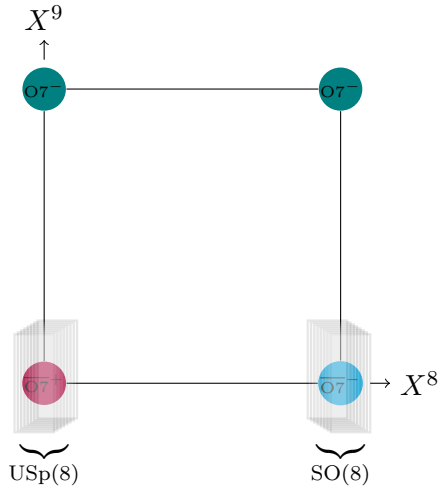

Our deformation of the CDP model starts from the observation that if one relaxes for a moment the requirement of stability of the open-string moduli and splits the 16 D7-branes into two stacks of eight D7-branes each, with one stack still on top of the $\overline{\text{O}7}^+$-plane at the origin and the second one placed on top of the $\mathrm{\overline{O7}^-}$-plane at $(\pi R_8,0)$, as depicted in Figure \ref{fig:usp8so8}, one realises a\footnote{See \cite{Angelantonj:2003hr}, \cite{carlo1} for similar, unstable, USp(8) $\times$ SO(8) constructions.  In \cite{Angelantonj:2003hr}, which involves both D-branes and $\overline{\text{D}}-$branes, the open-string Bose-Fermi matching is at the massless level only, giving rise to a suppression of the open-string one-loop vacuum energy in the large $R$ limit.  In \cite{carlo1}, whose spacetime interpretation is not known to our understanding, the open-string Bose-Fermi matching is at all mass levels.  The instabilities in \cite{Angelantonj:2003hr} are associated both with placing $\overline{\text{D}}$-brane stacks on O$^-$-planes and with having D/$\overline{\text{D}}$-pairs, whereas in \cite{carlo1} and our USp(8) $\times$ SO(8) configuration they are associated with placing D-brane stacks on $\overline{\text{O}}^-$-planes.} $\mathrm{USp(8)\times SO(8)}$ 
 configuration with an \emph{exact Bose-Fermi degeneracy at every
mass level in the open sector}. At one-loop, this is a sufficient condition for the open-sector
to not contribute to the cosmological constant. 

To support this claim, let us write down the amplitudes describing two stacks of $N_1$ and $N_2$ D7-branes respectively on top of the $\overline{\text{O}7}^+$ and $\overline{\text{O}7}^-$, with $N_1+N_2=16$. In the T-dual language of Wilson lines this brane configuration corresponds to ${\mathcal{W}}^i=\text{diag}(e^{2\pi ia^i_{\alpha}},e^{-2\pi i a_\alpha^i};\alpha=1,\dots,8)$, with $(a^8_{\alpha},a^9_{\alpha})=(0,0)$, for $\alpha=1,\dots,N_1/2$, and $(a^8_{\alpha},a_{\alpha}^9)=(1/2,0)$, for $\alpha=N_1/2+1,\dots,8$. It is  very convenient to start from the transverse channel where the geometric information is neatly encoded in the definition of the projectors (\ref{transverseamplpos}), (\ref{projdef}). For the D-brane/O-plane geometry at hand, the transverse amplitudes are
\begin{equation}\label{transverseUSpSO8}
\begin{split}
\tilde{\mathcal{K}}&=\frac{2^5\alpha'}{8 R_8 R_9}\int_0^\infty d\ell\, \sum_{\vec{m}} \frac{V_8-S_8}{\eta^8}P_{m_8}P_{m_9}\, ,\\
    \tilde{\mathcal{A}}&=\frac{2^{-5}\alpha'}{2 R_8R_9}\int_0^\infty d\ell\, \sum_{\vec{m}}\,(N_1+N_2(-1)^{m_8})^2\frac{V_8-S_8}{\eta^8}P_{m_8}P_{m_9}\,,\\
    \tilde{\mathcal{M}}&=\frac{\alpha'}{2R_8R_9}\int_0^\infty d\ell\, \sum_{\vec{m}}\left(\left(N_1\Pi_{\nsns}^{\text{USp}}+N_2\Pi_{\nsns}^{\text{SO}}\right)\hat V_8-\left(N_1\Pi_\rr^{\text{USp}}+N_2\Pi_\rr^{\text{SO}}\right)\hat{S}_8\right)\frac{1}{\hat{\eta}^8}P_{m_8}P_{m_9}\, ,
  \end{split}  
\end{equation}
where in the M\"obius amplitude, which is the only non-supersymmetric one, we have defined the projectors for the O-planes interacting with the USp($N_1$) and  SO($N_2$) stacks, respectively, as
\begin{equation}\label{projUSpSO8}
    \begin{split}
        \Pi_\nsns^{\text{USp}}&=1-(-1)^{m_9}-(-1)^{m_8}-(-1)^{m_8+m_9}\, ,\\
        \Pi_\rr^{\text{USp}}&=(-1)^{m_8}\Pi_\nsns^{\text{USp}}=\Pi_{\nsns}^{\text{SO}}\, ,\\
        \Pi_\nsns^{\text{SO}}&=-1+(-1)^{m_8}-(-1)^{m_9}-(-1)^{m_8+m_9}\, ,\\
        \Pi_\rr^{\text{SO}}&=(-1)^{m_8}\Pi_\nsns^{\text{SO}}\, .\\
    \end{split}
\end{equation}
The RR and NSNS tadpole cancellation conditions from (\ref{transverseUSpSO8}) now obviously read
\begin{equation}\label{tadpoleUSPSO8}
    \frac{2^{-5}}{2}(N_1+N_2-2^4)^2=0\implies N_1+N_2=16\, .
\end{equation}
The direct-channel amplitudes are readily found to be
\begin{equation}\label{directampliUSpSO8}
\begin{split}
\mathcal{K}&=\frac{1}{2}\int_0^\infty\frac{d\tau_2}{\tau_2^5}\, \sum_{\vec{n}}\, W_{2n_8}W_{2n_9}\frac{V_8-S_8}{\eta^8}\, ,\\
\mathcal{A}&=\frac{1}{2}\int_0^\infty\frac{d\tau_2}{\tau_2^5}\, \sum_{\vec{n}}\, \left((N_1^2+N_2^2)W_{n_8}+2N_1N_2 W_{n_8+1/2}\right)W_{n_9}\frac{V_8-S_8}{\eta^8}\, ,\\
\mathcal{M}&=\frac{1}{2}\int_0^\infty \frac{d\tau_2}{\tau_2^5}\, \sum_{\vec{n}}\, (N_1-(-1)^{n_9}N_2)(W_{2n_8}(-1)^{n_9}-W_{2n_8+1})W_{n_9}\frac{\hat{V}_8+(-1)^{n_9}\hat{S}_8}{\hat{\eta}^8}\, ,
\end{split}
\end{equation}
and, once the tadpole cancellation (\ref{tadpoleUSPSO8})  is taken into account, at the massless level they describe a non-supersymmetric USp($N_1$) $\times$ SO($16-N_1$) gauge theory, with gauge bosons and two scalars in the adjoint, alongside fermions, respectively, in the rank-2 antisymmetric and symmetric representations  $\left(\frac{N_1(N_1-1)}{2},1\right)\oplus\left(1,\frac{N_2(N_2+1)}{2}\right)$.

Apart from the closed-string torus contribution, only the M\"obius amplitude in (\ref{directampliUSpSO8}) can {\it{a priori}}  contribute to the one-loop vacuum energy, since the annulus and Klein-bottle amplitudes vanish supersymmetrically.  Crucially, the prefactor of the non-supersymmetric contribution in the M\"obius amplitude,  arising from winding states with even $n_9$ and arbitrary $n_8$, vanishes identically whenever $N_1=N_2$. This means that \emph{the open-string sector does not contribute to the one-loop cosmological constant when the USp and the SO stacks count the same number of branes, i.e. $N_1=N_2=8$.}

The origin of this cancellation  is clear from (\ref{projUSpSO8}): due to the specific O-planes tension and charge assignments, and the chosen  brane positions, the projector in the RR sector for the $\usp$ worldvolume configuration coincides with the projector in the NSNS sector for the $\So$ worldvolume configuration and vice versa,
\begin{equation}\label{idproj}
    \Pi_\rr^{\usp}=\Pi_\nsns^\So\,,\quad  \Pi_\rr^\So=\Pi_\nsns^\usp\,,
\end{equation}
so that  $\tilde{\mathcal{M}}$ in \eqref{transverseUSpSO8} indeed vanishes \textit{identically} for the choice $N_1=N_2$, via the Jacobi identity $V_8=S_8$.

 This cancellation suggests that the open-string spectrum resulting from the $\mathrm{USp(8)}\times \mathrm{SO(8)}$ configuration (see Figure  \ref{fig:usp8so8}) enjoys an exact Bose-Fermi degeneracy at every mass level, even though the spectrum is clearly non-supersymmetric.  At the massless level, the spectrum is identical to the model displayed in Section 4 of the precursor paper \cite{Angelantonj:2003hr}.  Since the model of \cite{Angelantonj:2003hr} was based on brane-antibrane pairs, whereas the one in our paper contains only branes, there are some differences at the massive level; \cite{Angelantonj:2003hr} has matching only at the massless level.  To make our structure explicit, let us rewrite
 the direct-channel annulus and M\"obius amplitudes, making explicit the bosonic and fermionic degrees of freedom at each mass level. We use the character decompositions $V_8=\sum_k d_{V,k}\,q^k$, $\hat{V}_8=\sum_k (-1)^k d_{V,k}\,q^k$, with $k$ summing on the string oscillator levels and $d_k$ their degeneracies, and similarly for $S_8$ and $\hat{S}_8$. Taking into account the alternating projections  at different $k$-levels due to the alternating sign in the degeneracies of the hatted characters in the M\"obius amplitude, and writing the resulting projections from the sum of the two amplitudes  in terms of the adjoint representations of the $\usp$ and $\So$ gauge groups, we find the full open-string spectrum to be
\begin{equation}
    \begin{split}
        \mathcal{A}+\mathcal{M}&\sim W_{2n_8}W_{2n_9}\biggl\{q^{2k}\left[d_{V,2k}\left(\adj^{\usp}(N_1)+\adj^\So(N_2)\right)-d_{S,2k}\left(\adj^{\So}(N_1)+\adj^\usp(N_2)\right)\right]\\
& \qquad\quad  +q^{2k+1}\left[d_{V,2k+1}\left(\adj^{\So}(N_1)+\adj^\usp(N_2)\right)-d_{S,2k+1}\left(\adj^{\usp}(N_1)+\adj^\So(N_2)\right)\right]\biggr\}\\
&+W_{2n_8}W_{2n_9+1}\biggl\{q^{2k}\left(\adj^{\So}(N_1)+\adj^\So(N_2)\right)(d_{V,2k}-d_{S,2k})\\
&\qquad\quad+q^{2k+1}\left(\adj^{\usp}(N_1)+\adj^\usp(N_2)\right)(d_{V,2k+1}-d_{S,2k+1})\biggr\}\\
&+W_{2n_8+1}W_{2n_9}\biggl\{q^{2k}\left[d_{V,2k}\left(\adj^{\So}(N_1)+\adj^\usp(N_2)\right)-d_{S,2k}\left(\adj^{\usp}(N_1)+\adj^\So(N_2)\right)\right]\\
&  \qquad \quad +q^{2k+1}\left[d_{V,2k+1}\left(\adj^{\usp}(N_1)+\adj^\So(N_2)\right)-d_{S,2k+1}\left(\adj^{\So}(N_1)+\adj^\usp(N_2)\right)\right]\biggr\}\\
&+W_{2n_8+1}W_{2n_9+1}\biggl\{q^{2k}\left(\adj^{\So}(N_1)+\adj^\So(N_2)\right)(d_{V,2k}-d_{S,2k})\\
& \qquad \quad +q^{2k+1}\left(\adj^{\usp}(N_1)+\adj^\usp(N_2)\right)(d_{V,2k+1}-d_{S,2k+1})\biggr\}\\
&+W_{n_8+1/2}W_{n_9}q^{k}(2N_1N_2)\,(d_{V,k}-d_{S,k})\, .
    \end{split}
\end{equation}
The states contributing positively and negatively are respectively bosons and fermions. On account of the Jacobi identity $d_{V,k}=d_{S,k}$ $\forall k$,
we clearly see that  for general $N_1$, $N_2$ the open spectrum is supersymmetric only in the winding sector with even $n_8$- odd $n_9$,  odd $n_8$- odd $n_8$ and  semi-integer $n_8$-integer $n_9$, and at any string oscillators level. By contrast,  supersymmetry is clearly broken, again at all oscillator levels, in the winding sector with arbitrary $n_8$-even $n_9$, as previously noted.   However, for the particular choice $N_1=N_2=8$, the SO(8) sector exhibits a Bose-Fermi degeneracy exactly opposite to that of the USp(8) sector. Consequently, despite complete string-scale supersymmetry breaking in each individual sector, the full open-string spectrum is exactly Bose-Fermi degenerate at every mass level, leading to a complete cancellation of its contribution to the one-loop vacuum energy.

Nevertheless, the $\text{USp}(8)\times \text{SO}(8)$ configuration is clearly unstable as the branes in the SO stack would flow back on the $\mathrm{\overline{O7}^+}$ plane at the origin, thus towards the USp(16) configuration, where a string-scale one-loop cosmological constant is unavoidably generated.  The scale associated with this instability is $M_s$, making it pathological for any cosmological application.

\subsubsection{\texorpdfstring{The $\mathrm{USp(8)\times SO(1)^8}$ 4d model}{The USp(8) x  SO(1) power 8 4d model}}\label{sec:uspso1}
To find a configuration with (perturbatively) fully stable open-string moduli, the key observation is that whenever a single D$p$-brane is placed on top of an $\text{O}p^-$- or $\overline{\text{O}p}^-$-plane, realising an SO(1) gauge group, the brane's position moduli are projected out by the orientifold and the brane must hence be rigid.  Therefore, for a single half D$p$-brane on top of an $\overline{\text{O}p}^-$, stability holds automatically.

To arrive at a configuration in which we can isolate each of the 8 D$p$-branes on a respective $\overline{\text{O}p}^-$-plane, we next compactify the  CDP 8d model \cite{Coudarchet:2021qwc} down to 4d on four more supersymmetric directions, which we then  T-dualise.  The 4d torus partition function is  straightforwardly obtained from the 8d one (\ref{toruscdp}) replacing, for each additional compact direction, a $\sqrt{\tau_2}$ factor  with the associated lattice sums, and it is given in Appendix \ref{A:ssvacuumenergy}. After the T-dualities, 
the  orientifold projection is given by the non-supersymmetric O3-projection
\begin{equation}\label{omega3cdp}
\Omega' =    \Omega \,\Pi_4\cdots\Pi_9(-1)^{F_L}(-\delta_{\omega_9})^F\, ,
\end{equation}
which introduces 64 fixed points, corresponding to $16\times (\mathrm{\overline{O3}^+},\mathrm{\overline{O3}^-})$ plane pairs and $32\times \mathrm{O3}^-$ planes, {\it{i.e.}} a replica of the CDP O-plane 2d geometry  all over the 6 internal directions. Of course, we have also 16 D3-branes to cancel all the tadpoles.

We  now place 8 D3-branes on top of the $\mathrm{\overline{O3}^+}$-plane at the origin and  stack $N_i$ D3-branes on top of the $i$-th $\mathrm{\overline{O3}^-}$-plane, with $\sum_{i=1}^{16} N_i=8$. The configuration that we obtain is thus USp(8)$\times \prod_i$ SO($N_i$). As we are about to show, this non-supersymmetric configuration still yields vanishing one-loop vacuum energy. The underlying intuition behind the cancellation is geometric:  the SO stacks are displaced in the additional four internal directions, which preserve supersymmetry. We therefore expect the transverse M\"obius amplitude to factorise, with the momenta $P_{8},P_9$ along the supersymmetry-breaking directions projected as in (\ref{transverseUSpSO8}) and the momenta $P_{4},\dots,P_7$ projected  supersymmetrically onto even KK states\footnote{We remind the reader that this is indeed the projection in the transverse M\"obius amplitude of the supersymmetric type I SO(32).}. In other words, the different SO stacks see the same O-plane distributions and contribute to the vacuum energy as an overall SO(8) stack which, as seen before, conspires with the USp(8) stack to give a vanishing one-loop cosmological constant. 

\begin{figure}[H]
    \centering
\begin{tikzpicture}[scale=1.3]
\def\a{2}        
\def\off{0.7}    

\tikzset{
  ominus/.style={fill=teal, circle, minimum size=1pt, inner sep=0pt, font=\tiny},
  obarplus/.style={fill=purple, circle, minimum size=1pt, inner sep=0pt,font=\tiny},   
  obarminus/.style={fill=cyan, circle, minimum size=1pt, inner sep=0pt,font=\tiny}  
}
\newcommand{\drawCube}[1]{%
    \pgfmathsetmacro{\shiftY}{#1*\a} 

    \node[obarplus] (FBL#1) at ($(-\off,-\off+\shiftY)$) {$\mathrm{\overline{O3}^{+}}$};
    \node[obarminus] (FBR#1) at ($( \a-\off,-\off+\shiftY)$) {$\mathrm{\overline{O3}^{-}}$};
    \node[obarplus] (FTL#1) at ($(-\off,\a-\off+\shiftY)$) {$\mathrm{\overline{O3}^{+}}$};
    \node[obarminus] (FTR#1) at ($( \a-\off,\a-\off+\shiftY)$) {$\mathrm{\overline{O3}^{-}}$};

    \draw[] (FBL#1.east) -- (FBR#1.west);
    \draw[] (FBR#1.north) -- (FTR#1.south);
    \draw[] (FTR#1.west) -- (FTL#1.east);
    \draw[] (FTL#1.south) -- (FBL#1.north);

    \node[ominus] (BBL#1) at ($(0,0+\shiftY)$) {$\mathrm{O3^-}$};
    \node[ominus] (BBR#1) at ($( \a,0+\shiftY)$) {$\mathrm{O3^-}$};
    \node[ominus] (BTL#1) at ($(0,\a+\shiftY)$) {$\mathrm{O3^-}$};
    \node[ominus] (BTR#1) at ($( \a,\a+\shiftY)$) {$\mathrm{O3^-}$};

    \draw[dashed] (BBL#1.east) -- (BBR#1.west);
    \draw[dashed] (BBR#1.north) -- (BTR#1.south);
    \draw[dashed] (BTR#1.west) -- (BTL#1.east);
    \draw[dashed] (BTL#1.south) -- (BBL#1.north);

    \draw[dashed] (FBL#1) -- (BBL#1);
    \draw[dashed] (FBR#1) -- (BBR#1);
    \draw[dashed] (FTL#1) -- (BTL#1);
    \draw[dashed] (FTR#1) -- (BTR#1);
}

\foreach \i in {0,1,2}{
    \drawCube{\i};
}

\draw[->] (FBR0.east) -- ++(0.8,0) node[right] {$X^8$};
\coordinate (X9vec) at ($ (FBL0) - (BBL0) $);

\draw[->,dashed] (BBL0) -- ($ (BBL0) - 0.8*(X9vec) $) node[left] {$X^9$};

\draw[->] (FTL2.north) -- ++(0,0.8) node[above] {$T^4$};

\foreach \nodeName in {FTR2,  BTR2, BTL2}{
    \node at ($(\nodeName.north) + (0,0.3)$) {$\vdots$};
}







\def\w{0.5}      
\def\h{1}      
\def\N{8}        
\def\phi{0}     
\def\dOffset{0.02} 

\begin{scope}[rotate around={\phi:(FBL0)}]
    \foreach \i in {0,...,7} {
        \coordinate (Pos) at ($ (FBL0)!{\i/30}!(FBR0) + (0,\i*\dOffset) $);

        \path[fill=gray!20!white, draw=black,thick, opacity=0.2]
             ($ (Pos) - (0.5,0.5)$) rectangle ++(\w,\h);
    }
\end{scope}



\def\braneAngle{-10}

\foreach \nodeName/\pivot in {
  FBR0/FBL0, FTR0/FBL0,
  FBR1/FBL1, FTR1/FBL1,
  FBR2/FBL2, FTR2/FBL2} { 
  \begin{scope}[rotate around={\braneAngle:(\pivot)}]

    \coordinate (P) at ($(\nodeName.north)+(0,0.1)$);

    \path[
      fill=gray!40!white,
      draw=black,
      thick,
      opacity=0.2
    ]
      ($ (P) - (0.2,0.8)$) rectangle ++(\w,\h);
  \end{scope}
}
\node[below=15pt] at (FBL0) {$\underbrace{}_{\mathrm{USp}(8)}$};
\node[below=15pt] at (FBR0) {$\underbrace{}_{\mathrm{SO}(1)}$};
\node[below=15pt] at (FTR0) {$\underbrace{}_{\mathrm{SO}(1)}$};
\node[below=15pt] at (FTR1) {$\underbrace{}_{\mathrm{SO}(1)}$};
\node[below=15pt] at (FTR2) {$\underbrace{}_{\mathrm{SO}(1)}$};

\end{tikzpicture}
 \caption{The O-plane/D-brane configuration for our USp(8) $\times$ SO(1)$^8$ construction, which has $\Lambda_{\text{open}}=0$.  After dimensionally reducing and T-dualising, there are 16 copies of the CDP distribution of O$^-$/$\overline{\text{O}}^\mp$-planes shown in Figure \ref{fig:USp16} (we show 4 of them).  Together with the USp(8) brane stack on one of the $\overline{\text{O}3}^+$-planes, there are eight single SO(1) branes placed on separate $\overline{\text{O}3}^-$-planes.  The numbers of bosons and fermions coming from the brane stacks match at all mass levels.  The configuration is moreover stable, as the SO(1) branes are stuck on the $\overline{\text{O}3}^-$-planes.}
    \label{fig:usp8so1}
\end{figure}
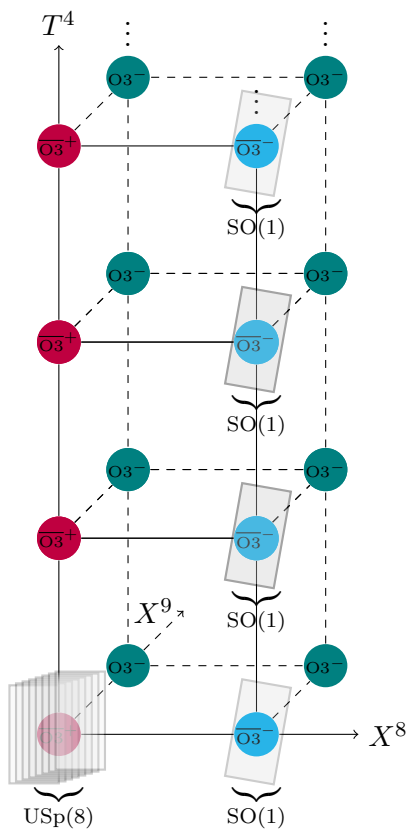

To confirm our expectation, we write down the amplitudes for the USp(8)$\times \prod_i$ SO($N_i$) configuration. We first need to assign a position to the D3-branes and the O3-planes. The position of a fixed
point along the direction $I \in\{4, \dots, 9\}$ is $X^I = 2\pi
\sqrt{G_{II}} z^I$
, with $z^
I = \{0, 1/2\}$. We enumerate the 64 fixed points with a label $A \in \{0, \dots , 63\}$ 
and place an $\mathrm{\overline{O3}^+}$ plane whenever $A=0\,\text{mod}\,4$, an $\mathrm{\overline{O3}^-}$-plane  whenever $A=2\,\text{mod}\,4$, and $\mathrm{O3^-}$ planes whenever $A=1,3\,\text{mod}\,4$. With this convention, the position vector $2\vec{z}$ of the fixed point labelled by $A$, hence of the O-plane sat there, is straightforwardly obtained by writing $A$ in binary\footnote{For example, within this ordering, to the 3rd fixed point it is associated $A=2=(0,0,0,0,1,0)$, so it is occupied by an $\mathrm{\overline{O3}^-}$ plane and its position vector is $\vec{z}=(0,0,0,0,1/2,0)$.
}. 
The transverse-channel amplitudes 
for the USp(8)$\times \prod_i$ SO($N_i$) configuration are
\begin{equation}
\begin{split}
\tilde{\mathcal{K}}&=\frac{2^5\alpha'^{3}}{8 R_4\cdot\cdot R_9}\int_0^\infty d\ell\sum_{\vec{m}}\left[\prod_{s=4}^7\frac{1+(-1)^{m_s}}{2}\right]P_{m_4}\cdot\cdot P_{m_9}\frac{V_8-S_8}{\eta^8}\\
\tilde{\mathcal{A}}&=\frac{2^{-5}\alpha'^{3}}{2 R_4\cdot\cdot R_9}\int_0^\infty d\ell\sum_{\vec{m}}\left(8+\sum_{i=1}^{16}N_i (-1)^{2\,\vec{m}\cdot \vec{z}_i}\right)^2P_{m_4}\cdot\cdot P_{m_9}\frac{V_8-S_8}{\eta^8}\\\
    \tilde{\mathcal{M}}&=\frac{\alpha'^{3}}{2R_4\cdot\cdot R_9}\int_0^\infty d\ell\sum_{\vec{m}}\left[\hat{V}_8\left(8\,\Pi_\nsns^{\text{USp}}+(\sum_{i=1}^{16} N_i)\Pi_{\nsns}^{\text{SO}}\right)-\hat{S}_8\left(8\,\Pi_\rr^{\text{USp}}+(\sum_{i=1}^{16} N_i)\Pi_{\rr}^{\text{SO}}\right)\right]\times\\
    &\qquad\qquad \times \left[\prod_{s=4}^7\frac{1+(-1)^{m_s}}{2}\right]\frac{1}{\hat{\eta}^8} P_{m_4}\cdot\cdot P_{m_9}
\end{split}
\end{equation}
with $\vec{z}_i$ the  position vector  of the $i$-th $\mathrm{\overline{O3}^-}$-plane, hence
of the $N_i$ branes stacked on top of it.
 The M\"obius amplitude indeed factorises. Using the projector identities (\ref{idproj}) we straightforwardly see that this amplitude vanishes for $\sum_{i=1}^{16} N_i=8$, which is also the condition to cancel the tadpoles.   Taking into account brane dynamics, which we study in detail in Appendix \ref{A:openstringmod}, we conclude that the only stable configuration which yields vanishing cosmological constant is $\mathrm{USp(8)\times SO(1)^8}$, depicted schematically in Figure \ref{fig:usp8so1}.  

The massless open-string spectrum contains gauge vectors and six scalars in the adjoint representation of $\mathrm{USp(8)}$, four Weyl fermions in the antisymmetric representation of  $\mathrm{USp(8)}$, and four Weyl fermions for each $\mathrm{SO(1)}$ factor. There is an excess of bosons in the non-Abelian 
$\mathrm{USp(8)}$ brane stack, whose contribution to vacuum energy is however exactly cancelled by the fermions coming from the $\mathrm{SO(1)}$ factors. Overall, there is therefore Bose-Fermi degeneracy at the massless level. As done for the 8d $\mathrm{USp(8)}\times \mathrm{SO(8)}$ configuration, it can be thus  confirmed that  the Bose-Fermi open string degeneracy continues to hold to all massive levels.  

The 4d  $\mathrm{USp(8)\times SO(1)^8}$ model we just described goes some way towards a  symmetry mechanism to address the Cosmological Constant Problem: the light gauge and matter fields coming from the principle USp(8) stack of D-branes --  whose relation with the Standard Model will be commented on in Section \ref{sec:freelyactingorbifold} and the outlook -- are not supersymmetric, and yet their contribution to the one-loop vacuum energy exactly cancels against that of hidden sectors with no mutual gauge-charges, sequestered from the visible sector by a geometrical separation, but guaranteed to be present by the orientifold symmetry.  Moreover, the interplay between the brane supersymmetry breaking and Scherk-Schwarz compactification provides a first string realisation of the Dark Dimension \cite{Montero:dd} and Supersymmetric Large Extra Dimensions \cite{Aghababaie:2003wz} proposals, while explaining how the open-string contributions to the one-loop vacuum energy exactly cancel without the need for light visible-sector superpartners; the closed-string contributions can then be at the scale of the observed Dark Energy when there are one or two large extra dimensions (which we will dynamically explain below).  Indeed, the torus partition function computed in Appendix \ref{A:ssvacuumenergy} gives, in the two large extra dimension case with  $R_8\simeq R_9\gg1$ and in string units (cfr. (\ref{V1loopR8R9largest}))\footnote{With an eye towards moduli stabilisation the computation has been carried out 
concretely for a $T^6/\mathbb{Z}'_2\times \mathbb{Z}'_2$ orbifold, where $n_b^{0}-n_f^0=16$.}
\begin{equation}\label{v1loop2}
V_{\text{1-loop}}=-M_s^4\cdot\lambda_{s,\text{2LED}}\cdot\frac{1}{(\text{Im}T'_3)^2}\mathcal{E}_{3}(U_3)\,,\quad \lambda_{s,\text{2LED}}\equiv\frac{(n_b^{0}-n_f^0)}{2^3\pi^7}\simeq(n_b^{0}-n_f^0)\cdot4.3\times10^{-5}\,,
\end{equation}
which is of Casimir-type with $n_b^0-n_f^0$ the difference in the number of bosons and fermions at the massless level in the closed-string spectrum, $\text{Im}(T'_3)=\sqrt{\det(g_3)}=R_8 R_9\sin(\omega_{89})$ and $U_3$ are respectively the  real part of the geometric K\"ahler modulus and the  complex structure modulus associated to  the susy-breaking torus (defined as in (\ref{UT - T}) and (\ref{UT - U})), and $\mathcal{E}_3 (U)$ is  the  non-holomorphic  $\Gamma_1(2)$ invariant weight-0 series
\begin{equation}\label{Epsilon3}
    \mathcal{E}_3(U)=\sum_{p,q\in\mathbb{Z}}\frac{\text{Im}(U)^3}{|p+(2q+1)U|^6}\, ;
\end{equation}
or, in the one large extra dimension case with $R_9\gg1$ (cfr. (\ref{V1loopR9largest}))
\begin{equation}\label{v1loop1}
     V_{\text{1-loop}}=-M_s^4\cdot\lambda_{s,\text{1LED}}\cdot\frac{1}{(R_9\sin\omega_{89})^4}\,,\quad \lambda_{s,\text{1LED}}\equiv \frac{93\zeta(5)}{2048\pi^5}\cdot (n_b^{0}-n_f^0)\simeq (n_b^{0}-n_f^0)\cdot1.5\times 10^{-4}\,.
\end{equation}
It is evident  that both (\ref{v1loop2}) and  (\ref{v1loop1})  agree with the scaling conjectured in the Supersymmetric Large Extra Dimension scenario \cite{Aghababaie:2003wz} and the Dark Dimension scenario \cite{Montero:dd}: $V\sim \lambda\,M_{\text{KK}}^4$, with $M_{\text{KK}}$ the KK scale of the dark dimensions, where either one or both the Scherk-Schwarz directions correspond to a dark dimension. We remark,  moreover, that the cancellation of the open-sector one-loop vacuum energy that we manage to achieve does not need any large-$R$ limit; rather it holds identically at any point of the closed-string moduli space.

Of course, it should be mentioned here that the gauge group USp(8) would need to be broken to that of the Standard Model and the Standard Model chiral fermions need to be identified: we comment on this in the next section. Another pressing problem is that the volume of the Scherk-Schwarz torus, together with the dilaton after transforming \eqref{v1loop2}-\eqref{v1loop1} to the 4d Einstein frame, have steep runaway potentials, which would be cosmologically unstable.  We will now turn to the stabilisation of these moduli -- and the other closed-string moduli -- in a way that 1) is consistent with perturbative $\gs-$ and $\alpha'$-expansions 2) gives large dark dimensions, and 3) provides a dynamical Dark Energy scenario {\it \`a la} hilltop quintessence.

\section{Moduli stabilisation}\label{S:modstab}
Having exactly cancelled the open-string contributions to the one-loop vacuum energy, and ensured that the open-string moduli are stable, we now turn in detail to the closed-string sector.   A powerful aspect of the worldsheet analysis used in the previous sections is that -- whilst we worked to one-loop in the string-coupling expansions -- our results for the vacuum energy and gravitino mass were exact to all finite orders in $\alpha'$.   To make further progress, in this section, we turn to an effective field theory (EFT) description.   We first show how the Scherk-Schwarz supersymmetry breaking in string theory can be described in the language of 4d $\mathcal{N}=1$ IIB supergravity. Thereafter, we shall present a moduli stabilisation scenario in which the interplay between spontaneous Scherk-Schwarz supersymmetry-breaking in the bulk, and further field-theoretic non-perturbative contributions with a clear  4d $\mathcal{N}=1$ description-- namely D(-1)-instantons and Euclidean D3-branes -- fix all the closed-string bulk moduli with controlled $\gs$- and $\alpha'$-expansions, large extra dimensions and at a de Sitter saddle point whose scale matches that of the observed Dark Energy without the need for fine-tuned cancellations.

Our requirement of 4d $\mathcal{N}=1$ supersymmetry, as a convenient tool,  forces the internal space to have $SU(2)$ holonomy.  In contrast, our 4d $\mathrm{USp(8)\times SO(1)^8}$  O3 orientifold model described above, where a Riemann-flat internal geometry is assumed,  has an underlying 4d $\mathcal{N}=4$ supersymmetry, broken spontaneously by the Scherk-Schwarz boundary conditions directly to 4d $\mathcal{N}=0$.   We will introduce an intermediate breaking of $\mathcal{N}=4$ to $\mathcal{N}=1$ via a further orbifolding.  It is important that this intermediate breaking does not spoil the exact cancellation of the open-string one-loop vacuum energy, {\it{e.g.}} due to the introduction of additional D-branes and O-planes as orbifold twisted sectors.  To this purpose, we assume our string construction to be replicated in a so-called shift-orbifold ${\mathbb T}^6/\mathbb{Z}'_2\times \mathbb{Z'}_2$ \cite{dudasshift}, which is the freely-acting counterpart of the $T^6/\mathbb{Z}_2\times \mathbb{Z}_2$ toroidal orbifold: each   of the $\mathbb{Z}_2'$ are made freely-acting thanks to the inclusion of momentum or winding shifts on top of the usual $\mathbb{Z}_2$ orbifold twists. As a consequence, the twisted sectors of  such orbifolds are lifted and no D7-brane/O7-plane sectors are introduced, in contrast to the standard $T^6/\mathbb{Z}_2\times \mathbb{Z}_2$ case. At  the same time the freely-acting orbifold retains the same salient features of its standard counterpart, in particular giving 4d $\mathcal{N}=1$ supersymmetry and sharing the same untwisted closed-string moduli.

\subsection{\texorpdfstring{The  4d $\mathcal{N}=1$ set-up}{The  4d N=1 set-up}}\label{sec:freelyactingorbifold}
We introduce our set-up for moduli stabilisation by presenting  an explicit implementation of a freely-acting $\mathbb{Z}'_2 \times \mathbb{Z}'_2$ operation that allows an intermediate 4d $\mathcal{N}=1$ supersymmetry, leaving for future work the detailed dovetailing to our construction for cancelling the open-string one-loop cosmological constant. We start from the standard $\mathbb{Z}_2\times\mathbb{Z}_2$ orbifold of ${\mathbb T}^6$, with the generators $\theta_1$ and $\theta_2$, corresponding to $\pi$-rotations, acting on the ${\mathbb T}^6$ complex coordinate $z_i$ $i=1,2,3$  as:
\begin{equation}\label{z2z2}
    \begin{split}
        \theta_1&:(z_1,z_2,z_3)\rightarrow (-z_1,-z_2,z_3)\, ,\\
        \theta_2&:(z_1,z_2,z_3)\rightarrow (z_1,-z_2,-z_3)\, ,\\
        \theta_1\,\theta_2&:(z_1,z_2,z_3)\rightarrow (-z_1,z_2,-z_3)\, .
    \end{split}
\end{equation}
 We also consider the combined momentum-winding shift 
 $\delta_{p_iw_i}$  acting as \cite{carlo2} 
 \begin{equation}
X_L^i \to X_L^i + \frac{\pi R_i}{2}
+ \frac{\pi \alpha'}{2R_i} \quad , \quad 
X_R^i \to X_R^i + \frac{\pi R_i}{2}
- \frac{\pi \alpha'}{2R_i} \ . \label{free-act1}
 \end{equation}
The corresponding action on lattice states is
\begin{equation}
\delta_{p_iw_i} \ |p_L,p_R \rangle
= (-1)^{m_i+n_i} \ |p_L,p_R \rangle
\ . \label{free-act2}
\end{equation}
We then define the three freely-acting operations according to
\begin{eqnarray}
&& g = (\delta_{p_4w_4} \delta_{p_5w_5},-1,- \delta_{p_8w_8} \delta_{p_9w_9}) \ , \nonumber \\
&& f = (- \delta_{p_4w_4} \delta_{p_5w_5}, \delta_{p_6w_6} \delta_{p_7w_7},-1) \ , \nonumber \\
&& h = (-1, - \delta_{p_6w_6} \delta_{p_7w_7}, \delta_{p_8w_8} \delta_{p_9w_9})  \ , \label{free-act3}
\end{eqnarray}
where $-1$ corresponds to a  $\pi$ rotation in the corresponding complex coordinate, whereas $- \delta_{p_8w_8} \delta_{p_9w_9}$ (for example) corresponds to a rotation by $\pi$ in the third torus, accompanied by a combined momentum and winding shift in the two coordinates of the same torus. 
Taking into account that our starting point is based on an orbifold $g' = (-1)^F \delta_{w_8} \delta_{p_9}$ and the orientifold projection $\Omega' = \Omega (-1)^{F_L} \Pi_{8}\Pi_{9} (-\delta_{w_9})^F$, then according to the analysis made in \cite{dudasshift} these operations are indeed freely-acting, in the sense that they have no fixed points, and therefore they do not induce additional O-planes. One can further check that the operations $g'g$, $g'f$ and $g'h$ are also freely-acting. Consequently, no additional background D-branes are needed for consistency either. The action of these orbifold operations on the zero-modes are like the standard $\mathbb{Z}_2 \times \mathbb{Z}_2$ orbifold operation, acting as a truncation of the closed-string sector and without adding a twisted sector\footnote{More precisely, the twisted sector is massive. However, according to the analysis in \cite{carlo2}, in the sectors $g'g$ and $g'h$ there are scalars that become tachyonic close to the self-dual radii of the first and second torus. If moduli stabilization happens for values not too close to these self-dual values, as will happen below, these states are massive.  In all other twisted sectors, twisted states are massive for all values of moduli fields.}.

Let us also analyse the action of the orbifold on the open string Chan-Paton degrees of freedom, in the specific model of interest with gauge group $\textrm{USp(8)} \times \textrm{SO(1)}^8$.
The orbifold breaks generically the symplectic gauge group $\textrm{USp(8)} \to 
\textrm{USp}(n_1) \times \textrm{USp}(n_2) \times \textrm{USp}(n_3) \times \textrm{USp}(n_4)$, with $n_1 +n_2 + n_3 + n_4=8$. No other conditions on $n_i$ arise, since the twisted sectors are massive. The action of the orbifold operations on the Chan-Paton factors are accordingly
\begin{eqnarray}
&& \gamma_g= (1_{n_1},1_{n_2},-1_{n_3},-1_{n_4}) \ , \nonumber \\
&& \gamma_f= (1_{n_1},-1_{n_2},1_{n_3},-1_{n_4}) \ , \nonumber \\
&& \gamma_h= (1_{n_1},-1_{n_2},-1_{n_3},1_{n_4})
\ . \label{free-act4}
\end{eqnarray}
We are interested for phenomenological reasons in two stacks with $n_1=2,n_2=6,n_3=n_4=0$, breaking therefore $\textrm{USp}(8) \to \textrm{USp}(2) \times \textrm{USp}(6)$.  The four-dimensional massless spectrum consists of gauge vectors in the adjoint representation of the two gauge groups,  Weyl fermions in the representation $(1,1)\oplus (1,15) \oplus (2,6)$ and  two real scalars in $(2,6)$. 
The massless spectrum on the hidden-sector $\textrm{SO}(1)^8$ gauge factors consists of eight Weyl fermions, one from each gauge factor. It is readily checked that the Bose-Fermi degeneracy is preserved by this orbifold, so the cancellation of the vacuum energy in the open sector is not spoiled.  It is possible to finally break $\textrm{USp}(6) \to \textrm{U}(3)$ by a Wilson line. The final gauge group would therefore be a Standard-Model like $\textrm{USp}(2) \times \textrm{U}(3) = \textrm{SU}(2) \times \textrm{SU}(3) \times \textrm{U}(1)$, accompanied by a hidden sector $\textrm{SO}(1)^8$. The final spectrum is however not exactly the one of the Standard Model. In addition, a na\"ive Wilson line breaking $\textrm{USp}(6) \to \textrm{U}(3)$ would spoil the one-loop vacuum energy cancellation in the open sector. We leave the construction of a more realistic model for future work.    
 
We can now focus on the closed-string moduli space. As we have discussed, the freely acting $T^6/\mathbb{Z}_2'\times \mathbb{Z}_2'$ orbifold acts on the closed-string spectrum as its standard counterpart $\mathbb{Z}_2\times \mathbb{Z}_2$, with the further bonus of not introducing closed-string twisted moduli at all. The resulting space is thus a smooth Calabi-Yau 3-fold  with  $h^{1,1}=h^{2,1}=3$, so vanishing Euler characteristic.  Since moreover the non-supersymmetric orientifold $\Omega'$ (\ref{omega3cdp}) acts as a standard O3 orientifold  on the zero-modes, it then follows that IIB compactified on  the $T^6/\mathbb{Z}_2'\times \mathbb{Z}_2'$ $\Omega'$ orientifold yields a low energy 4d $\mathcal{N}=1$ supergravity theory identical to that arising from the untwisted sector of $T^6/\mathbb{Z}_2\times \mathbb{Z}_2$.  The latter is very well-known (reviewed in Appendix \ref{A:IIBCY3}) and will provide the framework for our EFT.

The 4d $\mathcal{N}=1$ low energy effective action from type IIB CY$_3$ orientifold compactifications has been thoroughly described in \cite{Grimm:2004uq}. The supergravity tree-level moduli space factorises into the complex structure, K\"ahler and axio-dilaton moduli space
\begin{equation}\label{modulispace}
\mathbb{M}=\mathbb{M}_{\text{cs}}\times\mathbb{M}_{\text{kah}}\times \mathbb{M}_{\text{dil}}\, ,
\end{equation}
and the good holomorphic coordinates on $\mathbb{M}$ are the axio-dilaton $S$, $h^{2,1}$ geometric complex structure moduli $U_i$ and $h^{1,1}$  K\"ahler moduli $T_i$, whose expressions for $T^6/\mathbb{Z}'_2\times \mathbb{Z}'_2$ are (see Appendix \ref{A:IIBCY3})
\begin{equation} \label{eq:sugramoduli}
    \begin{split}
        S&=e^{-\phi}+i \,C_0\equiv \frac{1}{2}s+i\,\theta_s\, ,\\
        U_i&=\frac{\sqrt{\det g_{(i)}}}{g_{(i)11}}+i\frac{g_{(i)12}}{g_{(i)11}}\equiv \frac{1}{2}u_i+i\,\theta_{u_i}\,,\quad\\
    T_i&=e^{-\phi}(\det g_{j}\cdot \det g_{k})^{1/2}+i a_i\equiv\frac{1}{2}t_i+i\,\theta_{t_i}\, ,
    \end{split}
\end{equation}
with $i=1,\dots,h^{1,1}=h^{2,1}=3$, $g_i$ the string-frame metric on the $i$th 2-torus $\mathbb{T}_i$ and $a_i$ the axions from the dimensional reduction of the RR $C_4$ potential.
The tree-level factorisation (\ref{modulispace}) implies  that the field space metric is block diagonal and the tree-level K\"ahler potential of the 4d $\mathcal{N}=1$ action is the sum of the three contributions,
\begin{equation} \label{eq:Ktree}
\begin{split}
    K_\tree&= K_{\text{cs}}+K_{\text{kah}}+K_{\text{dil}}\\
    &=-\sum_{i=1}^3\log (u_i)-\sum_{i=i}^3\log(t_i)-\log(s)\, .
 \end{split}   
\end{equation}
For the following it is useful to keep in mind how the $U_i$ and $T_i$ moduli are related to the string-frame radii $(R_4,\dots,R_9)$ of the compactification 
\begin{equation}\label{moduliradi}
    \begin{split}
        &U_1=i\frac{R_5}{R_4}e^{-i\,\omega_{45}}\,,\quad U_2=i\frac{R_7}{R_6}e^{-i\,\omega_{67}}\,,\quad U_3=i\frac{R_9}{R_8}e^{-i\,\omega _{89}}\,,\\
        &t_1= 2\,e^{-\phi}R_{6}R_7R_8R_9\sin \omega_{67}\sin\omega_{89}\,,\\
        &t_2= 2\,e^{-\phi}R_{4}R_5R_8R_9\sin \omega_{45}\sin\omega_{89}\,,\\
         &t_3= 2\,e^{-\phi}R_{4}R_5R_6R_7\sin \omega_{45}\sin\omega_{67}\, .
    \end{split}
\end{equation}
Here, and in the remainder of this section, $R_a$ ($a=4,\dots,9$) are dimensionless and  in units of\footnote{In our EFT conventions the gravitino mass  from antiperiodic boundary conditions along direction $9$ reads $M_{3/2}=M_s/(4\pi R_9\sin \omega_{89})$ and the string-frame volume of the internal $\mathbb{T}^6$ is $\mathcal{V} = \ell_s^6 \sin\omega_{45}\sin\omega_{67}\sin\omega_{89}\prod_{a=4}^9 R_a$. The Weyl rescaling to the Einstein frame is done in 4d.  See \cite{ValeixoBento:2025emh} for a further discussion of  conventions.} $\ell_s=2\pi \sqrt{\alpha'}$, while the string theory computations in Appendix \ref{A:ssvacuumenergy} assumed the more worldsheet-friendly conventions of radii in units of $\sqrt{\alpha'}$: accounting for the $(2\pi)$-factors we thus find that the prefactor $\lambda_s$ of the Casimir-energy, computed in Appendix \ref{A:ssvacuumenergy}, within the EFT conventions is
\begin{equation}\label{eftconv1}
\lambda_s^{\ell_s}:=\frac{\lambda_s}{(2\pi)^4}\, .    
\end{equation}  
The 4d Planck mass and the string scale are related in terms of the moduli as
\begin{equation}\label{planckmass4dn1}
    \mpl^2=\frac{1}{4\pi}(st_1t_2t_3)^{1/2}\ms^2\,.
\end{equation}
We aim to stabilise these seven complex moduli $S, T_i,U_i$, $i=1,2,3$.
We remind the reader that for the  two dark dimensions/Supersymmetric Large Extra Dimensions scenario to be reproduced,  the stabilisation should fix $t_1,t_2$ at exponentially large scales and $u_3$ at $\mathcal{O}(1)$, with $R_8,R_9$ the radii of the would-be mesoscopic dark dimensions corresponding to the two directions inside $\mathbb{T}^2_{3}$. For just one dark dimension, instead, we need a stabilisation of $t_1$, $t_2$ -- and also $u_3$ -- at exponentially large values, with $R_9$ then the radius of the Dark Dimension corresponding to direction 9 in $\mathbb{T}^2_{\text{3}}$.

\subsection{Scherk-Schwarz mechanism as a no-scale IIB supergravity}
\label{s:SSsugra}
We have seen that Scherk-Schwarz supersymmetry-breaking leads -- at string tree-level -- to a vanishing cosmological constant, and the scale of supersymmetry-breaking, {\it{i.e.}} the gravitino mass $M_{3/2}= \ms/(4\pi R_{\text{9}}\sin\omega_{89}) \sim M_{\text{KK}}^{(9)}$ undetermined but tied to the KK scale of the 9th-direction. Moreover, at one-loop, a runaway potential of order $\mathcal{O}(M^4_{3/2})$ is generated.  Vanishing tree-level cosmological constant despite supersymmetry-breaking, with the scale of the latter unfixed at tree-level, is what defines  \textit{no-scale models} \cite{noscale-cremmer}, \cite{noscale-ellis}. The appearance of a (runaway) potential at one-loop then indicates that the tree-level no-scale cancellation is lifted by one-loop corrections.

Strictly speaking, since the gravitino mass is of the same order as the KK cutoff, it ought to be integrated out when dimensionally-reducing, yielding a non-supersymmetric 4d effective field theory. 
In this subsection we show that, when the gravitino is integrated in, a precise type of stringy Scherk-Schwarz supersymmetry-breaking can be effectively realised as an F-term supersymmetry-breaking of a 4d $\mathcal{N}=1$  EFT that we will identify,  in accordance with other known examples in type IIA and heterotic supergravity \cite{Ferrara:1988jx,Villadoro:2005cu,Derendinger:2004jn, noscale5, Zwirner:2025ohv}. The resulting 4d $\mathcal{N}=1$ theory exhibits, at tree-level, the structure of a no-scale supergravity; the no-scale structure is then broken at one-loop by a correction to the tree-level K\"ahler potential that reproduces the one-loop Scherk-Schwarz runaway potential  (recall that the superpotential is protected from perturbative corrections by non-renormalisation theorems \cite{Dine:nonreno,quevedononreno, GarciadelMoral:2017vnz}).
As we will comment on thoroughly, the type of supersymmetry-breaking that emerges from this EFT description is a Scherk-Schwarz like mechanism with some qualitative but not quantitative differences with the mechanism employed in the string computations in Section \ref{S:CCeq0}, which will allow us to conclude our moduli stabilisation programme.

In order to proceed with our discussion, it is convenient to recall the main features of a no-scale model in 4d $\mathcal{N}=1$ supergravity. In a no-scale supergravity, the moduli split into two sets, $\Phi^A=(\Phi^a, \Phi^\alpha)$, for which the K\"ahler metric takes a block-diagonal form.  The F-term scalar potential is thus given by the formula 
\begin{equation}\label{scalarpot4dn1}
    V_{\text{F}}=e^K(K^{A\bar B}D_AW\overline{D_{\bar B}W}-3|W|^2)\, ,
\end{equation}
where $K^{A\bar B}$
is the inverse of the K\"ahler metric $K_{A\bar B} := \partial_A\partial_{\bar B}K$ derived from the K\"ahler potential, $K$;  $D_A W: = \partial_A W +
K_A W$ is the K\"ahler covariant derivative of the superpotential $W$; $K_A := \partial_A K$ and $\partial_A$ refers to the derivative
with respect to the moduli $\Phi^A$.  The no-scale moduli, $\Phi^a$, appear in the tree-level K\"ahler potential such that:
\begin{equation}\label{treenoscale}
    K_\tree^{a\bar b}K_{\tree\,a}K_{\tree\,\bar {b}}=3\, .
\end{equation}   
Provided that $\Phi^a$ moreover do not appear in the tree-level superpotential, the scalar potential reduces to:
\begin{equation}\label{eq:Vtreenoscale}
    V_{\text{tree}} = e^{K_{\text{tree}}} K_{\text{tree}}^{\alpha\bar{\beta}}D_\alpha W_{\text{tree}} D_{\bar{\beta}}\bar{W}_{\text{tree}}\,.
\end{equation}
 Being \eqref{eq:Vtreenoscale} semi-positive definite, it has a minimum when $\langle D_\alpha W_{\text{tree}}\rangle = 0$ and $\langle W_\tree \rangle \neq 0$, where $\langle V_{\text{tree}}\rangle=0$ and the no-scale moduli $\Phi^a$ become flat-directions of the potential, which, however, break supersymmetry via the non-vanishing F-terms $\langle D_a W_{\text{tree}}\rangle = \langle K_{\text{tree}\,a} W_{\text{tree}}\rangle$.  The non-vanishing gravitino mass is then given, on-shell, by the formula
 \begin{equation}\label{gravmassformula}
     M_{3/2} = \langle e^{K_{\text{tree}}/2}|W_{\text{tree}}|\rangle\,\mpl\, .
 \end{equation} 
Since we already know the tree-level K\"ahler potential, the first step to write down our EFT is to understand which tree-level superpotential $W_\tree$ realises a
no-scale model and reproduces the gravitino mass from the Scherk-Schwarz supersymmetry breaking. A way to reverse engineer $W_\tree$ is to notice that, as is customary
in no-scale models, it should not depend on the moduli setting the supersymmetry
breaking scale.
 Recalling \eqref{eq:SSm32} and using (\ref{planckmass4dn1}) and (\ref{moduliradi}), we notice that the gravitino mass $M_{3/2}$ that results from the Scherk-Schwarz twist can be reproduced from the moduli dependence
\begin{equation}\label{gravmassanti}
    M_{3/2}\sim\frac{\ms}{2R_9\sin\omega_{89}}\sim  \frac{M_{\text{pl}}}{\sqrt{t_1 t_2 u_3}} \, .
\end{equation}
We thus infer that the tree-level superpotential must be $W_\tree=W_\tree (S,U_1,U_2,T_3)$, leaving $T_1$, $T_2$, $U_3$ as the no-scale moduli; indeed from \eqref{eq:Ktree}, we see that $e^{-K_\tree}$ is a homogeneous function of degree 3 in the variables $t_1,t_2,u_3$ and the no-scale condition \eqref{treenoscale} is satisfied.
  
On the one hand, a dependence on a K\"ahler modulus ($T_3$ in this case) of the superpotential in IIB $\mathcal{N}=1$ O3 compactifications  seems unusual at least, since the 4d effective potential generated by NSNS $H_3$ and RR $F_3$ 3-form fluxes famously depends on the axio-dilaton $S$ and the complex structure moduli $U_i$ but not on the K\"ahler moduli $T_i$. Indeed, this is evident if we recall that the flux-induced superpotential is of the Gukov-Vafa-Witten type
\begin{equation}
    W_{\text{GVW}}:=\sqrt{\frac{2}{\pi}}\frac{1}{(2\pi)^2\alpha'^2}\int (F_3-i\,S\,H_3)\wedge\Omega\, ,
\end{equation}
with explicit dependence on the axio-dilaton and implicit dependence of the complex structure moduli  through the definition of the holomorphic 3-form $\Omega=\Omega(U)$. On the other hand, it is also known that  \emph{non-geometric $Q$ fluxes} can introduce a linear dependence on the K\"ahler moduli in the 4d effective potential\footnote{Non-geometric fluxes have been considered in IIB flux compactifications, in the context of de Sitter vacua, {\it{e.g.}} in \cite{nongeoref1,nongeoref2,nongeoref3,nongeoref4,nongeoref5,nongeoref6,nongeoref7}.}. 
Therefore, the EFT description of the Scherk-Schwarz supersymmetry-breaking  we are looking for is that of a flux-induced superpotential from standard 3-form fluxes and the non-geometric $Q$-flux. When non-geometric $Q$-fluxes are switched on too, the induced superpotential is known to be an extension of the GVW superpotential \cite{nongeoflux}
\begin{equation}
    W_{\text{tree}}=\sqrt{\frac{2}{\pi}}\frac{1}{(2\pi)^2\alpha'}\int \left(F_3-i\,S\,H_3-i(Q\circ\tilde{\omega}^j)T_j\right)\wedge\Omega\,,
\end{equation}
where the $Q$ flux is defined through its action on the basis $\{\tilde{\omega}\}_{i=1}^{h^{1,1}=3}$ of the 4-th cohomology $H^4$  given in Appendix \ref{A:IIBCY3}. The most general expression for $W_\tree$ in terms of the moduli $S,U_i,T_i$ is given in eq. (\ref{wfluxesgeneral}), where the coefficients are the quantised flux numbers from  flux quantisation conditions along the 3-cycles (\ref{quantisationfluxz2z2}). Setting to zero some specific fluxes\footnote{\label{footnotefluxnoscale}The flux choice is $h_3=h^1=h^2=h^0={q_{1}}^k={q_{2}}^k=q^{0l}=q^{1l}=q^{2l}=q^{31}=q^{32}=0$.} then renders $W_\tree$ independent of $T_1,T_2,U_3$, as we require. On top of this, we find it quite convenient to also
switch off the $F_3$ flux completely, such that $W_\tree$ is generated by some components of $H_3$ and $Q$ only. This is to avoid an $\int H_3\wedge F_3$ contribution to the $C_4$- tadpole, {\it{i.e.}} to the background D3 charge, and also a $\int Q\circ F_3$ contribution to the $C_8$-tadpole {\it{i.e.}} to the background D7-charge \cite{nongeoflux}; setting $F_3=0$ then allows us to turn on arbitrary $H_3$ and $Q$ fluxes without introducing additional charge into the background and thus without spoiling the RR tadpole cancellation conditions achieved in our string D-brane/O-plane configuration of Section \ref{sec:uspso1}.

With the above restrictions, the most general $W_\tree$ depending only on $S,U_1,U_2,T_3$ turns out to be
\begin{equation} \label{eq:Wtree}
\begin{split}
   \sqrt{\frac{\pi}{2}}W_{\tree}&= S(i\,h_0-h_1 \,U_1-h_2 \,U_2+i\, h^3 \,U_1 U_2)+T_3\left(-i\,{q_0}^3-U_1\,{q_1}^3-U_2\,{q_{2}}^3+i\,{q_{12}}^{3}\, U_1\,U_2\right)\,,
 \end{split}   
\end{equation}
with the coefficients $h$ and $q$ the integer quantised flux numbers\footnote{Flux quantisation in our orbifold/orientifold space requires the flux numbers to be even integers. On top of that, there are additional consistency conditions on the $H_3$ flux numbers $h$ due to the presence of the (anti) ${\mathrm{O3}}^+$-planes, as the integral of the non-vanishing discrete NSNS B-field on the $\mathbb{RP}^2$ cycles surrounding the $\overline{{\mathrm{O3}}}^+$-planes contributes to the $H_3$ flux threading the 3-cycle whose boundary is  given by the $\mathbb{RP}^2$-cycles \cite{Frey:2002hf,Cascales:2003zp}. Since in our case the 3-cycles dual to the $H_3$ fluxes always intersect an even number of $\overline{\mathrm{O3}}^+$-planes, one requires $h=2\mathbb{Z}$ to have an overall even integer flux number as dictated by the flux quantisation conditions. Thus the flux choices in Footnote \ref{footnotefluxnoscale}, as well as in Table \ref{t:solex} and Table {\ref{t:solex2}}, are consistent.}.  We can then impose the  F-term equations $D_{\alpha}W_\tree=0$.  These further impose $h^3={q_0}^3=0$ and give as solutions:
\begin{equation}\label{heavymoduli}
\begin{split}
&\braket{\theta_s}=\braket{\theta_{t_3}}=\braket{\theta_{u_1}}=\braket{{\theta_{u_2}}}=0\,,\\
      &\braket{u_1}=2\sqrt{\frac{h_0\,{q_{2}}^3}{h_1\,{{q_{12}}^{3}}}}\,,\quad
    \braket{u_2}=2\sqrt{\frac{h_0\,{q_{1}}^3}{h_2\,{q_{12}}^{3}}}\,,\quad
    \braket{z}:=\frac{\braket{t_3}}{\braket{s}}=\sqrt{\frac{h_1\,h_2}{{q_{1}}^3\,{q_{2}}^3}}\, .
\end{split}    
\end{equation}
When the stabilised fields are set to their {\it{vevs}}, the on-shell value of the flux superpotential $W_{\tree}$ becomes linear in the axio-dilaton, with $W_0$ an overall complex constant depending on the flux numbers:
\begin{equation}\label{vevWtree}
\begin{split}
     &W_{\text{tree}}(S)=W_0\,S\,\quad \text{with} \quad W_0:=\sqrt{\frac{h_0}{{q_{12}}^{3}}}\left(i\sqrt{h_0\cdot {q_{12}}^{3}}-\sqrt{h_2\cdot {q_1}^3}-\sqrt{h_1\cdot {q_2}^3}\right)\,.
 \end{split}    
\end{equation}
Thus, the 4d $\mathcal{N}=1$ formula for the gravitino mass (\ref{gravmassformula})  gives, on-shell and at tree-level,
\begin{equation}\label{onshellm32}
\begin{split}
    & M_{3/2}=\frac{1}{\sqrt{2\pi}}\sqrt{\frac{{q_{12}}^3}{h_0}}\frac{|W_0|}{\sqrt{t_1t_2u_3}}\mpl=\frac{\left((\sqrt{h_2 {q_1}^3}+\sqrt{h_1{q_2}^3})^2+h_0\,{q_{12
}}^{3}\right)^{1/2}}{4\pi R_9\sin\omega_{89}}\ms\,,\\
 \end{split}   
\end{equation}
where in the last equality we switched to string units using (\ref{planckmass4dn1}) and \eqref{moduliradi}.

At this point, it is important to note that Eqs \eqref{heavymoduli} and \eqref{onshellm32} immediately spell out a qualitative difference between the EFT at hand and the Scherk-Schwarz mechanism employed in our string set-up of Section \ref{S:CCeq0}. There, because of the $e^{\pi i\,F}=(-1)^F$ twist acting on the fields' boundary conditions along the direction $9$ in the $\mathbb{T}^2_3$, only fermions acquired tree-level masses of order the gravitino mass 
\begin{equation} \label{eq:M32-1F}
 M^{(-1)^F}_{3/2}:=\frac{M_s}{(4\pi R_9\sin \omega_{89})}  \,. 
\end{equation}
Instead, in the EFT formulation just described, not only is the gravitino massive but also some scalars (\ref{heavymoduli}) acquire tree-level masses via the F-term conditions, which are clearly of order $M_{3/2}$. This  suggests that
 the EFT at hand is actually a 4d $\mathcal{N}=1$ realisation of an \textit{R-symmetry Scherk-Schwarz breaking}, where the boundary conditions of fields are twisted by $e^{\pi i Q_\mathcal{R}}$, with $Q_\mathcal{R}$ the R-charge of the field under an $R$-symmetry group. Of course, the gravitino is charged under any $R$-symmetry, hence its zero mode acquires a mass of order $M_{3/2}=Q_\mathcal{R}^{3/2}/(4\pi R_9\sin \omega _{89})$: from (\ref{onshellm32}) we then identify $Q_{\mathcal{R}}^{3/2}\equiv((\sqrt{h_2 {q_1}^3}+\sqrt{h_1{q_2}^3})^2+h_0{q_{12
}}^{3})^{1/2}$. Besides a massive gravitino, the tree-level spectrum of such theories also contains massive scalar fields provided their $R$-charge is non-trivial, with mass terms of the order $M_{3/2}$; the tree-level potential giving masses to these scalars is still of the no-scale type.

We will leave for future work the dovetailing of a detailed string construction in the style of Section \ref{S:CCeq0} and moduli stabilisation within a 4d $\mathcal{N}=1$ EFT.  For now, we will proceed with the EFT just described, and seek reasonable values for the parameters by matching the gravitino mass and one-loop vacuum energy with those derived from our string computations.  To this purpose, we can indeed match
 \eqref{onshellm32} and \eqref{eq:M32-1F}, $M_{3/2}=M_{3/2}^{(-1)^F}$, with the flux choice $h_1=h_2=q_{1}^3=q_2^3=0$ and $|h_0 q_{12}^3|=1$.  This stabilises the axions $\theta_s$, $\theta_{t_3}$, $\theta_{u_1}$, $\theta_{u_2}$ and only one combination of the saxions $s/(t_3u_1u_2)$. Turning on the remaining fluxes\footnote{We will assume that the backreaction of any further fluxes -- and the non-perturbative effects introduced below -- is sufficiently small that our worldsheet computations of the vacuum energy still hold.} allows us to stabilise $u_1$, $u_2$ and $z$, as in (\ref{heavymoduli}), with the gravitino mass clearly obtaining further contributions from those fluxes.  Then, at tree-level and with general fluxes,  the complex structure $U_1$ and $U_2$ are fully stabilised, with their vanishing axions implying that  the tori $\mathbb{T}^{2}_{\,1}$, $\mathbb{T}^{2}_{\,2}$  are square $(\omega_{45}=\omega_{67}=\frac{\pi}{2})$.  The $S$- and $T_3$-axions are also fixed, together with one combination of the corresponding saxions.  Notice that the string coupling, $e^{-\phi}$, remains a flat-direction, as indeed is the case at tree-level in the string results in Section \ref{S:CCeq0}.
  
Having identified the tree-level data $(K_{\tree},W_\tree)$, we now need to find a one-loop correction $\delta K_{\text{SS}}$ to the tree-level K\"ahler potential that lifts the tree-level no-scale cancellation and generates a one-loop potential that matches the string Scherk-Schwarz one-loop effective potential;
either (\ref{v1loop2}) for 2 LED case or (\ref{v1loop1}) for the 1 LED case. As seen in  Appendix \ref{A:ssvacuumenergy}, the modular dependence of the 2 LED one-loop potential (\ref{v1loop2}), through the $\Gamma_1(2)$-invariant modular series $\mathcal{E}_3$, stabilises the complex structure modulus $U_3$ at the (fixed) points $U_3=\frac{1}{2}(1-i)+i\,\mathbb{Z}$; these critical points will only be shifted  when non-perturbative effects are turned on in the next section to stabilise the other moduli. In contrast, the 1 LED one-loop potential (\ref{v1loop1}) lacks such structure  and an alternative moduli stabilisation mechanism -- which results in exponentially large $u_3$ -- needs to be found. For  now, we will focus on the 2 LED scenario, and hence look for a correction $\delta K_{\text{SS}}$ that matches the 2 LED potential (\ref{v1loop2}). In Planck units and in terms of the 4-cycle volumes, the latter reads
\begin{equation}\label{sspot}
   \begin{aligned}
       &{V_{\text{ss}}=-\lambda_{\text{pl}}\frac{\mathcal{E}_3(iU_3)}{t_1^2t_2^2}\mpl^4}\,,&&&& \lambda_{\text{pl}}:=(4\pi)^2\lambda^{\ell_s}_{s,\text{2LED}}\simeq6.68\times10^{-5}\,,
   \end{aligned}
\end{equation}
where $\lambda_{s,\text{2LED}}$ is given in (\ref{v1loop1}) and computed in Appendix \ref{A:ssvacuumenergy}, and we consider the critical point $\braket{U_3} = \frac12(1+i)$.
The simplest dependence on the no-scale moduli that matches the Scherk-Schwarz scaling for the one-loop vacuum energy is
\begin{equation} \label{eq:Kcorr}
    K=K_{\tree}+\delta K_{\text{SS}}\,,
\end{equation}
where the correction, which we were able to determine in\footnote{Note the difference between the definitions of the complex structure moduli in the string computation, $U_{\text{string}}$ \eqref{UT - U}, and in the supergravity basis, $U_{\text{sugra}}$ \eqref{eq:sugramoduli}, with $U_{\text{string}}=i U_{\text{sugra}}$.} Appendix \ref{A:extendednoscale}, has the form
\begin{equation} \label{eq:deltaK}
    \delta K_{\text{SS}}=k_1\frac{(U_3+\bar{U}_3)\mathcal{E}_3(iU_3)}{(T_1+\bar{T}_1)(T_2+\bar{T}_2)}\, ,
\end{equation}
with $k_1$ a constant.
Indeed, running the formula (\ref{scalarpot4dn1}) and fixing the heavy fields to their {\it{vevs}} with the above relations,  we find the on-shell value of the scalar potential to match precisely (\ref{sspot}),
up to large volume corrections, with $k_1$ fixed as
\begin{equation}
    k_1=\frac{\lambda_{{\text{pl}}}}{6}\cdot \frac{|W_0|^{-2}}{\braket{u_1}\braket{u_2}\braket{t_3}} .
\end{equation}

To summarise, the one-loop corrected K\"ahler potential \eqref{eq:Kcorr}, \eqref{eq:Ktree} and \eqref{eq:deltaK}, and tree-level superpotential \eqref{eq:Wtree} are able to exactly match the Scherk-Schwarz supersymmetry-breaking gravitino mass and one-loop vacuum energy.  The moduli $u_1$, $u_2$, $u_3$, $\theta_{u_1}$, $\theta_{u_2}$, $\theta_{u_3}$, $\theta_s$ and $\theta_{t_3}$ are fixed, together with one combination of $s$ and $t_3$.  The moduli $t_1$ and $t_2$ are runaway directions at one-loop, whilst the axions $\theta_{t_1}$ and $\theta_{t_2}$, together with the orthogonal combination of $s$ and $t_3$, are flat-directions.  We now turn to the stabilisation of these remaining moduli.  It might be tempting to implement a two-step process at this stage, setting the tree-level stabilised moduli at their {\it{vevs}} in $K$ and $W$ and proceeding with a low-energy effective field theory for the light and runaway moduli.  However, although indeed the tree-level stabilised moduli {\it{vevs}} will hardly be affected by subleading corrections, their dynamics can couple strongly to the light and runaway moduli\footnote{This is in contrast to {\it{e.g.}} KKLT \cite{Kachru:2003aw} and LVS \cite{LVS}.  In KKLT, supersymmetry ensures consistency of the two-step moduli stabilisation: assuming that the heavy moduli adjust to ensure that $D_{\text{heavy}} W_{\text{full}} = 0$, one can then use also $D_{\text{light}} W_{\text{full}} = 0$ and the fact that the supersymmetry conditions guarantee a solution to the equations of motion, to argue the two-step procedure gives a good approximate solution.  In LVS, it can be verified that the large volume expansion helps ensure consistency (see {\it{e.g.}} \cite{Gallego:2011jm}  and references therein).}.  Hence, we will keep all the moduli for now.

\subsection{EFT corrections and full moduli stabilisation}

We next incorporate a series of well-motivated  non-perturbative corrections that will fix all the remaining moduli, at weak string-coupling, a controlled $\alpha'$-expansion, with two large extra dimensions and towards the scale of the observed Dark Energy.  We will make some comments about the possibility of moduli stabilisation with one large extra dimension in the outlook.

The 4d scalar potential we consider for the full stabilisation is the F-term potential
\begin{equation}\label{fullv}
    V_{\text{F}} = V_{\text{ss}}+V_{\text{np}}
\end{equation}
 generated from the Scherk-Schwarz corrected K\"ahler potential and non-perturbative corrections to the superpotential, and it will hence contain the one-loop Scherk-Schwarz contribution.
 We assume that both non-perturbative D(-1)-instantons and Euclidean D3-branes (ED3) wrapping 4-cycles in the compactification make non-trivial contributions to the superpotential.  This requires that their worldvolume theory has exactly two unsaturated fermion zero modes (for a review see  \cite{Blumenhageninstantons}).  Then:
\begin{equation}\label{wnp}
    W_{\text{np}}=W_{\text{D(-1)}}+W_{\text{ED3}} =A (1-a\, U_3) \,e^{-\alpha\,S}+\sum_{i=1}^3 B_i\,\left(1-b_i \,S-c_i \, S \, U_3\right)\,e^{-\beta_i\,T_i}\, ,
\end{equation}
where we have included some  polynomial moduli dependence in the Pfaffians, as is generally expected, and otherwise reduced $A_i, B_i, a, b_i$ and $c_i$ to real constants for simplicity\footnote{The specific Pfaffian dependence is chosen so that the non-perturbative corrections do not destabilise $U_3$ and do stabilise $S$ and $T_{1,2}$.}$^,$\footnote{In principle, we might expect the Pfaffians to also depend on $U_1$ and $U_2$, but since these have been stabilised already at tree-level, their {\it{vevs}} will hardly shift by including that dependence.}.  Assuming we have the leading terms in the instanton expansions, we take $\alpha=\pi$ and $\beta_3=2\pi$ and, since we will be aiming for moduli stabilisation with $t_1 \simeq t_2 \gg t_3$, we will at first neglect the terms proportional to $B_{1,2}$.
  For convenience, we collect together here the complete K\"ahler potential, $K=K_{\tree}+\delta K_{\text{SS}}$, and superpotential, $W=W_{\tree} + W_{\np}$, considered:
\begin{equation} \label{eq:fullKW}
    \begin{split}
    &K=-\sum_{i=1}^3\log (U_i+\bar{U}_i)-\sum_{i=1}^3\log(T_i+\bar{T}_i)-\log(S+\bar{S})   + k_1\frac{(U_3+\bar{U}_3)\mathcal{E}_3(iU_3)}{(T_1+\bar{T}_1)(T_2+\bar{T}_2)}\, \\ \\
    &W \simeq\,S(i\,h_0-h_1 \,U_1-h_2 \,U_2+i\, h^3 \,U_1 U_2)+T_3\left(-i\,{q_0}^3-U_1\,{q_1}^3-U_2\,{q_{2}}^3+i\,{q_{12}}^{3}\, U_1\,U_2\right)\\& \qquad\qquad\qquad+A(1-a\,U_3) \,e^{-\alpha\,S}+ B_3\,\left(1-b_3 \,U_3 - c_3\,S\,U_3\right)\,e^{-\beta_3\,T_3}\,.
    \end{split}
\end{equation}
The resulting F-term potential can be analysed in a large volume ($t_1,t_2 \gg 1$) and weak string coupling ($s \gg 1$) expansion, which we can organise as a perturbative expansion in $\delta K_{\text{SS}}$ and $W_{\np}$, with, recall, a no-scale cancellation at tree-level.  
As we derive in Appendix \ref{A:extendednoscale},
up to second order corrections in the K\"ahler metric, the F-term scalar potential exhibits the following structure  (see \eqref{structurev1storder}))
\begin{equation} \label{eq:Vexp}
\begin{split}
   V_{\text{F}}&= V_0+\delta V+\mathcal{O}(\delta^2V)\\
&=e^{K_\text{tree}}\left(K_\text{tree}^{\alpha\bar{\beta}}(1+\delta K_{\text{SS}})D_\alpha^{(0)}W_{\np}D_{\bar{\beta}}^{(0)}\overline{W}_\np-\,\delta K_{\text{SS}}|W_\tree+W_{\text{np}}|^2\right)+\mathcal{O}(\delta^2K_{\text{ss}})\, .
 \end{split}   
\end{equation}
We want to explore the region of theory space where the following hierarchy holds
\begin{equation}
   \ab W_\tree\ab\sim\mathcal{O}(1)\,,\quad \text{and}\,\quad \ab W_\tree\ab\gg \delta K_{\text{SS}}\sim \ab W_{\text{np}}\ab^2\,.
\end{equation}
In other words, we require the large volume suppression in $\delta K_{\text{SS}}$  to be able to compete with the non-perturbative suppression in $W_{\text{np}}$. 
At leading order then, the F-term potential  is well-approximated by the first two terms in (\ref {eq:Vexp})
\begin{equation}\label{VFlead}
    V_{\text{F}}\simeq e^{K_\tree}\left(K_{\tree}^{\alpha{\bar\beta}}D^{(0)}_\alpha W_{\text{np}}\overline{D}^{(0)}_{\bar{\beta}} \overline{W}_{\text{np}}-\delta K_{\text{SS}}\,|W_{\tree}|^2\right)\, .
\end{equation}
Moreover, having derived the F-term scalar potential  from the full $K$ and $W$ in \eqref{eq:fullKW}, we can assume that the tree-level stabilised moduli lie close to their tree-level {\it{vevs}} (\ref{heavymoduli}) and hence can be consistently integrated out.
All in all, the 4d scalar potential for the light fields (\ref{VFlead}) boils down to
\begin{equation}\label{vstu}
    V_{\text{F}}=\frac{1}{t_1t_2u_3}f_\np(s,u_3,\theta_{u_3})-\frac{\lambda_{\text{pl}}}{t_1^2t_2^2}\mathcal{E}_3(iU_3)\,,
\end{equation}
where we conveniently defined the non-perturbative function $f_{\np}(s,u_3,\theta_{u_3}) \sim \mathcal{O}(W_\np^2)$.
We omit to write down the rather lengthy full expression for $f_{\np}(s,u_3,\theta_{u_3})$, but note that -- assuming no hierarchy in the Pfaffian parameters such as $A$ and $B_3$ -- the following terms dominate in the large $s$ limit 
\begin{equation}
   f_{\np}(s,u_3,\theta_{u_3}) \approx -\frac{u_3}{2sz}\left(2 A a e^{-\frac{\alpha s}{2}} + B_3 e^{-\frac12 \beta_3 z s}(2 b_3 + c_3 s)\right)\,.
\end{equation}
Recalling that $\beta_3=2\alpha$, and still assuming no hierarchy in the Pfaffians, this implies a runaway instability in $s$ unless\footnote{Notice that this value is well within the lower bound on $\braket{z}$ from requiring control of the $\alpha'$ expansion. Indeed from its definition $\braket{z}=(2\pi)^{-4}R^{{(\alpha')}}_4R^{{(\alpha')}}_5R^{{(\alpha')}}_6R^{{(\alpha')}}_7$, where $R^{{(\alpha')}}_i:=2\pi R_i$ are the dimensionless radii in units of $\alpha'$, so a controlled $\alpha'$ expansion requires $\braket{z}\gg \frac{1}{(2\pi)^4}\simeq 6.4 \times 10^{-4}\, .$} $\braket{z}\approx\frac12$.  With $\braket{z}=\frac12$, the leading contribution to $f_\np(s,u_3,\theta_{u_3})$ becomes
\begin{equation}
    f_{\np}(s,u_3,\theta_{u_3}) \approx -\frac{e^{-\frac{\alpha s}{2}}}{s} u_3\left(2 A a+2 B_3 b_3 + B_3 c_3 s\right)\,.
\end{equation}
We see that our choice of Pfaffians implies that the non-perturbative corrections do not lift the one-loop stabilisation $\braket{U_3}=\frac12(1+i)$ and further lead to a stabilisation of $s$ at:
\begin{equation}
    \braket{s}=\frac{-2 A a - 2 B_3 b_3 \pm \sqrt{(2 A a + 2 B_3 b_3)^2-\frac{4 B_3(4Aa+4B_3 b_3)c_3}{\alpha}}}{2 B_3 c_3}\,.
\end{equation}
Finally, we can turn to the stabilisation of $t_1$ and $t_2$.  Notice that only the saxion combination $t^2:=t_1 t_2$ enters the scalar potential, which is the one saxion that will be stabilised at this order.  It is easy to see that indeed \eqref{vstu} then stabilises $t^2:=t_1t_2$ at values that are exponentially large in $s=1/g_s$
\begin{equation}\label{tvev}
\begin{split}
    &\braket{t^2}=\frac{2\,\lambda_{\text{pl}}\,{\braket{u_3}\braket{\mathcal{E}_3}}}{{\braket{f_\np}}}\,,
   \end{split} 
\end{equation}
giving rise to two large extra dimensions.  Since, with $s$ and $U_3$ set at their {\it{vevs}}, the two-term potential $V(t)$ from \eqref{vstu} approaches zero from above for large $t$, we conclude that the extremum in $t$ is a de Sitter maximum, with cosmological constant $\Lambda=\braket{V}$ given by
\begin{equation}\label{lambdavev}
\begin{split}
    &{\Lambda=\frac{1}{4}\cdot  \frac{\braket{f_\np}^2}{\lambda_{\text{pl}}\braket{u_3}^2\braket{\mathcal{E}_3}}}\mpl^4\,.
\end{split}
\end{equation}
Having stabilised the product of the K\"ahler moduli $t^2=t_1t_2$, we are now left with the stabilisation of the remaining K\"ahler modulus, let us say $t_1$, as well as the axions $\theta_{t_1}$ and $\theta_{t_2}$,  which are still flat directions. To do so, we consider ED3-instantons on the largest 4-cycles, which have so far been neglected in the regime $t_1\sim t_2\gg t_3$ we are interested in. Neglecting for simplicity any Pfaffian dependence in the already heavy moduli, their superpotential is given by
\begin{equation}
    W_{\np}\supset B_1 \,e^{-i \beta_1\,(t_1/2+i\theta_{t_1})}+ B_2\,e^{-i\beta_2(t_2/2+i\theta_{t_2})}\, .
\end{equation}
 Running the 4d $\mathcal{N}=1$ formula with $K=K_\text{tree}+\delta K_{\text{SS}}$ and the full $W=W_\tree+W_\np$, and then integrating out all the heavy moduli, which now  also include $t$, $u_3$ and $\theta_{u_3}$, the leading order potential for the K\"ahler modulus $t_1$ and the axions $\theta_{t_1}$, $\theta_{t_2}$ is found to read
\begin{equation}
\begin{split}
    V(t_1,\theta_{t_1},\theta_{t_2})\simeq|W_0|\biggl(&B_1\,\beta_1\,t_1\,e^{-\frac{t_1\beta_1}{2}}\cos(\beta_1\theta_{t_1}+\theta_{W_0})\\
    &+{B_2\,\beta_2}\frac{\braket{t^2}}{t_1}e^{-\frac{\braket{t^2}\beta_2}{2t_1}}\cos(\beta_2\theta_{t_2}+\theta_{W_0})\biggr)\, ,
 \end{split}   
\end{equation}
where  $W_0\equiv|W_0|e^{i\theta_{W_0}}$ is the flux-dependent prefactor of the tree-level superpotential at its {\it{vev}} \eqref{vevWtree}.  We see that the axions are stabilised by the cosine potentials at
\begin{equation}
    \theta_{t_i}=\frac{1}{b_i}\left(-\theta_{W_0}+(2\mathbb{Z}+1)\pi\right)\, ,
\end{equation}
where note the cosine potentials give an overall minus sign. The critical point condition for $t_1$ is then found to be
\begin{equation}\label{criticalt1}
\begin{split}
   &B_1\beta_1\,(-2+t_1\beta_1)\,e^{-\frac{t_1\beta_1}{2}}+{B_2\,\beta_2}\braket{t^2}\frac{(2t_1-\braket{t^2}\beta_2)}{t_1^3}e^{-\frac{\braket{t^2}\beta_2}{2t_1}}=0\,.
 \end{split}   
\end{equation}
Note that in the large volume limits $t_1\beta_1\gg1$ and $\braket{t}\beta_2\gg t_1$ the two terms have opposite signs and can compete. Exploiting the symmetry of the background, it is reasonable to assume $B_1=B_2$, $\beta_1=\beta_2$, after which it is easy to see that a solution to (\ref{criticalt1}) at large volume is\footnote{The same conclusion can be reached more rigorously by rewriting the large volume limit of eq. (\ref{criticalt1}) as the transcendental equation $4\log t_1+\frac{c_2}{t_1}-c_1 t_1=\log c_3$, with coefficients  $c_1:=\frac{\beta_1}{2}$, $c_2:=\frac{\braket{t^2}\beta_2}{2}$, $c_3:=\braket{t^2}^2\frac{\beta_2^2}{\beta_1^2}\frac{|B_2|}{|B_1|}$, and then solving for $t_1$ perturbatively in a large-$\sqrt{c_1/c_1}\sim \sqrt{\braket{t^2}}$ expansion, to obtain $\braket{t_1}=\left(\frac{\braket{t^2}\beta_2}{\beta_1}\right)^{1/2}+\frac{1}{\beta_1}\log \frac{|B_1|}{|B_2|}+\mathcal{O}(\braket{t^2}^{-1/2})$.}
\begin{equation}
    \braket{t_1}=\braket{t_2}=\braket{t^2}^{1/2}
\end{equation}
such that we end up with the two K\"ahler moduli $t_1$, $t_2$ stabilised at the same scale, as we wanted. All moduli have now been stabilised.

\subsection{Addressing the phenomenological scales}

\begin{table}[t]
    \centering
    \begin{tabularx}{\textwidth}{|C|C|C|C|C|}
    \hline
        & Upper bound on size of 2 LEDs $(\mu m)$& Model independent?& Satisfied?& Does it apply? \\
    \hline\hline
     Table-top    & $30$ \cite{tabletop3} & Yes& Yes& Yes\\[1em]
    Neutrino-burst SN 1987A    &    \makecell{$1$ \cite{Hannestad:2003yd}\\ $0.33$ \cite{Hardy:2025ajb}}& Yes &No&Yes   \\
    Old neutron-star excess heat &\makecell{\\$0.059$ \cite{Hannestad:2003yd}\\$75$ \cite{Hardy:2025ajb}}& No &No&No   \\
    \hline
    \end{tabularx}
    \caption{Laboratory and observational upper bounds on the size of extra dimension for $n=2$ extra dimensional models due to the presence of (light) KK gravitons in the spectrum.}
    \label{tab:bounds}
\end{table}

Let us now discuss how close our solution's cosmological constant can be to the observed value \cite{DESI:2025zgx}
\begin{equation}\label{bounds}
\begin{split}
\Lambda_{\text{obs}}&\simeq(2.26\,\text{meV})^4\simeq (87 \mu m)^{-4}\simeq 7.4 \times 10^{-121}\mpl^4\, ,
\end{split}    
\end{equation}
once we take into account the experimental and observational constraints  on the size of the extra dimensions in $n=2$ large extra dimensional models \cite{tabletop1,tabletop2,tabletop3,tabletop4,tabletop5,Hannestad:2003yd,Hardy:2025ajb},  and on the string mass $M_s$ \cite{CMS:2019gwf, ATLAS:2019fgd, ParticleDataGroup:2024cfk}. 

At leading order, the cosmological constant resulting from our solution is related to the product of the 4-cycle volumes $t^2\equiv t_1t_2$ as 
\begin{equation}
    \Lambda=\lambda_\text{pl}\braket{\mathcal{E}_3}\frac{1}{\braket{t^2}^2}\mpl^4\, ,
\end{equation}
or, in units of $M_s$,
\begin{equation}\label{lambdarho9}
    \Lambda=\frac{\lambda_s^{\ell_s}}{4}\cdot\frac{\braket{\mathcal{E}_3}\braket{u_3}^2}{\braket{\sin\omega_{89}}^4}\cdot \frac{M_s^4}{R_9^4
    }={2.62}\times10^{-3}\cdot\lambda_s\cdot \frac{1}{\tilde{R}_9^4}\,,
\end{equation}
where we have introduced the dimensionful radius $\tilde R_9^{-1}:=  R_9^{-1} M_s$. For $\lambda_{s}=6.6\times 10^{-4}$, we can match $\Lambda_{\text{obs}}$ for 
\begin{equation}\label{rho9}
    \tilde{R}_9= 87 \cdot (2.62 \times 10^{-3})^{1/4}\cdot (6.6\times10^{-4})^{1/4}\mu m\simeq {3.18} \,\mu m\, ,
\end{equation}
and, using $\braket{U_3}=\frac12(1+i)$, the radius of the second large extra dimension is $\tilde{R}_8=4.49\,\mu m$. Notice, however, that the radii $\tilde{R}_{8,9}$ are not invariant under the shifts $U_3 \rightarrow U_3 + i \mathbb{Z}$; rather the physical parameters of the corresponding 2-torus (see \ref{eq:2torusmod}) are the invariant area, $\braket{\mathcal{A}}\equiv \sqrt{\text{det}\braket{g_{(3)}}}$, and the modular equivalence class of $\braket{U_3}$.  The invariant KK-masses can then be written as\footnote{This matches the 2$\pi$ conventions used by experimental and observational constraints, which have $M_{\text{KK}}=1/\tilde{R}$ and $\text{Vol}=(2\pi \tilde{R})^n$.} $M_{\text{KK}}^2=\braket{g_{(3)}^{ab}}m_am_b=\frac{1}{\braket{\mathcal{A}} \,\braket{\text{Re}U_3}}|m_2 - \braket{U_3}m_1|^2$.  Recalling also that the area $\mathcal{A}$ is reduced by a factor 4 by the freely-acting $\mathbb{Z}_2\times \mathbb{Z}_2$, the values above then correspond to a KK mass gap of $M_{\text{KK}} = \ell_{\text{KK}}^{-1} = (1.6 \mu m)^{-1}$.  We now check if $\ell_{\text{KK}}=1.6\mu m$ is allowed or excluded by the current upper bounds.

The cleanest and most model-independent bounds on large extra dimensions arise from table-top tests of Newton's inverse square law, given the deviations that would be induced by light KK gravitons \cite{tabletop1,tabletop2,tabletop3,tabletop4,tabletop5}. The resulting constraints depend only weakly on the number $n$ of extra dimensions, varying by a factor of a few, and the bound for the case $n=2$ is approximately $\ell_{\text{KK}}\lesssim 30\, \mu m$ ($M_{\text{KK}}\gtrsim 4\,\text{meV}$).

For the $n=2$ case, however, the most stringent bounds arise from astrophysical observations involving the production and decay of KK gravitons in supernovae (SN), which can lead to anomalous energy loss or gain\footnote{In the $n=1$ case these turn out to be weaker than the laboratory bounds.}.  A standard reference for such constraints is \cite{Hannestad:2003yd}, while a more recent analysis incorporating additional production and decay channels has been presented in \cite{Hardy:2025ajb}. In particular, the constraints on the size of the extra dimensions that we will quote are derived from the observed duration of the neutrino burst from SN~1987A and from measurements of the surface temperatures of old neutron stars (NS)\footnote{\label{sn}During a core-collapse supernova (SN), whose   explosion results in a burst of neutrinos, KK gravitons would also be produced. Those produced with sufficient kinetic energy would escape the core contributing to energy-loss, thus shortening the duration of the neutrino burst. The size of extra dimensions can be bounded by requiring that the KK graviton emission does not spoil the duration of the signal from the observed SN 1987A neutrino burst (the only SN neutrino burst detected to date), which lasted approximately 10 seconds. A second bound arises from KK gravitons that are produced near the kinematic mass threshold in the SN core. These would remain gravitationally trapped in a cloud of KK gravitons surrounding the newborn neutron-star (NS). In this case, the size of extra dimensions is bounded by the requirement that the excess heat produced from the decay of KK gravitons into photons does not spoil the observed surface temperature of old NS.}. Constraints on two micron-size dark dimensions from the astrophysical bounds have also been reviewed recently in \cite{Burgess:2023pnk, antoniadis2DD}. We summarise both laboratory and astrophysical bounds in Table \ref{tab:bounds}.

As can easily be seen from Table \ref{tab:bounds}, our solution  $\ell_{\text{KK}} \sim 1.6 \mu m$ is well within the upper bound coming from table-top experiments. There is however a tension with the upper bounds from the supernovae observations. The largest tension is with the bound from old NS excess-heat  given in ref. \cite{Hannestad:2003yd}, while  the tension disappears if we consider the old NS excess-heat bound given in \cite{Hardy:2025ajb}. The sizable difference between the two values resides in the different assumption on KK number conservation. For perfectly toroidal compactifications, as assumed in \cite{Hannestad:2003yd}, translational symmetry along extra dimensions implies KK number conservation; this prevents KK graviton decays into ligher KK gravitons, so they decay only into photons, and one has the very strong bound of \cite{Hannestad:2003yd}. If KK number conservation is not assumed -- as in \cite{Hardy:2025ajb} and as is the case in our construction given that the SM branes are perpendicular to the large extra dimensions
-- the intermediate decay into light KK gravitons slows the decay into photons and one finds the much less stringent bound of \cite{Hardy:2025ajb}, which is then even less constraining than table-top bounds. 

In addition to assumptions around KK number conservation, there is a further model-dependence on the bound from old NS excess-heat since it concerns KK gravitons' decay into photons (see footnote \ref{sn}). As originally noted in the ADD paper \cite{add}, limits on KK graviton decay into SM degrees of freedom can be  evaded if hidden branes are present in addition to the SM branes, since there will be additional decay channels for the KK gravitons. This is precisely the situation in our string construction, with the SO(1)-branes playing the role of hidden sectors.  We hence conclude that the old NS excess-heat bound does not apply in our case.

With this proviso, the only remaining tension is with SN 1987A bound. Remarkably, this tension is only  an order one factor if the bound of \cite{Hannestad:2003yd} is used, while it increases mildly to  approximately  a factor 5 if the revisited bound of \cite{Hardy:2025ajb} is adopted.  

One can wonder what are the consequences on  our parameter space if the more restrictive of the SN 1987A bound is implemented. For $\ell_{\text{KK}}\lesssim 0.33\mu m$ ($\ell_{\text{KK}}\lesssim 1\mu m$) we find from (\ref{lambdarho9}) that 
\begin{equation}
    \Lambda_{\text{SN 1987A}}\gtrsim  0.014\cdot\lambda_s\cdot \mu m^{-4} \quad (\Lambda_{\text{SN 1987A}}\gtrsim  1.6 \times 10^{-4} \cdot\lambda_s\cdot \mu m^{-4})
    \,.
\end{equation}
 To match $\Lambda_{\text{obs}}$ we thus need a coefficient $\lambda_s\lesssim 1.2 \times 10^{-6}$ ($\lambda_s\lesssim  1.0 \times 10^{-4}$).
 In our case, however, the coefficient $\lambda_s$, stemming from the computation of the string one-loop Casimir energy, reads $\lambda_s=(n_b^{0}-n_f^0)/2^3\pi^7$, {\it{i.e.}} $\lambda_s=6.6\times 10^{-4}$ for the $T^6/\mathbb{Z}'_2\times \mathbb{Z}'_2$ case at hand where $n_b^0-n_f^0=16$.
This suggests that a one-loop factor might not be sufficient to both accommodate the SN 1987A bound \cite{Hardy:2025ajb} and, at the same time, recover the observed value of the cosmological constant via a  Casimir energy contribution, and it may be that a  higher-loop suppression is needed. Indeed, it would be possible to respect the bound  and reproduce $\Lambda_{\text{obs}}$ if somehow the one-loop vacuum energy from the closed-string sector were vanishing too, which would mean that $\Lambda$ is set by a two-loop closed-string diagram with suppression factor given by
\begin{equation}
    \lambda_s^{\text{2-loop}}\sim \lambda_s\cdot \frac{g_s^2}{16\pi^2} \,,
\end{equation}
{{\it{i.e.}} $\lambda_s^{\text{2-loop}}\lesssim 1.2 \times 10^{-6}$  for a string coupling $g_s\lesssim 0.5$}. This  could be realised by {\it{e.g.}} the construction of a Scherk-Schwarz like \emph{super no-scale model} such that $n_b^0-n_f^0=0$ in the closed-string spectrum, with a brane supersymmetry breaking like non-supersymmetric open-string sector that still does not contribute to the one-loop vacuum energy. 

We should note that the possibility of matching the observed  cosmological constant for a fixed $\ell_{\text{KK}}$ and $\tilde{R}_9\ll \Lambda_{\text{obs}}^{-1/4}$ by compensating the missing suppression factors with the smallness of the prefactor $\lambda_s$ is  somewhat non-trivial  since it requires $\tilde{R}_9$ itself to not scale with $\lambda_s$ in (\ref{lambdarho9}). However, this is in general not guaranteed since $\tilde{R}_9$ is a (runaway) modulus that has to be stabilised, and $\lambda_s$ is then a parameter of the scalar potential that stabilises the moduli, with $\Lambda$ the on-shell value of this potential at the point where the moduli are stabilised. Hence, on general grounds, a dependence $\tilde{R}_9(\lambda_s)$ is expected once $\tilde{R}_9$ acquires a {\it{vev}}.  This can be seen concretely in our solution. Indeed from (\ref{planckmass4dn1}) $M_s\sim \braket{t^2}^{-1/4}$ and $R_9\sim\braket{t^2}^{1/4}$, hence $\tilde{R}_9\sim R_9 M_s^{-1}\sim \braket{t^2}^{1/2}\sim \lambda_s^{1/2}f_{\text{np}}^{-1/2}$ using (\ref{tvev}), which means that $\tilde{R}_9$ does not scale with $\lambda_s$ only if we also scale the non-perturbative factor, which is another parameter in the solution, as $f_{\text{np}}\sim \lambda_s$; with this choice, $M_s$ does not scale while $\Lambda$ scales linearly with $\lambda_s$, as can also be seen using (\ref{lambdavev}) $\Lambda\sim f_{\text{np}}^2\lambda_s^{-1}\sim \lambda_s$.

Having determined that $\ell_{\text{KK}} \simeq 1.6\mu m$ is needed to reproduce $\Lambda_{\text{obs}}$, we should now check under which conditions the fundamental string scale $M_s=R_9 \tilde{R}_9^{-1}$ associated to this solution meets the experimental constraint
\begin{equation}\label{Msbounds}
\begin{split}
M_s \gtrsim 8 \text{TeV}\,,
\end{split}    
\end{equation}
arising from the absence of string resonances at the LHC \cite{CMS:2019gwf, ATLAS:2019fgd, ParticleDataGroup:2024cfk}.  
 Substituting $t=t(\Lambda)$ and using (\ref{lambdavev}) in the relation between the 4d  Planck mass and the string scale \eqref{planckmass4dn1}, using moreover the observed value $\mpl\simeq2.435\times 10^{15}\text{TeV}$, we can express $M_s$ as function of $\Lambda$ in Planck units
\begin{equation}\label{mslambda}
\begin{split}
    M_s(\Lambda)&=2.435\times10^{15}\cdot(4\pi)^{1/2}\cdot\frac{1}{\braket{s\,t_3}^{1/4}}\frac{1}{\braket{\mathcal{E}_3}^{1/8}\lambda_\text{pl}^{1/8}}\left(\frac{\Lambda}{\mpl^4}\right)^{1/8}\text{TeV}\\
    &=1.70 \times 10^{16}\cdot g_s^{1/2}\left(\frac{\Lambda}{\mpl^4}\right)^{1/8}\text{TeV}\,.
 \end{split}   
\end{equation}
Then, to meet the experimental constraint  $M_s(\Lambda_{\text{obs}})\gtrsim8\text{TeV}$,  $g_s$ must be bounded from below  by 
\begin{equation}\label{zupperbound}
     g_s \gtrsim 0.17\, ,
\end{equation}
 and so $\braket{s} \lesssim 12$.
 
Meanwhile, from \eqref{lambdavev} we see that to find $\Lambda\simeq 7\times 10^{-121}$ we need 
\begin{equation}
 f_{\text{np}} \simeq {3} \times 10^{-62} \,.
\end{equation}
 For Pfaffians that are not hierarchically large or small, {\it{e.g.}} $A,B_3=\mathcal{O}(10^{n})$ with $-4 \lesssim n \lesssim 4$, to recover a non-perturbative suppression of the strength above we need a string coupling of roughly
\begin{equation}\label{sizegsnp}
 g_s\simeq \frac{\pi}{(30+n)\log10}\simeq \frac{1.3}{30+n}\simeq 0.038\Doteq0.05 \,,
\end{equation}
{\it{i.e.}} $40\lesssim \braket{s}\lesssim 53$.  This is in tension with the previous constraint $\braket{s} \lesssim 12$ from $M_s > 8\,$TeV.  We conclude that to hope to simultaneously match the experimental constraints on $M_s$, $\ell_{\text{KK}}$ and $\Lambda$, we would need to consider some hierarchy in the Pfaffians and/or suppression in $\lambda_s$.

\begin{table}[t!]
\centering
\renewcommand{\arraystretch}{1.2}
\begin{tabularx}{\textwidth}{|C|C|C|C|C|C|}
\hline
\multicolumn{6}{|c|}{\text{Flux numbers}}\\
 $h_0$ & $h_1$ & $h_2$ & $q_1^3$ & $q_{2}^3$  & $q_{12}^3$ \\
 \hline
 $2$ & 2&  2 & 4 & 4 & $4$ \\ 
 \hline 
 \end{tabularx}\par\vskip-1.4pt
 \begin{tabularx}{\textwidth}{|C|C|C|C|C|C|C|}
\hline
\multicolumn{7}{|c|}{Non-perturbative parameters}\\
 $A$ & $B_1$ & $B_2$ & $B_3$ & $\alpha$  & $\beta_{1,2,3}$ & $a, b_3, c_3$ \\ 
 \hline 
 $-0.43$  & $1$& $1$ & $0.01$ & $\pi$  & $2\pi$ & $-1$\\ 
 \hline 
 \hline
\end{tabularx}\par\vskip-1.4pt
\begin{tabularx}{\textwidth}{|C|C|C|C|C|C|C|}
\multicolumn{7}{|c|}{Moduli {\it{vevs}} -- saxions}\\
$s$ & $ t_1$ & $t_2$ & $t_3$ & $u_1$ & $u_2$ & $u_3$ \\ 
 \hline 
$84$ & $1.4\times10^{29}$ & $1.4\times10^{29}$ & $42$ & $2$ & $2$ & $1$ \\ 
 \hline 
 \end{tabularx}\par\vskip-1.4pt
 \begin{tabularx}{\textwidth}{|C|C|C|C|C|C|C|}
\multicolumn{7}{|c|}{Moduli {\it{vevs}} -- axions}\\
 $\theta_s$ & $\theta_{t_1}$ & $\theta_{t_2}$ & $\theta_{t_3}$ & $\theta_{u_1}$ & $\theta_{u_2}$ & $\theta_{u_3}$ \\ 
 \hline 
 $0$ & 0.85$\pi$ & 0.85$\pi$ & 0 & $0$ & 0 & 0.5 \\ 
 \hline 
 \hline
 \end{tabularx}\par\vskip-1.4pt
\begin{tabularx}{\textwidth}{|C|C|}
Cosmological Constant & $\Lambda = 7.4\times 10^{-121}\mpl^4$ \\ \hline
String coupling & $\gs =0.024 $ \\ \hline
String scale & $\ms =3.01\,\text{TeV}  $ \\ \hline
KK mass scale & $M_{\text{KK}} =0.12\,\text{eV}$ \\ \hline
KK length scale for the 2 dark dimensions &$\ell_{\text{KK}}=1.58\, \mu m$   \\ 
 \hline 
KK length scale for the 4 susy dimensions ($\alpha'$ units)  & $\ell_{\text{KK}i}^{(\alpha')}=5.28$, $i=1,2$ \\
 \hline
 \hline
 \end{tabularx}\par\vskip-1.4pt
\begin{tabularx}{\textwidth}{|C|C|C|C|C|C|C|}
\hline
\multicolumn{7}{|c|}{Mass-squared eigenvalues in eV$^2$}\\
\hline
0.0065 & 0.0031 & 0.0031 & 0.00094 & 0.00062 & 0.00062 & $3.0 \times 10^{-9}$\\
\hline
$-2.6 \times 10^{-61}$ & $-7.8 \times 10^{-65}$ & $-3.5 \times 10^{-65}$ & $2.5 \times 10^{-65}$ &
\multicolumn{3}{c|}{\parbox[c]{0.38\textwidth}{\centering
three values going as $\sim 10^{54}\mathrm{poly}(t)e^{-\pi t}$\\
where $t \sim 1.4 \times 10^{29}$
}}\\
\hline
\end{tabularx}
\caption{A first illustrative example of a de Sitter saddle point.  In addition to the flux numbers and non-pertubative parameters listed, we have used the parameters for the Scherk-Schwarz one-loop vacuum energy $\lambda_{\text{s}}=6.6 \times 10^{-4}$.  The dark dimensions are exponentially large in the inverse string coupling and the cosmological constant is exponentially suppressed, without fine-tuned cancellations.  Note that we should confront these solutions with the observational constraints from Table \ref{tab:bounds}, that is $\ms \gtrsim 8$TeV and $\ell_{\text{KK}} \lesssim 0.33 \mu m$, together with $\Lambda_{\text{obs}} \approx 7 \times 10^{-121}\mpl^4$, so a fully realistic model would require some further adjustment of parameters.  For reference, we also quote a value for the Hubble constant squared (assuming $\Lambda$CDM) as $H_0^{\,2}\sim 2.1 \times 10^{-66}\text{eV}^2$ \cite{DESI:2025zgx}.}\label{t:solex}
\end{table}
\vspace{4mm}

We conclude by presenting two concrete solutions to  the moduli stabilisation equations,  summarised in Tables \ref{t:solex} and \ref{t:solex2}. The choice of flux integers made in Table \ref{t:solex} gives $\braket{z}=1/2$ and $R_{4}=R_5=R_6=R_7=\braket{z}^{1/4}=0.84$, hence $ \ell_{\text{KK}i}^{\alpha'}=5.28$, $i=1,2$, and we have good numerical control of the $\alpha'$ expansion.  Note that this first solution
reproduces the size of the observed cosmological constant. As expected, however, the string scale $M_s$ and the size of the dark  dimensions are just outside the experimental bounds.  In Table \ref{t:solex2}, we kept the same flux choices as for our first solution, but adjusted the Pfaffians slightly resulting, overall, in  a slightly smaller non-perturbative suppression. The cosmological constant of this solution is correspondingly larger than the observed one by roughly  three orders of magnitude, whilst the dark dimensions  and the string scale $M_s$ fall within the experimentally allowed range.

\begin{table}[t]
\centering
\renewcommand{\arraystretch}{1.2}
\begin{tabularx}{\textwidth}{|C|C|C|C|C|C|}
\hline
\multicolumn{6}{|c|}{\text{Flux numbers}}\\
 $h_0$ & $h_1$ & $h_2$ & $q_1^3$ & $q_{2}^3$  & $q_{12}^3$ \\
 \hline
 $2$ & 2&  2 & 4 & 4 & $4$ \\ 
 \hline 
 \end{tabularx}\par\vskip-1.4pt
 \begin{tabularx}{\textwidth}{|C|C|C|C|C|C|C|}
\hline
\multicolumn{7}{|c|}{Non-perturbative parameters}\\
 $A$ & $B_1$ & $B_2$ & $B_3$ & $\alpha$  & $\beta_{1,2,3}$ & $a, b_3, c_3$ \\ 
 \hline 
 $-0.42$  & $1$& $1$ & $0.01$ & $\pi$  & $2\pi$ & $-1$\\ 
 \hline 
 \hline
\end{tabularx}\par\vskip-1.4pt
\begin{tabularx}{\textwidth}{|C|C|C|C|C|C|C|}
\multicolumn{7}{|c|}{Moduli {\it{vevs}} -- saxions}\\
$s$ & $ t_1$ & $t_2$ & $t_3$ & $u_1$ & $u_2$ & $u_3$ \\ 
 \hline 
 82&$2.0\times 10^{28}$ & $2.0\times 10^{28}$& 41 & $2$ & $2$ & $1$ \\ 
 \hline 
 \end{tabularx}\par\vskip-1.4pt
 \begin{tabularx}{\textwidth}{|C|C|C|C|C|C|C|}
\multicolumn{7}{|c|}{Moduli {\it{vevs}} -- axions}\\
 $\theta_s$ & $\theta_{t_1}$ & $\theta_{t_2}$ & $\theta_{t_3}$ & $\theta_{u_1}$ & $\theta_{u_2}$ & $\theta_{u_3}$ \\ 
 \hline 
 $0$ & 0.85$\pi$ & 0.85$\pi$ & 0 & $0$ & 0 & 0.5 \\ 
 \hline 
 \hline
 \end{tabularx}\par\vskip-1.4pt
\begin{tabularx}{\textwidth}{|C|C|}
Cosmological Constant & $\Lambda = 1.7\times 10^{-117}\mpl^4$ \\ \hline
String coupling & $\gs =0.024 $ \\ \hline
String scale & $\ms =8.0\,\text{TeV}  $ \\ \hline
KK mass scale & $M_{\text{KK}} =0.86\,\text{eV}$ \\ \hline
 KK length scale for the 2 dark dimensions &  $\ell_{\text{KK}} = 0.23 \mu m $ \\ 
 \hline 
 KK length scale for the 4 susy dimensions ($\alpha'$ units)  & $\ell_{\text{KK}i}^{(\alpha')}=5.28$, $i=1,2$ \\
 \hline
 \hline
 \end{tabularx}\par\vskip-1.4pt
\begin{tabularx}{\textwidth}{|C|C|C|C|C|C|C|}
\multicolumn{7}{|c|}{Mass-squared eigenstates in eV$^2$}\\
0.30 & 0.15 & 0.15 & 0.045 & 0.029 & 0.029 &  $2.7 \times 10^{-6}$ \\ \hline
$2.1 \times 10^{-6}$ & $-1.7 \times 10^{-61}$ & $-7.9 \times 10^{-62}$ & $5.6 \times 10^-{62}$ & \multicolumn{3}{c|}{\parbox[c]{0.38\textwidth}{\centering
three values going as $\sim 10^{54}\mathrm{poly}(t)e^{-\pi t}$\\
where $t \sim 2.0 \times 10^{28}$}}  \\
\hline
\end{tabularx}
\caption{A second illustrative example of a dS saddle ({\it{c.f.}} Table \ref{t:solex}).   Here $\Lambda\simeq \mathcal{O}(10^3)\,\Lambda_{\text{obs}}$, whilst $M_s\sim8 \text{TeV}$  and $\ell_{\text{KK}}$ are consistent with experimental constraints}.
\label{t:solex2}
\end{table}
\vspace{4mm}

\section{Conclusions and outlook}

\label{S:concl}

We have presented an explicit non-supersymmetric string theory construction in which symmetries enforce an exact cancellation of the open-string gauge and matter sectors to the cosmological constant, $\Lambda_{\text{open}}=0$, without introducing new light states in the would-be visible sector.  Symmetry also suppresses the contributions to the cosmological constant from the closed-string gravitational sector and we moreover propose a dynamical mechanism that renders them exponentially small, allowing the total one-loop vacuum energy to match the observed Dark Energy scale, $\Lambda_{\text{closed}}=\Lambda_{\text{DE}}$.  Our model has many features in common with the pioneering papers \cite{Angelantonj:2003hr} and \cite{carlo1}, whilst moreover providing a clear spacetime interpretation, removing open-string instabilities and including a mechanism to stabilise all closed-string moduli at the required scale.
 
 Our string construction involves an interplay between Brane Supersymmetry Breaking and Scherk-Schwarz supersymmetry breaking.  In particular, we envoke a Scherk-Schwarz orbifolding and orientifolding of type II string theory that produces a configuration of D-branes and O-planes that are non-mutually supersymmetric, and yet do not result in the usual NSNS disk-tadpole instability and associated tree-level vacuum energy $\sim M_{s}^4$; rather, the tree-level vacuum energy exactly cancels \cite{Coudarchet:2021qwc}.  By distributing the D-branes on the O-planes appropriately, this exact cancellation can persist at one-loop, thanks to an exact matching between bosonic and fermionic degrees of freedom, $n_F=n_B$, at all mass-levels.  Despite this matching, supersymmetry is broken explicitly on the individual brane stacks, with the localised spectra exhibiting a misaligned supersymmetry throughout the string towers.  
 
 In detail, the construction we present has 16 $\times$ (O3$^-$, O3$^{-}$, $\overline{\text{O}3}^+$,$\overline{\text{O}3}^-$)-planes and 16 (half-)D3-branes.  A non-supersymmetric USp(8) eight D3-brane stack on one of the $\overline{\text{O}3}^+$ and eight isolated SO(1) D3-branes on eight of the $\overline{\text{O}3}^-$-planes, realises $n_F=n_B$, and is moreover stable with respect to the open-string moduli.  The origin of the $n_F=n_B$ matching can be seen via an intermediate unstable\footnote{Note that this instability is too strong to be cosmologically viable.} USp(8)-SO(8) configuration (see \cite{Angelantonj:2003hr} and \cite{carlo1} for similar USp(8)-SO(8) constructions).  Here, at the massless level, the matching is clear from the fact that -- under interchange of orthogonal and symplectic structures -- symmetric and antisymmetric rank-two tensors exchange roles, with dim(adj$_\text{USp(8)}$)=dim(Sym$^2_\text{SO(8)}$) and dim(adj$_{\text{SO(8)}}$)=dim($\wedge^2_\text{USp(8)}$);  for the massive levels, there is a similar interchange between projectors in the NSNS/RR sectors for the USp/SO branes. The matching of degrees of freedom in the USp(8)-SO(8) set-up then survives a displacing of the SO(8) D3-branes along supersymmetric directions to a stable configuration.

 Whilst the open-string contributions to the one-loop vacuum energy exactly cancel, the closed-string contributions are set by the Scherk-Schwarz supersymmetry breaking scale, and are computed to go as $M_{s}^4/R_{\text{SS}}^4$, assuming the Scherk-Schwarz directions, $R_{\text{SS}}$ in string units, are the largest ones.  Therefore -- at one-loop -- $R_{\text{SS}}$ suffers a runaway instability, as does the dilaton after Weyl rescaling to the Einstein frame.  It is worth noting that, up to this point, our worldsheet computations allow us to work to all finite orders in $\alpha'$, summing contributions from the full perturbative string towers, with results tamed by the worldsheet modular symmetries.  We next turn to an $\mathcal{N}=1$ EFT description, via a further orbifolding with a freely-acting discrete symmetry, to allow us to study the effects of non-perturbative corrections.  We find that the interplay between the Scherk-Schwarz supersymmetry breaking -- described within supergravity as a no-scale supersymmetry-breaking flux background -- and standard non-perturbative D(-1)-instantons and ED3-instantons, can stabilise all closed string moduli at weak coupling and in a dS saddle.  Remarkably, the balancing between the Scherk-Schwarz $V_{\text{SS}} \sim -\mpl^4/R_{\text{SS}}^8$ and the non-perturbative $V_{\text{np}} \sim \mpl^4 e^{-1/g_{s}}f(g_{s})/R_{\text{SS}}^4$, dynamically sets the size of the Scherk-Schwarz dimensions to be exponentially large in $1/g_s$, and the vacuum energy to be correspondingly exponentially small.

 Our construction demonstrates how supersymmetry breaking in string theory can evade the traditional lore that supersymmetry can only protect the cosmological constant down to $ \gtrsim \mathcal{O}(\text{TeV})$ scales, given that superpartners to the visible sector have not been observed up to $\lesssim \mathcal{O}(\text{TeV})$.  It relies on a non-supersymmetric orientifolding $\Omega \Pi_4 \dots \Pi_9 (-1)^{F_L}(-\delta_{w_9})^F$ -- the gauging of a discrete symmetry that reverses the sign of the coordinates transverse to the corresponding O-planes and flips the orientation of the string, together with a winding shift for the spacetime fermions -- which induces a Brane Supersymmetry Breaking where the usual tadpole and tachyon instabilities can be avoided in certain regions of moduli space.  From the spacetime perspective, the $n_F=n_B$ matching is ensured by geometrically sequestered sectors that carry no mutual gauge charges.  Whilst supersymmetry is completely absent in the open-string sectors, it is only spontaneously broken in the closed-string sector, with the orientifold projection combining consistently with the Scherk-Schwarz orbifolding $(-1)^F \delta_{w_8}\delta_{p_9}$.   Thus the non-supersymmetric open-string sectors are coupled to a closed-string sector with supersymmetry-breaking scale $M_{3/2} \sim M_{\text{KK}} \sim \Lambda^{1/4}$.  This interplay between open-string and closed-string supersymmetry breaking scales and their backreaction on the vacuum energy is reminiscent of the Supersymmetric Large Extra Dimensions scenario \cite{Aghababaie:2003wz, Burgess:2004ib}.  The suppression of the closed-string vacuum energy from $M_{s}^4$ down to $M_{\text{KK}}^4$ has been further motivated by Swampland principles in the Dark Dimensions scenario \cite{Montero:dd}. 

 Our study opens up several important questions.  First and foremost is whether the one-loop cancellation $\Lambda_{\text{open}}=0$ persists to higher loops.  Reference \cite{carlo1} puts forward some qualitative arguments that higher genus amplitudes do also vanish for their USp(8)/SO(8) configuration; it is tempting then to further speculate that separating the SO(8) branes along supersymmetric directions might preserve the purported cancellation, as we have seen happen at one-loop.  Although we have put forward a successful moduli stabilisation scenario, it remains to dovetail this scenario onto an explicit string construction with $\Lambda_{\text{open}}=0$.  Most notably, whilst the supergravity description of $(-1)^R$ Scherk-Schwarz orbifolds is relatively well-understood, that of $(-1)^F$ Scherk-Schwarz orbifolds is not.  Our moduli stabilisation scenario turns out to favour two Supersymmetric Large Extra Dimensions, rather than a single Dark Dimension.  Whilst Supersymmetric Large Extra Dimensions have the added attraction of addressing the gauge hierarchy problem, it would also be interesting to identify dynamics that give rise to the Dark Dimension.  Observational and experimental bounds on one or two large extra dimensions have recently been reviewed in \cite{Burgess:2023pnk, antoniadis2DD}.  Given moduli stabilisation, it would be important to consider the implications of non-perturbative instabilities, such as those considered in \cite{Angelantonj:2007ts, Buratti:2018onj}. 
 
 It is intriguing that the USp(8) group that emerges from our construction can be broken to the Standard Model gauge group SU(3) $\times$ SU(2) $\times$ U(1), but more structure ({\it{e.g.}} further orbifolding or magnetised D-branes) would have to be added to induce chiral fermions and the matter representations of the Standard Model, all without spoiling the $\Lambda_{\text{open}}=0$.  The model of Dark Energy that emerges from our construction is a hilltop quintessence model.  Cosmological tests of hilltop quintessence against the wealth of available cosmological data have recently been made in \cite{Bhattacharya:2024kxp}, with a preference for the $w_0w_a$-parameterisation not yet statistically significant, but perhaps signalling some further interesting physics like Dark sector interactions \cite{Khoury:2025txd}. Hilltop quintessence also presents long-standing phenomenological problems: not least, unobserved fifth forces from the light quintessence field (and we have further even lighter moduli), which may suggest some kind of screening mechanism \cite{Burgess:2021qti, Brax:2023qyp}, and a fine-tuning of initial conditions.  Another question is -- given a cancellation of the vacuum energy between visible and hidden sectors -- what happens after a phase transition in the visible sector?  In this respect, it is interesting to ask if a low energy EFT description of our cancellation mechanism exists, perhaps related to ideas such as \cite{Burgess:2021obw}.  It remains to be seen whether our proposed solution to the Cosmological Constant Problem and Dark Energy can be used as a guide towards a more complete picture of particle physics and cosmology.

 We hope to report on some of these open questions in the near future.

\acknowledgments

We thank Henry Stubbs for helpful correspondence on the astrophysical constraints on large extra dimensions, and Bruno Bento and Joaquim Gomes for useful discussions.  We are also grateful to Carlo Angelantonj, Miguel Montero, Cliff Burgess and Fernando Quevedo for comments on the first version of this paper.  The work of SLP is partially funded by STFC grant ST/X000699/1.  MS was funded by a H.G. Baggs Fellowship, University of Liverpool.

\appendix

\section{One-loop Scherk-Schwarz vacuum energy}\label{A:ssvacuumenergy}
In this appendix we present the computation of the one-loop effective potential resulting from  
the Scherk-Schwarz supersymmetry breaking in the bulk in the string model presented in Section \ref{sec:uspso1}, further adapted to the freely-acting orbifold $T^6/\mathbb{Z}'_2\times \mathbb{Z}'_2$ described in Section \ref{sec:freelyactingorbifold}.

In general, the one-loop potential of a given orientifold model is given by
\begin{equation}\label{v1loopamplitudes}
    \mathcal{V}_{\text{1-loop}}=-\frac{M_s^4}{2(2\pi)^4}(\mathcal{T}+\mathcal{K}+\mathcal{A}+\mathcal{M})\, .
\end{equation}
A salient feature of the 4d open-string model constructed in Section \ref{sec:uspso1} is that $\mathcal{K}=\mathcal{A}=\mathcal{M}=0$ identically. Therefore the one-loop effective potential receives its only contribution from the torus partition function
\begin{equation}\label{toruscdp4d}
\begin{split}
    \mathcal{T}=\frac{1}{2}\int_{\mathcal{F}} \frac{d^2\tau}{\tau_2^3}\frac{1}{|\eta^8|^2}\sum_{\vec{m},\vec{n}}\biggl(&\Lambda_{\vec{m},\vec{n}}|V_8-S_8|^2+(-1)^{m_9+n_8}\Lambda_{\vec{m},\vec{n}}|V_8+S_8|^2\\
    &+\Lambda_{\vec{m}+\vec{\delta},\vec{n}+\vec{\epsilon}}\,|O_8-C_8|^2+(-1)^{m_9+n_8}\Lambda_{\vec{m}+\vec{\delta},\vec{n}+\vec{\epsilon}}\,|O_8+C_8|^2\biggr)\, ,
 \end{split}   
\end{equation}
with $\vec{\delta}:=(0,0,0,0,0,1/2)$, $\vec{\epsilon}=(0,0,0,0,1/2,0)$ the shift vectors. 

As commented in Section \ref{sec:freelyactingorbifold}, this feature is expected to hold in the freely-acting orbifold $\mathbb{T}^6/\mathbb{Z}'_2\times \mathbb{Z}'_2$ as the latter basically acts by projecting out closed-string states as a standard $\mathbb{Z}_2\times \mathbb{Z}_2$ orbifold but with a twisted sector which is now massive. Therefore, once the orientifold and the $\mathbb{Z}_2'\times \mathbb{Z}_2'$  projectors, with $P_{\mathbb{Z}_2'\times \mathbb{Z}_2'}=\frac{1}{4}(1+g+f+h)$ as defined in Eq. \eqref{free-act3}, are inserted into the trace of the torus partition function, we have
\begin{eqnarray}
        \mathcal{T}=\mathcal{T}_{\text{untw}}+\mathcal{T}_{\text{tw}}\,,
\end{eqnarray}
where $\mathcal{T}_{\text{untw}}$ is essentially the amplitude in (\ref{toruscdp4d}) with an overall $1/8$ factor in front of it. Crucially, because the twisted sector is massive, we expect the contribution of $\mathcal{T}_{\text{tw}}$ to the vacuum energy to be completely negligible at large volume; hence, the one-loop effective potential receives its only significant contribution from the untwisted torus amplitude  $\mathcal{T}_{\text{untw}}$ that we shall now focus on.

The integration of the torus partition function is usually carried out 
by performing the so-called unfolding technique (UT) \cite{Dixon:1990pc,Kiritsis:1997hf,Trapletti:2002uk}, to switch from an integral over the $SL(2,\mathbb{Z})$ fundamental domain to a more suitable integral over the strip $\mathcal{S}$. 
The starting point of the UT is to recall that the (untwisted) torus amplitude\footnote{From now on we drop the subscript.} can be rewritten as the orbit of its non-trivial $T$-invariant element 
\begin{equation}\label{torusUTg}
    \mathcal{T}=\frac{1}{16}\int_\mathcal{F}\frac{d^2\tau}{\tau_2^2}\frac{1}{\tau_2^4|\eta|^{16}}\sum_{\vec{m},\vec{n}}\sum_{g\in \mathcal{G}}g\,\circ\left[\tau_2^3(-1)^{m_9+n_8}\Lambda_{6,6}\left|V_8+S_8\right|^2\right]\, ,
\end{equation}
under $\mathcal{G}=\{1,S,TS\}$, subgroup of $SL(2,\mathbb{Z})$, whose action ensures modular invariance of the amplitude under the full modular group, with the phase $(-1)^{m_9+n_8}$ resulting from the action of the Scherk-Schwarz orbifold.
The freely-acting orbifold  factorises the internal 6-torus $\mathbb{T}^6=\mathbb{T}^2_{1}\times \mathbb{T}^2_2\times \mathbb{T}^3_3$. As a consequence, the 6d internal lattice $\Lambda_{6,6}\equiv \Lambda_{\vec{m},\vec{n}}$ factorises as well
\begin{equation}
    \Lambda_{6,6}=\prod_{k=1}^3\Lambda_{2,2}^{(k)}\,,
\end{equation}
where for each 2-torus $\mathbb{T}^2_k$ the associated 2d lattice has the standard expression in terms of the left and right momenta
\begin{equation}
\begin{split}
    p^{}_{(k)L\,i}=m^{(k)}_i+(g_{(k)}-b_{(k)})_{ij}n^{(k)}_j&\,,\quad p_{(k)R\,i}=m^{(k)}_i-(g_{(k)}+b_{(k)})_{ij}n^{(k)}_j\,,\quad i,j=1,2 \\[4mm]
    \Lambda^{(k)}_{2,2}:=q^{\frac{1}{4}p_{(k)L}^2}\bar{q}^{\frac{1}{4}p_{(k)R}^2}&=e^{\frac{i\pi}{2} \tau_1(p_{(k)L}^2-p_{(k)R}^2)}e^{-\frac{\pi}{2} \tau_2(p_{(k)L}^2+p_{(k)R}^2)} \, ,
\end{split}    
\end{equation}
where {\it{e.g.}} $(m_9,n_8)\equiv (m^{(3)}_2,n^{(3)}_{1})$ in this notation,  
 $g_{(k)}$ is the metric on the 2-torus $\mathbb{T}^2_{k}$ and for the sake of generality we also included on each 2-torus  a quantised background of the antisymmetric NSNS 2-form $B_2$, with components $b_{(k)ij}\in\mathbb{Z}/2$,  allowed by the orientifold projection. As we are interested in the moduli dependence of the one-loop effective potential, we conveniently recast the background data of each torus, metric and background $B_2$-field, in terms of  geometric K\"ahler and complex structure moduli\footnote{We stress that $b_{(k)12}$ are simply (quantised) background data and not moduli from the 4d point of view.}$^,$\footnote{Note the difference between the definitions of the complex structure moduli in the present string computation, $U_k$ \eqref{UT - U}, and in the supergravity basis, $U_i$ \eqref{eq:sugramoduli}, with $U_i=-i U_k$.  \label{foot:Ubases}}
\begin{subequations}
\begin{align}
    T'_k &  = b_{(k)12} + i\sqrt{\mathrm{det} \, g_{(k)}}, \label{UT - T} \\
    U_k & =  \dfrac{1}{g_{(k)11}} \bigl[ g_{(k)12} + i \sqrt{\mathrm{det} \, g_{(k)}} \bigr]. \label{UT - U}
\end{align}
\end{subequations}
Then, $p_{L,R}^{(k)}=p_{L,R}^{(k)}(T'_k,U_k)$, but for our purposes we keep this dependence implicit  and simply recall the familiar identity from level-matching that will be used later on
\begin{equation}\label{levelmatch}
    p_{(k)L}^2-p_{(k)R}^2= 4\,m^{(k)}_i n^{(k)}_i\, ,
\end{equation}
with sum on $i=1,2$ understood.

We are interested in the region of moduli space where the Scherk-Schwarzed torus $\mathbb{T}^2_{3}$ is the largest torus in the compactification, {\it{i.e.}} the radii $R_{8}\equiv \sqrt{{g_{(3)_{11}}}}$ and $R_9\equiv \sqrt{{g_{(3)_{22}}}}$ are the largest radii. We will thus compute the one-loop effective potential in two cases a) two large extra (susy-breaking) dimensions, {\it{i.e.}} $R_8,R_9\gg 1$ with no further hierarchy among them and b) one large extra (susy-breaking) dimension, {\it{i.e.}} $R_9\gg R_8\gg1$. Both cases are well-defined as they include the supersymmetric limit $R_9\rightarrow \infty$, where the torus amplitude is tachyon-free and hence stays finite.

\paragraph*{Two large extra dimensions}

Convergence in the large radii regime requires the unfolding technique to start with a Poisson resummation  on the KK numbers $m^{(3)}_1\equiv m_8$ and $m^{(3)}_2\equiv m_9$,  which recasts the lattice in (\ref{torusUTg}) in the following form
\begin{equation}\label{poissonresumm2}
    \sum_{m^{(3)}_i,n^{(3)}_i\in \mathbb{Z}}(-1)^{m^{(3)}_2+n^{(3)}_1}\Lambda^{(3)}_{2,2}=\frac{\text{Im}T_3'}{\tau_2}\sum_{\ell^{(3)}_i,n^{(3)}_i\in\mathbb{Z}}e^{-\frac{\pi}{\tau_2}(g_{(3)}+b_{(3)})_{ij}(\tilde{\ell}^{(3)}_i+n^{(3)}_i\tau)(\tilde{\ell}^{(3)}_j+n^{(3)}_j\bar{\tau})+i\pi n^{(3)}_1}\, ,
\end{equation}
where ${\ell^{(3)}_i\in \mathbb{Z}}$, $i=1,2$, are the resummed KK-numbers and we define
\begin{equation}
    \tilde{\ell}^{(3)}=(\ell_1^{(3)},\ell_2^{(3)}+1/2)\,,
\end{equation}
where we can clearly see the customary effect of the  Scherk-Schwarz phase $(-1)^{m^{(3)}_{2}}$ in producing a Poisson resummed half-integer, instead of integer, KK number. 
From the RHS of (\ref{poissonresumm2}) it is then possible to find a quite convenient expression of the resummed lattice in terms of K\"ahler and complex structure moduli (\ref{UT - T}), (\ref{UT - U})
\begin{equation} \label{UT identity}
 \sum_{m^{(3)}_i,n^{(3)}_i\in \mathbb{Z}}(-1)^{m^{(3)}_2+n^{(3)}_1}\Lambda^{(3)}_{2,2}=\frac{\text{Im}T_3'}{\tau_2}\sum_{\set{A}}e^{{i\,\frac{\pi}{2} (1-T'_3) \, \mathrm{det} \, A - \frac{\pi}{4\tau_2}\frac{\text{Im}(T'_3)}{\text{Im }U_3} \biggl| (1, \, U_3) \, A 
   \left({\substack{\tau \\[2mm] 1}}\right) 
     \biggr|^2}}\, ,
\end{equation}
where the sum is now on matrices $A\in\text{Mat}_{2\times 2}(\mathbb{Z})$ of the form 
\begin{equation}\label{formA}
    A=\begin{pmatrix}
        2n^{(3)}_1& 2\ell^{(3)}_1\\
        2n^{(3)}_2 &2 \ell_2^{(3)}+1
    \end{pmatrix}\, .
\end{equation}
 The torus amplitude thus becomes 
\begin{equation} \label{torus UT1final}
    \mathcal{T} = \frac{\text{Im} T'_3}{16} \int_{\mathcal{F}} \frac{d^2\tau}{\tau_2^2}\frac{1}{\tau_2^4|\eta|^{16}} \, \sum_{g \in G} g \Biggl[\tau_2^2\,\Lambda^{(1)}_{2,2}\,\Lambda^{(2)}_{2,2}\sum_{\set{A}}e^{i\,\frac{\pi}{2} (1-T'_3) \, \mathrm{det} \, A - \frac{\pi}{4\tau_2}\frac{\text{Im }T'_3}{\text{Im }U_3} \left| (1, \, U_3) \, A 
    \left({\substack{\tau \\ 1}}\right) 
     \right|^2}|{V_8 + S_8}|^2 \Biggr]\, .
\end{equation}
It is now straightforward to see that, for any $2\times2$ matrix $A$ the right multiplication $A M$ by $M\in{SL}(2,\mathbb{Z})$ in
\begin{equation} \label{UT identity 2}
    \frac{1}{\tau_2} \biggl| (1, \, U) \, A M \left({\substack{\tau \\[2mm] 1}}\right)  \biggr|^2 = \dfrac{1}{\tau'_2} \biggl| (1, \, U) \, A \left({\substack{\tau' \\[2mm] 1}}\right)  \biggr|^2
\end{equation}
acts as a $SL(2,\mathbb{Z)}$ modular transformation on $\tau$,
with   $\tau' = M \tau$ defined by the usual M\"obius action. Of course we also have $\textrm{det}(A M)=\text{det}A$. A quick check then shows that the matrix $A M$ is still of the form (\ref{formA}) if $M$ belongs to the congruence subgroup of $SL(2,\mathbb{Z})$
\begin{equation}
\begin{split}
    \Gamma_1 (2) &= \left\{ \begingroup\renewcommand{\arraystretch}{0.9}\scalebox{0.8}{$\begin{pmatrix}
        a & b \\
        c & d
    \end{pmatrix}$}\endgroup \in \mathrm{SL}(2,\mathbb{Z}) : \, c \equiv 0 \pmod{2} \, \wedge \, d \equiv 1 \pmod{2} \right\}\,.
 \end{split} 
\end{equation}
Hence, we can trade $\sum_{\set{A}}A=\sum_{\set{\tilde{A}}}\sum_{\gamma\in \Gamma_1(2)} \tilde{A}\gamma$, where the sum is on some representative matrices $\tilde{A}_i$ which are not connected each other via modular transformations. We are thus considering the orbits of the set of matrices (\ref{formA}) in $SL(2,\mathbb{Z})$ and there are only two orbits whose generators are the representative matrices we are looking for.  The first is the so-called degenerate orbit, generated by matrices with vanishing determinants of the form
    \begin{equation*}
        \tilde{A}_0 = \matr{0}{2p}{0}{2q+1} , \qquad p, q \in \mathbb{Z}.
    \end{equation*}
     Due to the invariance $\tilde{A}_0 = \tilde{A}_0 T$, they actually generate  orbits for $\Gamma_1(2)/\mathrm{T}$.  The second so-called  non-degenerate orbit is instead represented by matrices with non-vanishing determinant of the form
    \begin{equation*}
        \tilde{A}_1 = \matr{2r}{2p}{0}{2q+1}\,, \qquad 2r>2p\ge 0\,,\quad q\in\mathbb{Z}\, .
    \end{equation*}
If we now remind ourselves that the combination $[\tau_2^2(\Lambda_{2,2})^2]$ is invariant under the full $SL(2,\mathbb{Z})$ modular group while the character combination $V_8+S_8$ is invariant only under $T$ and $S T^2 S$ transformations, which generate the congruence subgroup $\Gamma_0 (2)\supset \Gamma_1 (2) $
\begin{equation}
\begin{split}
     \Gamma_0 (2) &= \left\{ \begingroup\renewcommand{\arraystretch}{0.9}\scalebox{0.8}{$\begin{pmatrix}
        a & b \\
        c & d
    \end{pmatrix}$}\endgroup
    \in \mathrm{SL}(2,\mathbb{Z}) : \, c \equiv 0 \pmod{2} \right\}\,,
 \end{split} 
\end{equation}
then the torus amplitude can be  written as the sum of the two orbits contributions
\begin{align}\label{tcdpm8m9noU}
    \mathcal{T} & = \begin{aligned}[t] \frac{\text{Im}T'_3}{16} \!\!\sum_{\substack{g \in G, \\ \gamma_* \in \Gamma_1 (2) / \mathrm{T}}}\!\! \int_{\mathcal{F}} \frac{d^2 \tau}{\tau_2^2} \, \frac{1}{\tau_2^4|\eta|^{16}} g\circ\gamma_\star\circ \left[\tau_2^2\Lambda^{(1)}_{2,2}\Lambda^{(2)}_{2,2}\sum_{\set{\tilde{A}_0}} e^{i\frac{\pi}{2}(1-T'_3) \, \mathrm{det} \, \tilde{A}_0 - \frac{\pi}{4 \tau_2} \frac{\text{Im}T'_3}{\text{Im}U_3} \, \left| (1, \,U_3) \, \tilde{A}_0 \, \left({\substack{\tau \\ 1}}\right) \right|^2} \biggl| \frac{\vartheta_2^4}{\eta^{4}}  \biggr|^2 \right]& \\
    + \frac{\text{Im}T'_3}{16} \!\!\sum_{\substack{g \in G, \\ \gamma \in \Gamma_1 (2)}}\!\! \int_{\mathcal{F}} \frac{d^2 \tau}{\tau_2^2} \frac{1}{\tau_2^4|\eta|^{16}} g\circ\gamma\circ\left[\tau_2^2\Lambda^{(1)}_{2,2}\Lambda^{(2)}_{2,2} \sum_{\set{\tilde{A}_1}} e^{i\frac{\pi}{2}(1-T'_3) \, \mathrm{det} \, \tilde{A}_1  - \frac{\pi}{4 \tau_2}\frac{\text{Im}T'_3}{\text{Im}U_3} \, \left| (1, \, U_3) \, \tilde{A}_1 \, \left({\substack{\tau \\ 1}}\right) \right|^2} \biggl|\dfrac{\vartheta_2^4}{\eta^{4}} \biggr|^2\right] & .
    \end{aligned}
\end{align}
Hence, given the identity $(1+S+TS)\circ\Gamma_1(2)=SL(2,\mathbb{Z)}$, it then  holds that
\begin{equation*}
    \sum_{\substack{g \in G, \\ \gamma_* \in \Gamma_1 (2) / {T}}}\!\! g \circ\gamma_* (\mathcal{F}) = \mathcal{S}, \qquad \sum_{\substack{g \in G, \\ \gamma \in \Gamma_1 (2)}}\!\! g\circ \gamma (\mathcal{F}) = 2\,\mathbb{C}^+,
\end{equation*}
and we see that after a change of variable the degenerate orbit unfolds into the strip, while the non-degenerate one unfolds into the double-cover of the upper-half complex plane, thus completing the UT procedure. The integral over the fundamental domain finally reduces to
\begin{align}
    \mathcal{T} = &\,16\cdot\text{Im}T'_3 \int_{-1/2}^{1/2}d\tau_1 \int_0^\infty\frac{d\tau_2}{\tau_2^{4}}\sum_{\substack{k,\bar{k}\in\mathbb{N}\\m_i,n_i\in \mathbb{Z}\\p,q\in\mathbb{Z}}}\,d_k d_{\bar{k}} e^{2i\pi\tau_1(k-\bar{k}+\sum_{j=1}^2m^{(j)}_in^{(j)}_i)}\nonumber\\[-10mm]
    &\qquad\qquad\qquad\qquad\qquad \qquad \qquad \qquad \qquad \times e^{- \frac{\pi}{4 \tau_2}\frac{\text{Im}T_3}{\text{Im}U_3}|2p+(2q+1)\,U_3|^2}e^{-\pi\tau_2\mathcal{M}^2}\nonumber  \\[5mm]
    + &32\cdot\text{Im}T'_3 \int_{-\infty}^{\infty}d\tau_1\int_{0}^\infty \frac{d\tau_2}{\tau_2^4}\sum_{\substack{k,\bar{k}\in\mathbb{N}\\m_i,n_i\in\mathbb{Z}\\0\le p<r\\q \in \mathbb{Z}}}(-1)^r 64\,d_k d_{\bar{k}}e^{-i\pi\text{Re}T'_3r(2q+1)}\nonumber\\[-10mm]
    &\qquad\qquad\qquad\qquad\qquad \qquad \qquad \times e^{-\frac{\pi}{\tau_2}\frac{\text{Im}T'_3}{\text{Im}U_3}r^2\tau_1^2-\frac{\pi}{\tau_2}\frac{\text{Im}T'_3}{\text{Im}U_3}\left(2rp+r(2q+1)\text{Re}U_3+\frac{i\pi}{2}(k-k'+\sum_{j=1}^2 m^{(j)}_in^{(j)}_i)\right)\tau_1}\nonumber\\
    &\qquad \qquad \qquad \qquad \qquad \qquad  \qquad \times e^{-\frac{\pi}{4\tau_2}\frac{\text{Im}T'_3}{\text{Im}U_3}|2p+(2q+1)U_3|^2}
    e^{-\pi \tau_2\left(\mathcal{M}^2+r^2\frac{\text{Im}T'_3}{\text{Im} U_3}\right)}\,,
\end{align}
where we have moreover inserted the $q$-expansion for the characters
\begin{equation}\label{eqA:qexp}
\begin{split}
    \frac{V_8}{\eta^8}[q]=\frac{S_8}{\eta^8}[q]\equiv\frac{\theta_2^4}{2\eta^{12}}[q]&=8\sum_{k=0}^{\infty} d_k q^{k}\,,\quad d_0=1\,,
\end{split}    
\end{equation}
and we have defined 
\begin{equation}
\mathcal{M}^2:=2(k+\bar{k})+\sum_{j=1}^2p_{(j)L}^2+p_{(j)R}^2\, ,
\end{equation}
which encodes the mass-square contributions from the string oscillators and  the KK and winding modes associated only to the smallest torii $\mathbb{T}^2_{1}$ and $\mathbb{T}^{2}_2$. Notice that the dependence of the one-loop vacuum energy on the K\"ahler and complex structure moduli $U_{1,2}$ and $T_{1,2}$ is implicit through $p_{(1,2)L}$ and $p_{(1,2)R}$ in $\mathcal{M}^2$.
Let us start from the degenerate orbit contribution. The $\tau_1$-integral simply imposes the level-matching conditions
\begin{equation}
    k-\bar{k}+\sum_{j=1}^{2}m_i^{(j)}n^{(j)}_i=0\, ,
\end{equation}
and we highlight that the KK and winding modes of the Scherk-Schwarz torus $m^{(3)}_i$ and $n^{(3)}_i$ do not appear in the equation.
The leftover $\tau_2$-integral is then of the well-known type
\begin{equation}\label{tau2int}
    \int_0^{+\infty} \frac{d\tau_2}{\tau_2^{1+\lambda}}e^{-c\,\tau_2-\frac{b}{\tau_2}}=\frac{2}{b^\lambda}(c\,b)^\frac{\lambda}{2}K_\lambda\left(2\sqrt{c\,b}\right)=\frac{1}{b^\lambda}\,\Gamma(\lambda)\,\mathcal{H}_\lambda\left(\sqrt{c\,b}\right)\, ,\quad c,b>0
\end{equation}
where $K_\lambda$ is the modified Bessel function of the second kind and $\mathcal{H}_\lambda(z)$, defined through the second equality above, has the following limiting behaviours 
\begin{equation}\label{limitingh}
    \mathcal{H}_\lambda(z)\sim \frac{\sqrt{\pi}}{\Gamma(\lambda)}\,z^{\lambda-\frac{1}{2}}\,e^{-2\,z}\,\quad \text{if}\,\, z\gg 1\,,\qquad \mathcal{H}_\lambda(z)= 1-\frac{z^2}{\lambda-1}+\mathcal{O}(z^4)\,\quad \text{if}\, \,\ab z\ab\ll 1\, .
\end{equation}
In our case we have
\begin{equation}
    \lambda=3\,,\quad b=\frac{\pi}{4}\frac{\text{Im}T'_3
    }{\text{Im}U_3}|2p+(2q+1)\,U_3|^2\,,\quad c=\pi \mathcal{M}^2\, ,
\end{equation}
hence we eventually obtain
\begin{equation}\label{Tdeg}
    \mathcal{T}_{\text{deg}}=16\cdot\frac{4^{3}\Gamma(3)}{\pi^3}\cdot\sum_{\substack{k,\bar{k}\in\mathbb{Z}\\m_i,n_i\in \mathbb{Z}\\p,q\in\mathbb{Z}}}
d_k d_{\bar{k}}\frac{1}{(\text{Im}T_3')^2}\frac{(\text{Im}U_3)^3}{|2p+(2q+1)\,U_3|^6}\mathcal{H}_3\left(\frac{\pi}{2}\left(\frac{\text{Im}T'_3}{\text{Im}U_3}\right)^{1/2}\mathcal{M}|2p+(2q+1)U_3|\right)\, .
\end{equation}
To integrate the non-degenerate orbit we start again from the $\tau_1$-integral which is now Gaussian
\begin{equation}
\begin{split}
    &\int_{-\infty}^{+\infty}d\tau_1e^{-\frac{\pi}{\tau_2}\frac{\text{Im}T'_3}{\text{Im}U_3}r^2\tau_1^2-\frac{\pi}{\tau_2}\frac{\text{Im}T'_3}{\text{Im}U_3}\left(2rp+r(2q+1)\text{Re}U_3+\frac{i\pi}{2}(k-\bar{k}+\sum_{j=1}^2m^{(j)}_in^{(j)}_i)\right)\tau_1}\\
    &=\left(\frac{\tau_2\text{Im}U_3}{r^2\text{Im}T'_3}\right)^{1/2}e^{\frac{\pi}{4\tau_2}\frac{\text{Im}T'_3}{\text{Im}U_3}\left(2p+(2q+1)\text{Re}U_3\right)^2}e^{-\pi\frac{\tau_2}{r^2}\frac{\text{Im}U_3}{\text{Im}T'_3}(k-\bar{k}+\sum_{j=1}^2m_i^{(j)}m_i^{(j)})^2}e^{-2i\pi\frac{p}{r}(k-\bar{k}+\sum_{j=1}^2m_i^{(j)}n_i^{(j)})}\\
    &\qquad\qquad\qquad \quad \times e^{-i\pi\frac{2q+1}{r}\text{Re}U_3(k-\bar{k}+\sum_{j=1}^2m_{i}^{(j)}m_i^{(j)})}\,.
\end{split}    
\end{equation}
At this point the sum on $p$ becomes trivial
\begin{equation}
    \sum_{p=0}^{r-1}e^{-2i\pi \frac{p}{r}(k-\bar{k}+\sum_{j=1}^2m^{(j)}_in^{(j)}_i)}=\begin{cases}
        &r\,\quad \text{if}\,\quad k-\bar{k}+\sum_{j=1}^2m^{(j)}_in^{(j)}_i\equiv 0\pmod r\\
        &0\,\quad \text{otherwise}\,
    \end{cases}
\end{equation}
and cancels out with the factor $r$ in the denominator from the Gaussian integration.

Therefore, replacing  $\bar{k}=k+\sum_{j=1}^2m^{(j)}_in^{(j)}_i+\ell\,r$, $\ell\in \mathbb{Z}$, and moreover using (\ref{levelmatch}) the $\tau_2$-integral becomes
\begin{equation}
    \int_0^{+\infty}\frac{d\tau_2}{\tau_2^{7/2}}e^{-\frac{\pi}{4\tau_2}(2q+1)^2\text{Im}U_3\text{Im}T'_3}e^{-\pi\tau_2\left(4k+\frac{1}{2}\sum_{j=1}^2(3p_{(j)L}^2+p_{(j)R}^2)+\left(\sqrt{\frac{\text{Re}U_3}{\text{Re}T'_3}}\ell+\sqrt{\frac{\text{Re}T'_3}{\text{Re}U_3}}r\right)^2 \right)}\,,
\end{equation}
again of the form (\ref{tau2int}), this time with parameters
\begin{equation}
    \lambda=\frac{5}{2}\,,\quad b=\frac{\pi}{4}\text{Im}T'_3\,
    {\text{Im}U_3}\,(2q+1)^2\,,\quad c=4\pi k+\frac{\pi}{2}\sum_{j=1}^2(3p_{(j)L}^2+p_{(j)R}^2)+\pi\left(\sqrt{\frac{\text{Re}U_3}{\text{Re}T'_3}}\ell+\sqrt{\frac{\text{Re}T'_3}{\text{Re}U_3}}r\right)^2\, .
\end{equation}
The non-degenerate orbit is thus found to give
\begin{equation}\label{Tnondeg}
\begin{split}
    \mathcal{T}_{\text{non-deg}}&=32\cdot\frac{4^{5/2}\Gamma(5/2)}{\pi^{5/2}}\cdot\sum_{\substack{k\in\mathbb{N}\\m_i,n_i\in\mathbb{Z}\\r>0\\\ell,q \in \mathbb{Z}}}\frac{(-1)^r
    \,d_k d_{|k+m_in_i+\ell r|}}{(\text{Im}T'_3\text{Im}U_3)^2}\frac{e^{-i\pi(2q+1)(\text{Re}U_3\ell+\text{Re}T'_3r)}}{|2q+1|^5}\\
    &\times\mathcal{H}_{5/2}\left({\pi}|2q+1|(\text{Im}T'_3\text{Im}U_3)^{1/2}\sqrt{k+\frac{1}{8}\sum_{j=1}^2(3p_{(j)L}^2+p_{(j)R}^2)+\frac{1}{4}\left(\sqrt{\frac{\text{Re}U_3}{\text{Re}T'_3}}\ell+\sqrt{\frac{\text{Re}T'_3}{\text{Re}U_3}}r\right)^2
    }\right)\, .
\end{split}    
\end{equation}
 We then note that the argument of $\mathcal{H}_3$ in (\ref{Tdeg}) is $\mathcal{O}(R_9)$,  except for the unique possibility $k=\bar{k}=m_i^{(1),(2)}=n_i^{(1),(2)}=0$, thus $\mathcal{M}=0$, in which case it vanishes;  the argument of $\mathcal{H}_{\frac{9}{2}}$ in (\ref{Tnondeg}) is instead always $\mathcal{O}(R_8R_9)$ because of the lower bound $r\ge 1$. Then, according to the behaviour of $\mathcal{H}$ for large argument, we realise that \emph{the leading order contribution to the one-loop effective potential comes exclusively from the KK modes of the massless string states along the susy-breaking directions $T^2_3$}, and from (\ref{v1loopamplitudes}) reads, after rescaling $R_9\rightarrow 2 R_9$,
\begin{equation}\label{V1loopR8R9largest}
    V_{\text{1-loop}}\simeq-M_s^4\cdot16\cdot\frac{1}{2^3\pi^7}\cdot\frac{1}{(\text{Im}T'_3)^2}\mathcal{E}_{3}(U_3)\, ,
\end{equation}
where we recognise $n_b^{0}-n_f^{0}=16$ to be the difference between numbers of massless bosons and fermions in the spectrum of type IIB on the $T^6/\mathbb{Z}'_2\times\mathbb{Z}'_2$ orientifold after the Scherk-Schwarz orbifold: $n_f^0=0$ since now all the fermions acquired a mass,  while the number of massless bosons is left untouched $n_b^{0}=2+2\times (h^{2,1}+h^{1,1})+2=16$, respectively  given by the symmetric-traceless part of the 4d metric, $h^{2,1}=3$ complex structure moduli $U_i$ and 3 real K\"ahler moduli $\text{Re}(T_i')$ from the internal metric, 3 axions from the dimensional reduction of the RR 4-form $C_4$, grouped with the former into $h^{1,1}=3$ complex K\"ahler moduli, lastly the dilaton and the RR $C_0$ axion forming the axio-dilaton.
We note  the expected power-like behaviour in the Scherk-Schwarz torus volume, 
while the dependence on the torus complex structure is encoded in  the  non-holomorphic  weight-0 Eisenstein-like series
\begin{equation}\label{eqA:Epsilon3}
    \mathcal{E}_3(U)=\sum_{p,q\in\mathbb{Z}}\frac{(\text{Im}U)^3}{|p+(2q+1)U|^6}\, ,
\end{equation}
which is readily found to be invariant under the congruence subgroup $\Gamma_1(2)$  of the target-space $SL(2,\mathbb{Z})_U$ modular group 

\begin{figure}[t]
\centering
\begin{minipage}{0.45\textwidth}
    \includegraphics[width=\linewidth]{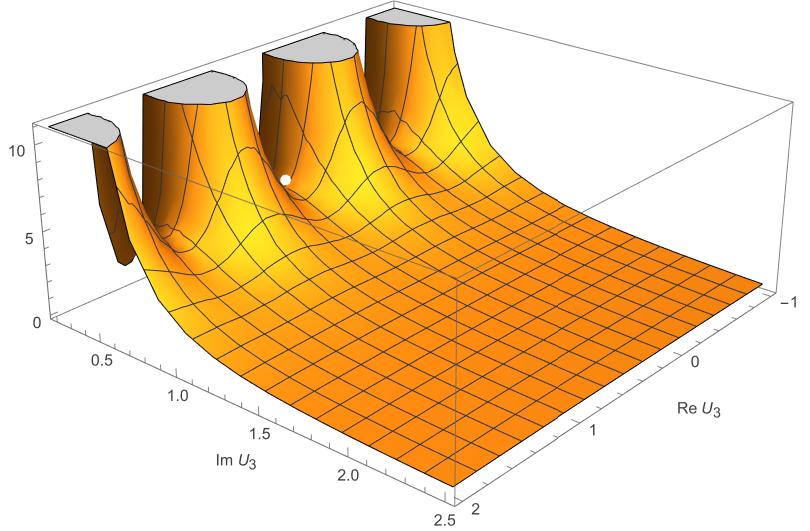}
\end{minipage}
\hfill
\begin{minipage}{0.45\textwidth}
    \includegraphics[width=\linewidth]{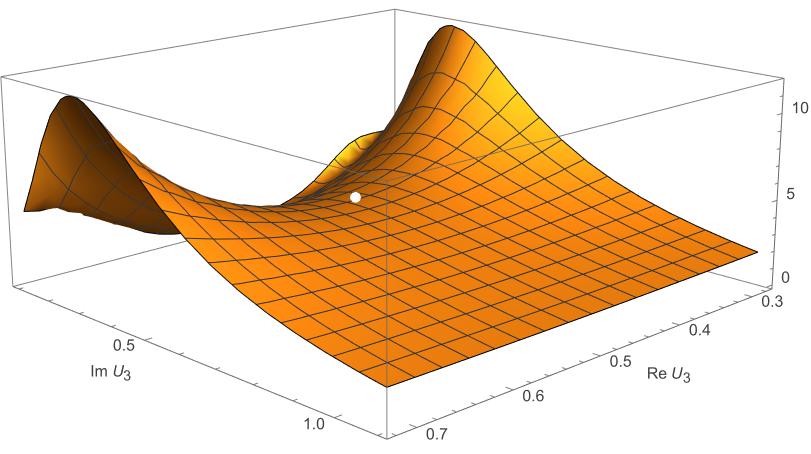}
\end{minipage}
\caption{The dependence of the leading order term in the one-loop effective potential on the complex structure $U_3$ through the function $\mathcal{E}_3(U_3)$, with its saddles and its manifest $\text{Re}\,U_3\rightarrow\text{Re}\,U_3+1$ symmetry. The saddle point $U_3^*=\frac{1}{2}(1+i)$ in the fundamental domain is highlighted with a white dot.}
\label{fig:E3}
\end{figure}
The $U_3$-dependence of the one-loop effective potential is plotted in Figure \ref{fig:E3}:
the function $\mathcal{E}_3(U)$ admits a critical point at $U^*=\frac{1}{2}(
1+i)$, which is found to be a saddle point; moreover, because of the $\Gamma_1(2)$-invariance, and being $T\in \Gamma_1(2)$, any point $U'=U^*+\mathbb{Z}$ is then also a saddle point. We thus conclude that the one-loop effective potential alone would fix the complex structure in the saddle $U_3=\frac{1}{2}(1+i)$, thus resulting in a negative runaway in the direction of $\text{Im} T'_3$.

All the other states in the degenerate orbit and all the states in the non-degenerate one -- {\it{i.e.}} the  massive string states and their KK/winding excitations along any directions, states with non trivial winding masses and non level-matched states -- are extremely massive in the large $R_8,R_9$ limit and therefore yield only exponentially suppressed contributions to the vacuum energy. This also implies that, up to exponentially suppressed terms, \emph{the K\"ahler and complex structure moduli $T_{1,2}$ and $U_{1,2}$ of the supersymmetric torii are flat directions of the one-loop effective potential}.

For our purposes, we shall notice  that $\mathcal{E}_3$ can be written as a linear combination of the non-holomorphic weight-0 Eisenstein series $E_3(U)$
\begin{equation}\label{relationwithEisenstein}
    \mathcal{E}_3(U)=\frac{1}{2^6}\left(2^3 E_3(U)-E_3(2U)\right)\, ,
\end{equation}
where
\begin{equation}
    E_s(U):=\sum_{(m,n)\in \mathbb{Z}/(0,0)}\frac{(\text{Im}U)^s}{|m+n \,U|^{2s}}\,,\quad s\in\mathbb{Z} ,
\end{equation}
 clearly invariant under the full $SL(2,\mathbb{Z})$.
It will also be useful to recall that $E_s(z)$ are eigenfunctions of the hyperbolic Laplacian with eigenvalue equation
\begin{equation}
    \Delta\,E_s(z)=s(1-s)\,E_s(z)\,, 
\end{equation}
where, for $z=x+iy$ the hyperbolic Laplacian is defined as
\begin{eqnarray}\label{hyperlaplacian}
    \Delta:=-y^2(\partial_x^2+\partial_y^2)\, .
\end{eqnarray}
From the identity (\ref{relationwithEisenstein}) and  the invariance of the operator $\Delta$  under the rescaling $z\rightarrow 2z$, we thus infer that $\mathcal{E}_{3}(z)$ is also an eigenfunction of $\Delta$ with eigenvalue equation
\begin{eqnarray}\label{eigenE3}
\Delta\,\mathcal{E}_3(z)=-6\,\mathcal{E}_3(z)\, .
\end{eqnarray}

\paragraph{One large extra dimension}
\vspace{2mm}

\noindent In this case the UT would start with a Poisson resummation over the KK number associated to the largest radius $R_9$ only, {\it{i.e.}} $m^{(3)}_2\equiv m_9$. The very same result, at least for what concerns the leading order term in the vacuum energy that we are interested in, can be obtained in a more straightforward way by discussing the $R_9\gg R_8$ limit, {\it{i.e.}} large $|U_3|$ limit, of the two large extra dimension result \eqref{V1loopR8R9largest}. This boils down to determining the large $|U_3|$ behaviour of $\mathcal{E}_3(U_3)$.

To this purpose we shall use the Gamma-function integral
\begin{eqnarray}
    \int_0^{+\infty} t^{s-1}e^{-\lambda\,t}dt=\frac{\Gamma(s)}{\lambda^s}\, ,
\end{eqnarray}
specialising $s=3$, $\lambda=|p+(2q+1)U|^2$, and permuting the sums with the integral, to  obtain an integral expression of our series
\begin{equation}
    \mathcal{E}_3(U_3)=\frac{(\text{Im}U_3)^3}{2^3\cdot\Gamma(3)}\sum_{p,q}\int_0^{+\infty}dt\,t^2\,e^{-((p+x_q)^2+y_q^2)t}\, ,
\end{equation}
where for convenience we have defined
\begin{equation}
    x_q:= (2q+1)\,\text{Re}U_3\,,\quad y_q:= (2q+1)\,\text{Im}U_3\, .
\end{equation}
We thus notice that, with a Poisson resummation on $p$ in the above integral 
\begin{equation}
   \mathcal{E}_{3}(U_3)= (\text{Im}U_3)^3\frac{\sqrt{\pi}}{2^3\cdot\Gamma(3)}\sum_{m,q\in\mathbb{Z}}e^{2i\pi m x_q}\int_0^{+\infty} dt\,t^{3/2}e^{-t y_q^2-\frac{\pi^2}{t}m^2}\,,
\end{equation}
we can link our series to a by now familiar integral: distinguishing the $m=0$ term integration from $m\ge 1$, where the integral is of the type
(\ref{tau2int}) with $\lambda=-5/2$, $b=\pi^2 m^2$ and $c=y^2_q$, we eventually find
\begin{equation}
  \mathcal{E}_3(U_3)= \frac{(\text{Im} U_3)^2}{2^3\cdot \Gamma(3)}\left(\sum_{q\in\mathbb{Z}}\frac{3\pi^2}{4}\frac{1}{|y_q|^5}+{\Gamma(-5/2)}\pi^{11/2}\sum_{m\neq 0}e^{2i\pi m x_q} |m|^5\mathcal{H}_{-5/2}(\pi |m| |y_q|)\right)\,.
\end{equation}
Since the $\mathcal{H}$-function yields only exponentially suppressed terms in the limit $\text{Im}U_3\gg1$, at leading order we thus have
\begin{equation}
    \mathcal{E}_3(U_3)\simeq\frac{1}{(\text{Im}U_3)^2}\frac{3\pi^2}{32\,\Gamma(3)}\sum_{q\in\mathbb{Z}}\frac{1}{|2q+1|^5}=\frac{1}{(\text{Im}U_3)^2}\frac{3\pi^2}{32\,\Gamma(3)}\frac{31\zeta(5)}{16}\, .
\end{equation}
Therefore, in the large $|U_3|\gg1$ limit, {\it{i.e.}} for the hierarchy $R_9\gg R_8\gg1$, (\ref{V1loopR8R9largest}) reduce to
    \begin{equation}\label{V1loopR9largest}
    V_{\text{1-loop}}=-\frac{93\,\zeta(5)}{32\pi^5}\frac{M_s^4}{(R_9\sin \omega_{89})^4}\,,
\end{equation}
which is the same result one would have obtained by performing the UT with a Poisson resummation on $m_9$ only.

\section{Open-string moduli stability}
\label{A:openstringmod}

In this appendix we study the stability at one-loop of the open-string moduli, {\it{i.e.}} the brane position moduli of the generic   $\usp(8)\times \prod_{i=1}^{16} \So(N_i)$ configuration in the 4d orientifold model described in Section \ref{sec:uspso1}.

In general, brane position moduli acquire masses whose squares are determined by the on-shell Hessian of the one-loop open-string effective potential
\begin{equation}
    \mathcal{V}_{\text{1-loop,open}}=-\frac{M_s^4}{2(2\pi)^4}(\mathcal{A}+\mathcal{M})\, ,
\end{equation}
where $\mathcal{A}, \mathcal{M}$ are respectively the annulus and M\"obius strip one-loop amplitudes.
We will show that, as expected, $\mathcal{V}_{\text{1-loop,open}}$ is extremised whenever all the branes sit on top of the O-planes at the fixed points of the orbifold/orientifold involution. Among these critical points only one is actually a minimum, {\it{i.e.}}  D-brane position moduli with positive mass-squareds,   where moreover the on-shell value of the one-loop open-string effective potential vanishes. This  happens when, besides the eight D3-branes on top of the $\overline{\text{O}3}_+$-plane at the origin, each of the remaining eight D3-branes is stuck on top of one of the 16 $\overline{\text{O}3}^-$-planes, giving an overall $\usp(8)\times \So(1)^8$ configuration. This one-loop result  confirms the tree-level expectation based on energetics considerations in terms of D-brane/O-plane repulsions and attractions. The $\usp(8)\times \So(1)^8$  configuration is hence the only one that yields stable brane positions and a vanishing cosmological constant.

For our purposes, let us  distinguish  between position moduli associated to the symplectic gauge group and the orthogonal gauge group factors. In general, the orientifold projection requires the D-brane positions to be symmetric under a  $\mathbb{Z}_2$ action,  which means that for a D-brane sitting at position $\vec{a}$ inside ${\mathbb T}^6$ there must be a mirror D-brane sitting at position $-\vec{a}$.  Consequently, when D-brane dynamics is allowed,  D-branes must move in pairs.  Therefore, we split the eight (half) D3-branes realising an  $\usp(8)$ gauge group at the origin into  a stack of $M=4$ D3-branes and their mirror $\overline{M}=4$ D3-branes and assign dynamical brane positions only to the first ones, {\it{i.e.}} $\vec{a}_\alpha=(a^4_\alpha,\dots,a^9_\alpha)$, where $\alpha=1,\dots,M=4$ runs on the independent degrees of freedom; the $\alpha$-th D3-brane in the dynamical stack would thus acquire position along direction $X^I$ given by $2\pi a^I_\alpha R_I$, with its  mirror D3-brane sitting at position $-2\pi a^I_\alpha R_I$. 
In the dual picture where all the ${\mathbb T}^6$ directions are T-dualised, D3-brane positions becomes D9-brane Wilson lines described by the following Wilson line matrix 
\begin{equation}\label{uspwl}
\begin{split}
    \mathcal{W}^I_{\text{USp}}&=\text{diag}\left(e^{2\pi i a^I_\alpha}, e^{-2\pi i a^I_\alpha};\alpha=1,\dots,4\right)\,.
  \end{split}  
\end{equation}

The D3-brane position moduli space associated to the orthogonal gauge group factors contains two disconnected components.
The first component sees an even number of (half) D-branes stacked on each of the 16 $\overline{\text{O}3}_-$-planes, {\it{i.e.}} $N_i$ even for some $i=1,\dots,16$.  Analogously to the $\usp$ case, then, we split into stack and mirror stack each of the $\So(N_i)$ D3-brane systems, with the $\beta_i$-th dynamical D3-brane now having position along $X^I$ given by $2\pi b^I_{\beta_i} R_I$ and its mirror D3-brane sitting at $-2\pi b^I_{\beta_i} R_I$, $\vec{b}_{\beta_i}=(b^4_{\beta_i},\dots,b^9_{\beta_i})$, $\beta_i=1,\dots,\frac{N_i}{2}$,  $i=1,\dots,16$. 
Upon switching to the T-dual picture, this component of the D3-brane position moduli space is described by the following D9-brane  Wilson line matrix 
\begin{equation}\label{sowl}
\begin{split}
    \mathcal{W}^I_{\text{SO}}&=\text{diag}\left( e^{2\pi i b^I_{\beta_i}},e^{-2\pi i b^I_{\beta_i}};\{\beta_i\}=1,\dots,N_i/2\right)\,,\quad \sum _{i=1}^{16} N_i=8\, .
  \end{split}  
\end{equation}
As said, the brane position moduli space of the orthogonal group factors admits a second component which is disconnected from the former, in the sense that it cannot be accessed  simply by varying the {\it{vev}} of some moduli in the first component.  For orthogonal gauge groups, in fact, it is possible to have branes with rigid positions, {\it{i.e.}} stuck at the orientifold fixed points. This indeed must be the case when some of the  brane stacks  have an odd number of branes {\it{i.e.}} $N_i$ odd for some $i=1,\dots,16$. To respect the orientifold projection, one half-brane in the stack must then be rigid; it is stuck on top of the orientifold plane and cannot move, because it does not come with a mirror pair.  This means that we can no longer associate a dynamical brane position to such a brane. In the T-dual picture, the component of D3-brane position moduli space when $k$ D3-branes are rigid is described by the following D9-brane Wilson line matrix: 
\begin{equation}\label{so2wl}
\begin{split}
    \mathcal{W}^I_{\So}=&\text{diag}\left(e^{2\pi i b^I_{\beta_i}},e^{-2\pi i b^I_{\beta_i}},\mathds{1}_k,-\mathds{1}_k,;\{\beta_i\}=1,\dots \left[N_i/2\right]\right)\,,\quad\sum_{i=1}^{16}\left[\frac{N_i}{2}\right]+k=4\,.
 \end{split}   
\end{equation}

Let us first discuss the cases (\ref{uspwl}) and (\ref{sowl}), as the case (\ref{so2wl}) follows through.  To obtain the one-loop potential for the brane position moduli/Wilson lines, we shall insert the dynamical brane positions in the direct-channel M\"obius amplitude\footnote{For convenience, we  drop all the prefactors as well as the integration measure, and keep the winding/KK sums implicit.}
\begin{equation}
\begin{split}
    \mathcal{M}&=\sum_{\alpha}(W_{2n_8+2a_\alpha^8}(-1)^{n_9}-W_{2n_8+1+2a_\alpha^8})W_{n_9+2a_\alpha^9}(\hat{V}_8+(-1)^{n_9}\hat{S}_8)W_{n_4+2a_\alpha^4}W_{n_5+2a_\alpha^5}W_{n_6+2a_\alpha^6}W_{n_7+2a_\alpha^7}\\
    &-\sum_{\{\beta_i\}}(W_{2n_8+2b^8_{\beta_i}}-(-1)^{n_9}W_{2n_8+1+2b^8_{\beta_i}})W_{n_9+2b^9_{\beta_i}}(\hat{V}_8+(-1)^{n_9}\hat{S}_8)W_{n_4+2b^4_{\beta_i}}W_{n_5+2b^5_{\beta_i}}W_{n_6+2b^6_{\beta_i}}W_{n_7+2b^7_\beta}\,.
\end{split}
\end{equation}
To perform our computation, it is actually more advantageous to work in the transverse-channel, where brane positions become Wilson lines and the partition function  reads
\begin{equation}
\begin{split}
    \tilde{\mathcal{M}}=\sum_{\alpha}&\left(\hat{V}_8-(-1)^{m_8}S_8\right)\left(P_{2m_9+1}e^{4i\pi(m_9+1/2)a^9_\alpha}-e^{4i\pi m_9a^9_\alpha}(-1)^{m_8}P_{2m_9}\right)\\
    & \times e^{4i\pi (m_4 a^4_\alpha+m_5 a^5_\alpha+ m_6 a^6_\alpha+ m_7 a^7_\alpha+m_8/2 a^8_\alpha)}P_{2m_4}P_{2m_5}P_{2m_6}P_{2m_7}P_{m_8}\\
     +\sum_{\{\beta_i\}}&\left(\hat{V}_8-(-1)^{m_8}S_8\right)(-1)^{m_8}\left(P_{2m_9+1}e^{4i\pi (m_9+1/2) b^9_{\beta_i}}-e^{4i\pi m_9 b^9_{\beta_i}}(-1)^{m_8}P_{2m_9}\right)\\
    & \times e^{4i\pi (m_4 b^4_{\beta_i}+m_5 b^5_{\beta_i}+ m_6 b^6_{\beta_i}+ m_7 b^7_{\beta_i}+m_8/{2} b^8_{\beta_i})}P_{2m_4}P_{2m_5}P_{2m_6}P_{2m_7}P_{m_8}\,.
\end{split}
\end{equation}
Exploiting the Jacobi identity $V_8=S_8$, and recalling that in our model the annulus amplitude is supersymmetric  $\mathcal{A}=0$, we can recast the effective potential for the open string moduli as
\begin{equation}\label{vopenf}
    \mathcal{V}_{\text{one-loop,open}}=-\hat{V}_8 \left( f(\vec{a})- f(\vec{b})\right)\,,
\end{equation}
where we have again dropped the prefactor $\sim M_s^4$ but keep the important minus sign and,  denoting  the brane positions collectively as $\vec{z}_r=\{\vec{a}_{\alpha},\vec{b}_{\{\beta_i\}_i}\}$, we have defined
\begin{equation}
\begin{split}
    f(\vec{z}):=\sum_{r}(P_{2m_9+1}e^{2i\pi z_r^9}+P_{2m_9})&e^{4i\pi(m_4 z^4_r+ m_5 z_r^5+m_6 z^6_r+ m_7 z_r^7+(m_8+1/2)z_r^8+m_9z_r^9)}\times\\
    &\times P_{2m_4}P_{2m_5}P_{2m_6}P_{2m_7}P_{2m_8+1}\,.
 \end{split}   
\end{equation}
It is now useful to express $f$ in terms of Jacobi $\vartheta$-functions
\begin{equation}
    \vartheta\left[\begin{smallmatrix}
        \alpha\\
        \\
        \beta
    \end{smallmatrix}\right](z,\tau):=\sum_n q[\tau]^{\frac{1}{2}(n+\alpha)^2}e^{2i\pi(n+\alpha)(z+\beta)}\,,
\end{equation}
using the following identities
\begin{equation}
    \begin{split}
        P_{2m+1}e^{2i\pi(2 m+1) z}&=q[2i\ell/\rho^2]^{\frac{1}{2}(m+1/2)^2}e^{4i\pi(m+1/2)z}\equiv \vartheta\left[\begin{smallmatrix}
        1/2\\
        \\
        0
    \end{smallmatrix}\right](2z,2i\ell/\rho^2)\\
    P_{2m}e^{4i\pi m z}&=q[2i\ell/\rho^2]^{\frac{1}{2}(m)^2}e^{4i\pi(m)z}\equiv \vartheta\left[\begin{smallmatrix}
        0\\
        \\
        0
    \end{smallmatrix}\right](2z,2i\ell/\rho^2)\,.
    \end{split}
\end{equation}
We can therefore write
\begin{equation}\label{fexpr}
\begin{split}
    f({\vec{z}})=\sum_r\vartheta\left[\begin{smallmatrix}
        1/2\\
        \\
        0
    \end{smallmatrix}\right](2z_r^8,\tau_8)&\left(\vartheta\left[\begin{smallmatrix}
        1/2\\
        \\
        0
    \end{smallmatrix}\right](2z_r^9,\tau_9)+\vartheta\left[\begin{smallmatrix}
        0\\
        \\
        0
    \end{smallmatrix}\right](2z_r^9,\tau_9)\right)\times\\
    &\times\vartheta\left[\begin{smallmatrix}
        0\\
        \\
        0
    \end{smallmatrix}\right](2z_r^5,\tau_5) \,\,\vartheta\left[\begin{smallmatrix}
        0\\
        \\
        0
    \end{smallmatrix}\right](2z_r^6,\tau_6) \,\,\vartheta\left[\begin{smallmatrix}
        0\\
        \\
        0
    \end{smallmatrix}\right](2z_r^7,\tau_7)\,,
 \end{split}   
\end{equation}
where we have conveniently defined  $\tau_I:=2i\ell/\rho_I^2\equiv2i\ell \alpha'/R_I^2$. 
Critical points  of the one-loop open string potential (\ref{vopenf}) hence correspond to solutions of
\begin{equation}
    \partial_{z^I_r} f(\vec{z})\vert_{{\bar{z}}_r^{ I}}=0\, .
\end{equation}
From
\begin{subequations}
\begin{equation}\label{crit1}
\begin{split}
    \partial_z \vartheta\left[\begin{smallmatrix}
        0\\
        \\
        0
    \end{smallmatrix}\right](2 z,\tau)&=4i\pi\sum_m \,m\,q^{\frac{1}{2}m^2}e^{4i\pi z m}\\
    &=-8i\pi \sum_{m>0}q^{1/2m^2}\,m\,\sin(4\pi z m)
    \\
    &=0 \iff z=\{0,1/2,1/4\}\, ,
 \end{split}   
\end{equation}
\begin{equation}\label{crit2}
\begin{split}
    \partial_z \vartheta\left[\begin{smallmatrix}
        1/2\\
        \\
        0
    \end{smallmatrix}\right](2 z,\tau)&=2i\pi \sum_m (2m+1)q^{\frac{1}{2}(m+1/2)^2}e^{2i\pi z(2m+1)}\\
    &=-4\pi \sum_{m\ge 0}(2m+1)q^{\frac{1}{2}(m+1/2)^2}\sin(2\pi z(2m+1)) \\
    &=0 \iff z=\{0,1/2\}\, ,
 \end{split}   
\end{equation}
\end{subequations}
we then find the critical points to be
\begin{equation}\label{critpoint}
    {z}_r^{\star I}=\{0,1/2\}\,,\quad\forall r\,,\forall I\,,
\end{equation}
which means that $\mathcal{V}_{\text{1-loop,open}}$ is extremised only when all the branes sit at the fixed points on top of the O-planes. 
To understand the nature of such critical points we have to inspect the mass matrix associated to the variation of the position of each brane in the different type of stacks
\begin{equation}
    M^{IJ}_{\vec{z}^\star_r}:=-(\partial_{ z_r^{I}}\partial_{ z^{J}_r}\tilde{\mathcal{M}})\vert_{{z}^{\star I}_r,{z}^{\star J}_r}\, .
\end{equation}
In our conventions then, the full mass matrix describing the whole brane configuration is  given by the block diagonal matrix
\begin{equation}
    M=\text{diag}\left(M^{IJ}_{\vec{z}^\star_r};\, r=(\alpha,\{\beta_i\}_i)\right)\,,\quad\alpha=1,\dots,4\,,\beta_i=1,\dots,N_i/2\,.
\end{equation}
It is easy to see that because of (\ref{critpoint}), all the off-diagonal terms in $ M^{IJ}_{\vec{z}^\star_r}$ vanish\footnote{For example, along the directions 8,9 and using (\ref{fexpr}), (\ref{crit1}), (\ref{crit2}) we have
$
\partial_{z^8_r}\partial_{z^9_r}f(\vec{z})\vert_{ {z}_r^{\star 8}, {z}_r^{\star 9}}\stackrel{(\ref{fexpr})}{=}\partial_{z^8_r}\vartheta\left[\begin{smallmatrix}
        1/2\\
        \\
        0
    \end{smallmatrix}\right](2z_r^8,\tau_8)\vert_{{z}_r^{\star 8}}\left(\partial_{z_r^9}\vartheta\left[\begin{smallmatrix}
        1/2\\
        \\
        0
    \end{smallmatrix}\right](2z_r^9,\tau_9)+\partial_{z_r^9}\vartheta\left[\begin{smallmatrix}
        0\\
        \\
        0
    \end{smallmatrix}\right](2z_r^9,\tau_9)\right)\vert_{{z}_r^{\star 9}}\times\cdot\cdot\cdot=0$
and similarly for the other off-diagonal terms.}; $ M^{IJ}_{\vec{z}_r}$ is then already diagonal 
\begin{equation}
M=\text{diag}\left(\lambda^{(4)}_{\vec{a}^*_\alpha},\dots,\lambda^{(9)}_{\vec{a}^*_\alpha},\lambda^{(4)}_{\vec{b}^*_{\beta_i}},\dots,\lambda^{(9)}_{\vec{b}^*_{\beta_i}} \right)\, ,
\end{equation}
where the eigenvalues are
\begin{equation}
    \lambda^{(I)}_{\vec{{a}}^\star_\alpha}=-\,\partial^2_{a^I_\alpha} f(\vec{a})\vert_{\vec{{a}}^\star_\alpha}\,,\quad  \lambda^{(I)}_{\vec{{b}}^\star_{\beta_i}}=+\,\partial^2_{b^I_{\beta_i}} f(\vec{b})\vert_{\vec{{b}}^\star_{\beta_i}}\, .
\end{equation}
We immediately notice the opposite signs in the definitions of the USp and SO Wilson lines eigenvalues. Therefore, critical Wilson lines configurations such that $\vec{a}^*_\alpha=\vec{b}^*_{\{\beta\}_i}$ are automatically saddle points. This is indeed the case of the  $\usp(8)\times \prod_i SO(N_i)$ configuration, described by the Wilson lines $\vec{a}_\alpha=\vec{b}_{\{\beta_i\}_i}=\vec{0}$ $\forall \,\alpha,\forall\, \beta_i$ and by (\ref{critpoint}) a critical point of $\mathcal{V}_{\text{1-loop,open}}$, and hence a saddle.
 This confirms the tree-level result from energetic considerations, since the $N_i$ branes are repelled by the $\overline{\text{O}3}_-$ planes and attracted by the $\overline{\text{O}3}_+$ planes.
 More generally, exploiting the identities
\begin{equation}\label{theta2}
\begin{split}
    -\partial^2_z\vartheta\left[\begin{smallmatrix}
        1/2\\
        \\
        0
    \end{smallmatrix}\right](2z,\tau)&=8\pi^2\sum_m (2m+1)^2q^{\frac{1}{2}(m+1/2)^2}2e^{2\pi i(2m+1)z}\\
     -\partial^2_z\vartheta\left[\begin{smallmatrix}
        0\\
        \\
        0
    \end{smallmatrix}\right](2z,\tau)&=16\pi^2\sum_m m^2q^{\frac{1}{2}m^2}e^{4\pi i m z}
 \end{split}   
\end{equation}
it is easy to verify that 
\begin{equation}\label{positiveeigen}
    \lambda^{(I)}_{\vec{0}}>0\quad\forall I\,,
\end{equation}
which confirms that also at one-loop Wilson lines associated to SO($N$), with $N$ even, are tachyonic, while those associated to USp($N$) are massive. Therefore, in this portion of moduli space, all the branes stacked on top of the $\overline{\text{O}3}_-$-planes are  expected to flow towards the nearest $\overline{\text{O}3}_+$-planes. This gives a $\usp(8+N_1)\times \Pi_{i=2}^{16} \usp(N_i)$ configuration, which is described by the Wilson lines $\vec{a}_\alpha=\vec{0}$, $\vec{b}_{\{\beta_i\}}=(0,0,0,1/2,0)$  $\forall \,\alpha,\forall\, \beta_i$. Indeed, from the first eq. in (\ref{theta2}) when $z=b^8=1/2$ it follows straightforwardly
\begin{equation}
\lambda^{(I)}_{0,\dots,1/2,0}=-\lambda^{(I)}_{\vec{0}} \,,
\end{equation}
hence this brane configuration is a local minimum of $\mathcal{V}_{\text{1-loop,open}}$.

In this local minimum, however, the vacuum energy does not cancel, rather it yields a contribution set by the string scale $\ms$. Actually, this local minimum is degenerate in energy with the $\usp(16)$ configuration with all the branes stacked on the same $\overline{\text{O}3}_+$-plane: indeed, it can be readily checked that  the M\"obius amplitudes of the two configurations are the same.  The USp(16)  configuration, corresponding to the choice $\vec{a}_{\alpha}=\vec{0}$ $\forall\alpha=1,\dots,8$, extremises the Wilson line potential because of (\ref{crit1}, \ref{crit2}) and it is a minimum because of (\ref{positiveeigen}), as expected.

The vacuum energy is instead identically vanishing for the $\usp(8)\times[\So(1)]^8$ configuration, which belongs to the disconnected component of the moduli space described by the Wilson line matrix \eqref{so2wl} for $k=4$. 
Since all the eight SO-branes are stuck at the fixed points in SO(1) configurations and cannot be given a dynamical position, we conclude that this is the only minimum in the open string moduli space where $\mathcal{V}_{\text{1-loop,open}}=0$.

\section{\texorpdfstring{IIB on $T^6/\mathbb{Z}_2\times \mathbb{Z}_2$ O3 orientifolds}{IIB on T6/Z2xZ2 O3 orientifolds}}\label{A:IIBCY3}
In order to be self-contained, we here condense the main features of IIB compactifications on $T^6/\mathbb{Z}_2\times \mathbb{Z}_2$ orientifolds.  We first motivate the well-known fact that the $T^6/\mathbb{Z}_2\times \mathbb{Z}_2$ orbifold  has the same properties as a CY$_3$; compactifications of type IIB string theory on such a space therefore  give  4d $\mathcal{N}=2$ supersymmetry, which is lowered to 4d $\mathcal{N}=1$ if an orientifold projection is applied on top of the orbifold. We then give the explicit $T^6/\mathbb{Z}_2\times \mathbb{Z}_2$ expression for the tree-level K\"ahler potential and the superpotential, starting from the general expression IIB CY$_3$ orientifolds \cite{Grimm:2004uq}.

The orbifold is constructed by acting with the $\mathbb{Z}_2$ 
 generators $\theta_1$ and $\theta_2$, corresponding to $\pi$-rotations,  on the ${\mathbb T}^6$ complex coordinate $z_i$ $i=1,2,3$  as:
\begin{equation}\label{eqA:z2z2}
    \begin{split}
        \theta_1&:(z_1,z_2,z_3)\rightarrow (-z_1,-z_2,z_3)\, ,\\
        \theta_2&:(z_1,z_2,z_3)\rightarrow (z_1,-z_2,-z_3)\, ,\\
        \theta_1\,\theta_2&:(z_1,z_2,z_3)\rightarrow (-z_1,z_2,-z_3)\, .
    \end{split}
\end{equation}
We clearly see that this results in the factorisation $\mathbb{T}^6=\mathbb{T}^2_{\,1}\times \mathbb{T}^2_{\,2}\times \mathbb{T}^2_{\,3}$, plus a total of $48$ fixed points constituting the orbifold twisted sector. The data of each 2-torus $\mathbb{T}_i^{\,2}$  can be re-expressed in terms of  a real geometric K\"ahler modulus $t'_i$ and a complex structure modulus $U_i$, measuring respectively the  volume  and the shape of the torus, with expressions
\begin{equation}
\begin{split} \label{eq:2torusmod}
    U_i&=\frac{\sqrt{\det g_{(i)}}}{g_{(i)11}}+i\frac{g_{(i)12}}{g_{(i)11}}\equiv \frac{1}{2}u_i+i\,\theta_{u_i}\,,\quad\\
    t'_i&=\sqrt{\det g_{(i)}}\, .
 \end{split}   
\end{equation}
 For each of the 2-tori we can thus introduce complex coordinates $z^i=x^i+i\,U_i y^i$, $i=1,2,3$ subjected to the identifications $z^i\sim z^i+1$ and $z^i\sim z^i+i\,U_i$. 
 
Let us then inspect the resulting untwisted cohomology for the orbifold space, ignoring the twisted sectors. We shall indicate with $H^p$ the $p$-th real cohomology group and use the complex decomposition $H^p=\bigoplus_{\substack{r=0}}^p H^{p-r,r}$, with the Hodge numbers $h^{p,q}:=\text{dim}(H^{p,q})$ having the obvious properties $h^{p,q}=h^{q,p}$ and $h^{p,q}=h^{3-p,3-q}$.
As in every CY$_3$, it is readily checked that there are no 1- or (dual) 5-forms invariant under the  orbifold action (\ref{eqA:z2z2}), hence $h^{1,0}=0$. It is instead possible to find
     eight invariant real 3-forms $(\alpha_K,\beta^K)$, $K=0,\dots,3$  defining a basis for $H^3$, which obeys the intersecting relation
\begin{equation}\label{eq:formsbasis}
    \int  \alpha_I\wedge \beta^J=\delta_I^J\,,
\end{equation}
where the integral is all over the internal space: an explicit choice for this basis is
\begin{equation}\label{alphabetaforms}
\begin{split}
&\alpha_0=d x^1 \wedge d x^2 \wedge d x^3,\,\qquad   \beta^0=+d y^1 \wedge d y^2 \wedge d y^3, \\
&\alpha_1=d y^1 \wedge d x^2 \wedge d x^3, \,\qquad \beta^1=-d x^1 \wedge d y^2 \wedge d y^3, \\
&\alpha_2=d x^1 \wedge d y^2 \wedge d x^3,  \,\qquad\beta^2=-d y^1 \wedge d x^2 \wedge d y^3, \\
&\alpha_3=d x^1 \wedge d x^2 \wedge d y^3,  \,\qquad \beta^3=-d y^1 \wedge d y^2 \wedge d x^3 .
\end{split}
\end{equation}
Once expressed in complex coordinates, the basis $(\alpha_K,\beta^K)$ results in one (3,0)-form and three $(2,1)$-forms, hence $h^{3,0}=1$ and $h^{2,1}=3$.
In complex coordinates, we can define the unique invariant holomorphic 3-form $\Omega$ as
\begin{equation}\label{omegacomplex}
    \Omega=dz^1\wedge dz^2\wedge dz^3\,,
\end{equation}
which clearly can be expanded in the real basis $(\alpha_K,\beta^K)$ with coefficients given by (products of) the complex structure moduli $U_i$
\begin{equation}\label{omega3basis}
\begin{aligned}
\Omega=\alpha_0 & +i\left(U_1 \alpha_1+U_2 \alpha_2+U_3 \alpha_3\right) \\
& +\left(U_2 U_3 \beta^1+U_3 U_1 \beta^2+U_1 U_2 \beta^3\right)-i\,U_1 U_2 U_3 \beta^0 .
\end{aligned}
\end{equation}
A quick check shows that, for the $H^2$ cohomology, it is not possible to define invariant $(2,0)$-forms, so $h^{2,0}=0$, but only $h^{1,1}=3$ invariant $(1,1)$-forms 
\begin{equation}\label{11form}
    \omega_i=\frac{1}{2\,i\,\mathrm{Re}\,U_i}dz^i\wedge d\bar{z}^i\,,\quad i=1,\dots,h^{1,1}=3\,.
\end{equation} 
From these, we can define the dual basis $\tilde\omega^i$ of invariant  $(2,2)$-forms for $H^4$
\begin{equation}\label{22form}
    \tilde\omega^1=-\omega_2\wedge\omega_3\,,\quad \tilde\omega^2=-\omega_3\wedge\omega_1\,,\quad \tilde\omega^3=-\omega_1\wedge\omega_2\, ,
\end{equation}
such that
\begin{equation}
    \int \omega_i\wedge \tilde\omega^j=\delta_i^j\, .
\end{equation}
We can therefore introduce the real (1,1) K\"ahler form $J$ as
\begin{equation}\label{Jform}
    J=\sum_{i=1}^{h^{1,1}} t_i'\,\omega_i\, ,
\end{equation}
which is related to the  6d volume $\mathcal{V}_s$ through the usual CY relation
\begin{equation}
    \mathcal{V}=\frac{1}{6}\int J\wedge J\wedge J=\frac{1}{6}\kappa_{ijk}\,t_i't_j't_k'\, ,
\end{equation}
with $\kappa_{ijk}$ the triple-intersection numbers
\begin{equation}
    \kappa_{ijk}:=\int \omega_i\wedge \omega_j\wedge \omega_k\, .
\end{equation}
From the explicit choice of the (1,1)-forms (\ref{11form}) we then see that, for $T^{6}/\mathbb{Z}_2\times\mathbb{Z}_2$ we have
\begin{equation}\label{tripleint}
    k_{ijk}=\begin{cases}
        1\,,\quad \text{if}\,\,\,\{i,j,k\}=\{1,2,3\}\\
        0\,,\quad \text{otherwise}\, .
    \end{cases}
\end{equation}
When the fixed points are taken into account too, the resulting space is thus a singular limit of a Calabi-Yau threefold CY$_3$ with $h_{\text{untw}}^{1,1}=3$ untwisted K\"ahler moduli, $h_{\text{unt}}^{2,1}=3$ untwisted complex structure moduli and $h^{1,1}_{\text{tw}}=48$ twisted K\"ahler moduli,  thus with Euler number
\begin{equation}
  \chi({\mathbb T}^6/{\mathbb Z}'_2\times {\mathbb Z}'_2)=2(h^{1,1}-h^{2,1})=96\,.
\end{equation}
In other words $T^6/\mathbb{Z}_2\times \mathbb{Z}_2$ has $SU(3)$ holonomy and thus type IIB compactifications on such space give 4d $\mathcal{N}=2$ supersymmetry. Applying an orientifold projection results in 4d $\mathcal{N}=1$. Here, we are interested in the O3 orientifold
\begin{equation}
    O3=\Omega (-1)^{F_L}\sigma\,,\quad \sigma:z_i\rightarrow-z_i\,,
\end{equation}
with $\sigma$ acting as an internal parity along all the internal coordinates of the torus. In the smooth CY$_3$ case, such a geometric action generalises into an action directly on the K\"ahler form $J$ and the holomorphic form $\Omega$ of the Calabi-Yau
\begin{equation}
    \sigma\,J=J\,,\quad \sigma\,\Omega=-\Omega\, .
\end{equation}
Because $\sigma$ acts holomorphically, the cohomology groups split into even and odd eigenspaces $H^{p,q}=H^{p,q}_+\oplus H^{p,q}_{-}$. It turns out that the forms in (\ref{alphabetaforms}) are all odd under $\sigma$, and from (\ref{omegacomplex})   $\Omega$ is correctly odd too, hence  $h^{2,1}_-=3$, $h^{2,1}_+=0$, $h^{3,0}_-=1$, $h^{3,0}_+=0$. The forms (\ref{11form}) and (\ref{22form}) are all even under $\sigma$, hence $h^{1,1}_+=3$, $h^{1,1}_-=0$, and the expression for $J$ in (\ref{Jform}) has the correct parity under $\sigma$.

Knowing which forms can be defined on this toroidal orientifold, we can now find which fields from the IIB closed-string bosonic spectrum survive the combined orbifold and orientifold projections. Given the combined action of the worldsheet parity and  $F_L$  on the NSNS and RR IIB sectors
\begin{equation}\label{omegafl}
    \Omega(-1)^{F_L}\{\phi,B_2\}=\{\phi,-B_2\}\,,\quad   \Omega(-1)^{F_L}\{C_0,C_2,C_4\}=\{C_0,-C_2,C_4\}\, ,
\end{equation}
we see that the fields have to be expanded in the following way to survive the orientifold projection
\begin{equation}
\begin{split}
\phi&=\phi(x)\cdot 1\,,\quad C_0=C_0(x)\cdot1\, ,\\
B_2&=\sum_{a=1}^{h^{1,1}_-} b_2(x)\cdot\omega_a\,,\quad C_2=\sum_{a=1}^{h^{1,1}_-} c_2(x)\cdot \omega_a\, ,\\
C_4&=\sum_{a=1}^{h^{2,1}_+}V^a_\mu(x)\cdot\omega_a+\sum_{i=1}^{h^{1,1}_+}a_i(x)\cdot\sigma_i\, .
\end{split}
\end{equation}
Therefore, since $h^{2,1}_+=h^{1,1}_-=0$, $B_2$ and $C_2$ are completely projected out from the spectrum, as well as the vectors $V_\mu^a$ from the dimensional reduction of $C_4$, while the 4d moduli that survive 
 are the  dilaton $\phi$, the axion $C_0$ and $h^{1,1}_+=3$  axions $a_i$ from the dimensional reduction of the  $C_4$, to which we add $h^{1,1}_+=3$ real K\"ahler moduli $t_i'$ and $h^{2,1}_-=3$ complex structure moduli $U_i$ from the allowed deformations of the internal metric. Pairing up the dilaton $\phi$ and the axion $C_0$ into  the axio-dilaton $S$ and  the real geometrical K\"ahler moduli $t_i$ and the $C_4$ axions $a_i$ into the complexified geometrical K\"ahler moduli $T_i'$ as
\begin{equation}\label{sti'}
\begin{split}
    S&=e^{-\phi}+i\,C_0\,,\\
    T_i'&= t_i'+i\,a_i\, ,
 \end{split}   
\end{equation}
we clearly see that the resulting spectrum is that of 4d $\mathcal{N}=1$ supergravity coupled to $h^{1,1}_+,h^{2,1}_-$, and the universal axio-dilaton chiral multiplets.

On the $T^6/\mathbb{Z}_2\times \mathbb{Z}_2$ orbifold it is also possible to consider fluxes of the NSNS and RR gauge potentials, $H_3=dB_2$ and $F_3=dC_2$, as well as non-geometric $Q$-fluxes. Because of (\ref{omegafl}), $H_3$ and $F_3$ are both odd under $\Omega(-1)^{F_L}$, hence they must be expanded in the basis of 3-forms (\ref{alphabetaforms})
\begin{equation}\label{h3f3forms}
  \,H_3=(2\pi)^2\alpha'(h^K\alpha_K+h_K\beta^K)\,, \quad F_3=(2\pi)^2\alpha'(f^K\alpha_K+f_K\beta^K)\, .
\end{equation}
Consistency of string theory demands a  flux quantisation condition\footnote{Strictly speaking, for the $\mathbb{Z}_2 \times \mathbb{Z}_2$ orbifold, fluxes are quantised in multiples of 8 due to the presence of twisted cycles. In the freely-acting $\mathbb{Z}'_2 \times \mathbb{Z}'_2$ orbifold of interest to us, there are no twisted cycles and fluxes obey the standard quantisation condition.} along a basis of 3-cycles $(A_K,B^K)$ Poincare dual to $(\alpha_K,\beta^K)$ 
\begin{equation}\label{quantisationfluxz2z2}
\begin{split}
   \frac{1}{(2\pi)^2\alpha'} \int_{A_K}H_3&=h^K\in \mathbb{Z}\,,\quad  \frac{1}{(2\pi)^2\alpha'}\int_{B_K}H_3=h_K\in \mathbb{Z}\, .\\
 \end{split}   
\end{equation}
The non-geometric $Q$-flux is also odd under $\Omega(-1)^{F_L}$ and its action on an element of the basis of (2,2)-forms $\tilde\omega^i$ reads \cite{nongeoflux}
\begin{equation}\label{Qflux}
    Q\circ\omega^i=(2\pi)^2\alpha'(q^{Ki}\alpha_K+{q_K}^i\beta^K)\, ,
\end{equation}
where  again the flux quanta $q^{Ki},{q_K}^i\in \mathbb{Z}$ because of quantisation conditions similar to (\ref{quantisationfluxz2z2}).

The tree-level moduli space of the 4d $\mathcal{N}=1$ supergravity resulting from IIB compactification on CY$_3$ O3 orientifolds factorises into the complex structure, K\"ahler and axio-dilaton moduli space \cite{Grimm:2004uq}. 
While the axio-dilaton $S$ and the geometric complex structures $U_i$ are incidentally good holomorphic coordinates on the moduli space, this does not hold for the geometric K\"ahler moduli $T_i'$ (\ref{sti'}); the correct K\"ahler coordinates are instead the (supergravity) K\"ahler moduli $T_i$
\begin{equation}
    T_i=\text{Re}(T_i)+i\,a_i\,,
\end{equation}
whose real parts are
\begin{equation}\label{4cycles}
    \text{Re}(T_i):=\frac{1}{2}e^{-\phi}\int J\wedge J\wedge \omega_i=\frac{1}{2}e^{-\phi}\kappa_{ijk}\,t_j'\,t'_k\, .
\end{equation}
The K\"ahler potential of the 4d $\mathcal{N}=1$ action is the sum of the three contributions
\begin{equation}\label{kahleriibo3}
\begin{split}
    K_\tree&= K_{\text{cs}}+K_{\text{kah}}+K_{\text{dil}}\\
    &=-\log\left(i\int\Omega\wedge \bar{\Omega}\right)-2\log (e^{-3 \phi/2}\mathcal{V})-\log(S+\bar S)\, .
 \end{split}   
\end{equation}
Note that $\mathcal{V}$ has explicit dependence on the $t_i'$ and only implicit dependence on $\text{Re}(T_i)$ through (\ref{4cycles}), which is in general not analytically invertible. For $T^6/\mathbb{Z}_2\times \mathbb{Z}_2$, an analytic expression $\mathcal{V}=\mathcal{V}(\text{Re}(T_i))$ from (\ref{4cycles}) can be actually found: using (\ref{tripleint}) we find 
\begin{equation}
\text{Re}(T_i)=e^{-\phi}(\det g_{j}\cdot \det g_{k})^{1/2}\, ,
\end{equation}
thus
\begin{equation}
    \mathcal{V}=e^{3\phi/2}\sqrt{
    \text{Re}(T_1)\text{Re}(T_2)\text{Re}(T_3)
    }\, \,.
\end{equation}
The explicit dependence of $K_\tree$ on the complex structure moduli $U_i$ from the first term in (\ref{kahleriibo3}) is found straightforwardly using the expansion for $\Omega$ in (\ref{omega3basis}) and the symplectic relation (\ref{eq:formsbasis}). All together, the K\"ahler potential for  $T^6/\mathbb{Z}_2\times\mathbb{Z}_2$ reads, up to an irrelevant constant term
\begin{equation} \label{A:eq:Ktree}
    K_\tree=-\sum_{i=1}^3\log (U_i+\bar{U}_i)-\sum_{i=i}^3\log(T_i+\bar{T}_i)-\log(S+\bar{S})\, .
\end{equation}
When 3-form fluxes $H_3$ and $F_3$ (and the non-geometric $Q$-flux) are switched on, the superpotential of the 4d $\mathcal{N}=1$ theory is (an extension) of the Gukov-Vafa-Witten form
\begin{equation}
    W=\sqrt{\frac{2}{\pi}}\frac{1}{4\pi^2\alpha'}\int(F_3-iSH_3-i(Q\circ \tilde\omega^j)T_j)\wedge\Omega\,.
\end{equation}
The explicit dependence on the $T^6/\mathbb{Z}_2\times \mathbb{Z}_2$ orientifold moduli is obtained by plugging in the decompositions (\ref{omega3basis}), (\ref{h3f3forms}) and (\ref{Qflux}) and then using   the intersection relations (\ref{eq:formsbasis}), to obtain eventually
    \begin{equation}\label{wfluxesgeneral}
\begin{aligned}
\sqrt{\frac{\pi}{2}}W_\tree=&f_0+\,i\,h_0S-i\,{q_0}^iT_i\\
&+i\,U_j\left(f_j+i\,h_jS+i\,{q_j}^kT_k\right)\\
&+\frac{1}{2}\sigma_{ljk}U_{j}U_k\left(f^l+i\,h^lS+i\,q^{lm}\,T_m\right)\\
&+\frac{i}{6}\sigma_{ljk}U_{l}U_{j}U_k\,(-f^0+i\,h^0S-i\,q^{0l}\,T_l)\,,
\end{aligned}
\end{equation}
where we split the index $K=(0,i)$, $i=1,2,3$ and $\sigma_{ijk}$ is a symmetric symbol  with non vanishing components $\sigma_{123}=+1$ and permutations thereof.

\section{Leading corrections to the tree-level no-scale potential and extended no-scale structure} \label{A:extendednoscale}
As we have discussed in Section \ref{s:SSsugra}, the tree-level K\"aher potential and superpotential that correspond to our Scherk-Schwarz compactification -- reproducing the gravitino mass and with vanishing tree-level cosmological constant -- are given by:
\begin{equation}
   \begin{split}
    &K_{\text{tree}}=-\sum_{i=1}^3\log (U_i+\bar{U}_i)-\sum_{i=1}^3\log(T_i+\bar{T}_i)-\log(S+\bar{S}) \,,\\
     &\sqrt{\frac{\pi}{2}}W_{\text{tree}}=S(i\,h_0-h_1 \,U_1-h_2 \,U_2 + i h^3 U_1 U_2)+T_3\left(-iq_0^{\;3}- q_1^{\;3}\,U_1-q_2^{\;3}\,U_2+i\,q_{12}^{\;\;3}\, U_1\,U_2\right) \,.
    \end{split}
\end{equation}
In this appendix, we determine the one-loop correction to the K\"ahler potential, $\delta K_{\text{SS}}$, which leads to a one-loop F-term potential that matches the one-loop vacuum energy \eqref{V1loopR8R9largest} computed for our string construction in Appendix \ref{A:ssvacuumenergy}.  At the same time, we include non-perturbative corrections to the superpotential, $W_{\text{np}}$, and thus derive the leading order corrections to the scalar potential from both $\delta K_{\text{SS}}$ and $W_{\text{np}}$.  Along the way, we also show that the potential has an extended no-scale structure that protects the scalar potential from the {\it{a priori}} dangerous $\mathcal{O}(\alpha'^3)$ correction to $K$, similarly to what famously happens for the string-loop corrections in the LVS scenario \cite{Cicoli:2007xp,LVS}.

Before beginning, let us recall that indeed $\mathcal{R}^4$ couplings in 10d are well-known to generate, via dimensional reduction, localised Einstein-Hilbert $\mathcal{R}$ terms in 4d that result in an $\mathcal{O}(\alpha'^3)$ correction to $K_\tree$ which shifts the argument of the $-2\log\mathcal{V}_{\text{E}}$ in \eqref{eq:Ktree} to \cite{Antoniadis:r4couplings,AntoniadisnonCompactCY}:
\begin{equation} \label{eq:logVxi}
    K_{\alpha'^3}=-2\log(\mathcal{V}_{\text{E}}+\xi)\,.
\end{equation}
For orbifolds the correction starts at one-loop and takes the form
\begin{equation}
\begin{split}
    {\xi}&:=-\frac{\zeta(2)\gs^{1/2}}{(2\pi^3)}\chi=-\frac{\pi^2\,\gs^{1/2}}{6(2\pi^3)}\chi\,,
 \end{split}   
\end{equation}
with $\chi$ the Euler number.  In our case, $\mathcal{\chi}({\mathbb T}^6/{\mathbb{Z}'_2\times\mathbb{Z}'_2})=0$, and so this $\alpha'$-correction is actually vanishing.  More generally, our worldsheet computation of the one-loop vacuum energy includes all finite-orders in $\alpha'$, so we would not expect these contributions to come into play.  However, it is worthwhile checking what happens when such corrections are present. Expanding \eqref{eq:logVxi} in large volume, we can write
\begin{equation}
    K_{\alpha'^3}=K_{\tree}+\delta K_{\alpha'}\, ,
    \end{equation}
    where, defining for convenience $\hat{\xi}:=\frac{\pi^2}{6(2\pi^3)}\chi$, the correction reads
    \begin{equation}\label{deltakalpha}
        \quad  \delta K_{\alpha'}=-2\,\hat{\xi}\frac{\gs^{1/2}}{\mathcal{V}_{\text{E}}}=-2\,\hat{\xi}\,\frac{1}{(s\,t_1t_2\,t_3)^{1/2}}\, .
\end{equation}
We will see that -- except for unreasonably  small values of the string coupling, $\gs$ -- at large volume, $\mathcal{V}$, the $\alpha'$ correction $\delta K_{\alpha'}$ dominates over the Scherk-Schwarz correction $\delta K_{\text{SS}}$ 
\begin{equation}\label{ordecorrections}
    \delta K_{\text{SS}}=\mathcal{O}(\mathcal{V}_{\text{E}}^{-2})\ll  \delta K_{\alpha'}=\mathcal{O}(\gs^{1/2}\mathcal{V}_{\text{E}}^{-1})\, .
\end{equation}
Then, the $\alpha'$-correction to the vacuum energy, $\sim\delta K_{\alpha'}\ab W_\tree\ab^2$ -- if it exists -- would unavoidably dominate over the Scherk-Schwarz contribution,  $\sim\delta K_{\text{SS}}\ab W_\tree\ab^2$, and spoil any matching of scales between the one-loop Scherk-Schwarz vacuum energy and the observed Dark Energy.  We will eventually find that the leading contribution from $\delta K_{\alpha'}$  to the scalar potential is actually vanishing, due to the property of $\delta K_{\alpha'}$  being a homogeneous function of degree $-1$ in the tree-level no-scale coordinates $t_1,t_2,u_3$. 

To proceed, recall our identification of the tree-level no-scale moduli $\Phi^a=\{T_1,T_2,U_3\}$, and the remaining moduli $\Phi^\alpha=\{S,U_1,U_2,T_3\}$ on which $W_\tree$ depends.  We have:
\begin{equation}
    \begin{split}
        K&=K_{\tree}(\Phi^a,\bar{\Phi}^{\bar{a}},\Phi^{\alpha},\bar{\Phi}^{\bar{\alpha}})+\delta K(\Phi^a,\bar{\Phi}^{\bar{a}},\Phi^{\alpha},\bar{\Phi}^{\bar{\alpha}})\\
        W&= W_{\tree}(\Phi^\alpha)+W_{\text{np}}(\Phi^a,\Phi^\alpha)
    \end{split}
\end{equation}
where the corrections to the K\"ahler potential have the structure
\begin{equation}
    \delta K(\Phi^a,\bar{\Phi}^{\bar{a}},\Phi^{\alpha},\bar{\Phi}^{\bar{\alpha}})= \delta K_{\text{SS}}(\Phi^a,\bar{\Phi}^{\bar{a}})+\delta K_{\alpha'}(\Phi^a,\bar{\Phi}^{\bar{a}},\Phi^{\alpha},\bar{\Phi}^{\bar{\alpha}})\, ,
\end{equation}
and we further assume, as in \eqref{wnp}, that the non-perturbative superpotential, $W_{\text{np}}(\Phi^a,\Phi^\alpha)$
\begin{equation}
   \begin{split}
    &W_{\text{np}}=A\,e^{-\alpha\,S}+ B_3\,(1-b_3\,S)\, e^{-\beta_3\,T_3} \quad \, .
   \end{split}
\end{equation}
is actually independent of $T_1$ and $T_2$ at leading order, since we are interested in the regime of moduli space where $t_1,t_2 \gg t_3$.  The $\mathcal{N}=1$ formula for the F-term scalar potential can then be organised as:
\begin{equation}\label{scalarcorr}
\begin{split}
    V&=e^K\left( K^{\alpha\bar \beta}D_\alpha W\overline{D_{\bar \beta}W}+2\text{Re}(K^{a\bar{\beta}}K_aW\overline{D_{\bar \beta}W})+(K^{a\bar b}K_aK_{\bar {b}}-3)\ab W\ab^2\right)\, ,
\end{split}
\end{equation}
Notice that the $\alpha'$-corrections to the K\"ahler potential lift the tree-level factorisation of the moduli space, and all the $\delta K$ corrections lift  the tree-level no-scale identity (\ref{treenoscale}).

Following \cite{Cicoli:2007xp}, we now proceed in a  systematic large-volume expansion of the scalar potential in powers of $\delta K$.
As a first step, we need to find the inverse of the corrected K\"ahler metric. In the regime $\delta K/K_\tree\ll 1$,
\begin{equation}
    K^{AB}=\left(K_{\tree}+\delta K\right)^{AB}=(K_{\tree}(I+K_{\tree}^{-1}\,\delta K))^{AB}=(I+K_{\tree}^{-1}\,\delta K)^{A}_{\,C} K_\tree^{CB} \, ,
\end{equation}
which can be expanded using the Neumann series
\begin{equation}
    (I+ K_\tree^{-1}\delta K)^{A}_{\,C}=\delta^A_C-K_\tree^{AD}\delta K_{DC}+K_\tree^{AD}\delta K_{DE}K_\tree^{EF}\delta K_{FC}+\mathcal{O}({\delta}^3)\, ,
\end{equation}
to find
\begin{equation}\label{invkahl}
    K^{AB}=K_{\tree}^{AB}-K_\tree^{AD}\delta K_{DC}K_\tree^{CB}+K_\tree^{AD}\delta {K}_{DE}K_\tree^{EF}\delta K_{FC}K_\tree^{CB}+\mathcal{O}({\delta}^3)\, .
\end{equation}
We organise the resulting scalar potential as an expansion  in powers of the correction $\delta K$ to the tree-level K\"ahler metric 
\begin{equation}\label{expv}
    V= V_0+\delta V+\delta^2V+\mathcal{O}(\delta^3)\, .
\end{equation}
Let us start from the zero-order term. Using (\ref{invkahl}) in (\ref{scalarcorr}) we find 
\begin{equation}
   V_0=e^{K_{\tree}}\left((K_\tree^{a{b}}K_{\tree\,a}K_{\tree\,{b}}-3)\ab W\ab^2+ K_\tree^{\alpha \beta}D^{(0)}_\alpha W{D^{(0)}_{ \beta}\overline{W}}\right)\,,
\end{equation}
where  $D_\alpha^{(0)}\equiv\partial_\alpha+\partial_\alpha K_{\tree}$. 
We now recall
 the tree-level no-scale cancellation (\ref{treenoscale}) and impose the tree-level F-term stabilisation $D_\alpha^{(0)}W_\tree=0$, to conclude that 
 \begin{equation}\label{zeroterm}
V_0=e^{K_{\tree}}K_{\tree}^{\alpha\beta}D_\alpha^{(0)}W_{\text{np}}{D^{(0)}_{\beta}\overline{W_{\text{np}}}}\,\,.
 \end{equation}
 For the first-order correction we find
\begin{equation}
    \begin{split}
\delta V=&\,e^{K_{\tree}}\delta K K_\tree^{\alpha\bar\beta}D_\alpha^{(0)}W D_{\bar\beta}^{(0)}\overline{W}\\
& -e^{K_{\tree}}K_{\tree}^{\alpha M}\delta K_{ML}K_\tree^{L\bar\beta}D_{\alpha}^{(0)}W D_{\bar\beta}^{(0)}\overline{W}+e^{K_{\tree}}\,2\,\text{Re}(K_\tree^{\alpha\bar\beta}\delta K_\alpha W{D_{\bar\beta}^{(0)} \overline{W}})\\
&-e^{K_{\tree}}\,2\,\text{Re}( K_\tree^{aM}\delta K_{ML}K_\tree^{L\bar\beta}K_{\tree\,a}W{D_{\bar\beta}^{(0)}\overline{W}})+e^{K_{\tree}}\,2\,\text{Re}(K_\tree^{a\bar\beta}\delta K_a W{D_{\bar\beta}^{(0)}\overline{W}})\\
&+e^{K_{\tree}}\,2\,\text{Re}(K_\tree^{a\bar\beta}K_{\tree\,a}\delta K_{\bar\beta} )\ab W\ab^2-e^{K_{\tree}}K_\tree^{aM}\delta K_{ML}K_\tree^{L\bar{b}}\,K_{\tree\,a}\,K_{\tree\,\bar{b}}\ab W\ab^2\\
&+e^{K_{\tree}}\,2\,\text{Re}(K_\tree^{a\bar b}\delta K_aK_{\tree\,\bar b})\ab W\ab^2\, .
    \end{split}
\end{equation}
We next use that $K_{\tree}^{a \bar\beta}=K_\tree^{\alpha \bar b}=0$  and that  $\delta K_{\text{SS}}=\delta K_{\text{SS}}(\Phi^a,\bar{\Phi}^{\bar{a}})$ depends exclusively on the no-scale moduli, $\Phi^a$. Moreover, we still impose the tree-level F-term condition $D_\alpha^{(0)}W_\tree=0$. The former equation then simplifies to
\begin{equation}\label{dV}
\begin{split}
    \delta V=&\,e^{K_{\tree}}\delta K K_\tree^{\alpha\bar\beta}D_\alpha^{(0)}W_{\text{np}}D_{\bar\beta}^{(0)}\overline{W}_{\text{np}}\\&- e^{K_{\tree}}K_\tree^{\alpha\bar \gamma}(\delta K_{\alpha'})_{\,\bar \gamma{\delta}}K_\tree^{\delta{\bar \beta}}D^{(0)}_\alpha W_{\text{np}}{D_{{\bar\beta}}^{(0)}\overline{W}_{\text{np}}}+e^{K_{\tree}}\,2\,\text{Re}(K_\tree^{\alpha\bar \beta}(\delta K_{\alpha'})_{\alpha}W{D_{\bar \beta}^{(0)}\overline{W}_{\text{np}}})\\
    &-e^{K_{\tree}}\,2\,\text{Re}(K_\tree^{a \bar b}(\delta K_{\alpha'})_{\bar b  \alpha}K_\tree^{\alpha\bar{\beta}}K_{\tree\,a}W{D_{\bar{\beta}}^{(0)}\overline{W}_{np}})\\
    &-e^{K_{\tree}}K_\tree^{a \bar c}\delta K_{\bar{c} d}K_\tree^{d \bar{b} }\, K_{\tree\,a} K_{\tree\, \bar{b}}\ab W\ab^2+e^{K_{\tree}}\,2\,\text{Re}(K_{\tree}^{a \bar b}\delta K_aK_{\tree\,\bar b})\ab W\ab^2\, .
    \end{split}
\end{equation}
It is clear by inspecting (\ref{dV}) that, at first order, the leading order correction to $V_0$ in (\ref{zeroterm}) comes from terms $\propto \delta K \ab W_\tree\ab^2\subset\delta K|W|^2$  in the last line, as all the other terms in (\ref{dV}) are more suppressed than the latter by the non-perturbative term $W_{\text{np}}$. 
Let us thus focus on these terms. Recalling the form of $K_{\tree}$, which depends only on the combinations $\phi^a=2\,\text{Re}(\Phi^a)$, we can make use of the following identities.
\begin{equation}\label{propktree}
\begin{split}
    K_{\tree}^{a\bar b} K_{\tree\, \bar b}&=- \phi^a \\
      K_{\tree}^{a\bar b} K_{\tree\,  a}&=-\phi^b
    \end{split}
    \end{equation}
to find that these terms simplify as
\begin{equation}\label{extnoscale}
\begin{split}
   \delta V&\supset-e^{K_\tree}K_\tree^{a\bar{c}}\delta K_{\bar{c} d}K_\tree^{d\bar{b}}\, K_{\tree\,a} K_{\tree\, \bar{b}}\ab W\ab^2+e^{K_\tree}\,2\,\text{Re}(K_{\tree}^{a \bar b}\delta K_{a}K_{\tree\,\bar b})\ab W\ab^2\\
    &=-\left(\phi^a\frac{\partial \delta K}{\partial\Phi^a}+\phi^a\frac{\partial \delta K}{\partial\bar{\Phi}^{\bar{a}}}+\phi^a\phi^b\frac{\partial^2\,\delta K}{\partial \Phi^a\partial \bar{\Phi}^{\bar{b}}}\right)e^{K_\tree}\ab W\ab^2 \, .
\end{split}    
\end{equation}
Let us thus define
\begin{equation} \label{eq:box}
    \Xi (\delta K):=-\left(\phi^a\frac{\partial \delta K}{\partial\Phi^a}+\phi^a\frac{\partial \delta K}{\partial\bar{\Phi}^{\bar{a}}}+\phi^a\phi^b\frac{\partial^2\,\delta K}{\partial \Phi^a\partial \bar{\Phi}^{\bar{b}}}\right).
\end{equation}
We will now consider the contributions to \eqref{extnoscale} from $\delta K_{\text{SS}}$ and $\delta K_{\alpha'}$ in turn.

\paragraph*{The Scherk-Schwarz correction}
Our goal is now to identify a $\delta K_{\text{SS}}$ that provides an on-shell matching with the one-loop Scherk-Schwarz potential \eqref{V1loopR8R9largest}.  Using \eqref{extnoscale} and \eqref{eq:box}, we therefore impose:
\begin{equation}
    \braket{\Xi(\delta K_{\text{SS}})e^{K_\tree}|W_\tree|^2}={V}_{\text{1-loop}}\,.
\end{equation}
Using also (\ref{onshellm32}), this equation boils down to  
\begin{equation}\label{eqdeltakss}
    \Xi(\delta K_{\text{SS}})=-\frac{u_3}{t_1t_2}\,\mathcal{E}_3(iU_3)\, .
\end{equation}
Assuming the following ansatz
\begin{eqnarray}
    \delta K_{\text{SS}}=\frac{(U_3+\bar{U}_3) \hat{f}(U_3,\bar{U}_3)}{(T_1+\bar{T}_1)(T_2+\bar T_2)}\,,
\end{eqnarray}
a straightforward computation gives
\begin{equation}
    \Xi(\delta K_{\text{SS}})=-\frac{1}{t_1t_2}u_3^3\partial_{U_3}\partial_{\bar U_3}\hat{f}\,.
\end{equation}
This allows us to simplify the equation (\ref{eqdeltakss}) to
\begin{equation}\label{eqfEpsilon}
    u_3^2\partial_{U_3}\partial_{\bar U_3}\hat{f}=\mathcal{E}_3(iU_3)\,.
\end{equation}
To solve this equation, we notice that for $U_3=x+i y$ ($u_3\equiv 2x$) we have the identity
\begin{equation}
    u_3^2\partial_{U_3}\partial_{{\bar U}_3}=x^2\left(\partial^2_x+\partial^2_y\right)=-\frac{x^2}{y^2}\Delta
\end{equation}
with the hyperbolic Laplacian defined in (\ref{hyperlaplacian}), and that, from the eigenvalue equation (\ref{eigenE3}), it follows 
\begin{equation}
    \Delta\,\mathcal{E}_3(i U_3)=-\frac{6\,y^2}{x^2}\mathcal{E}_3(iU_3)\, .
\end{equation}
The equation for $f$ then assumes the illuminating form
\begin{equation}
    \Delta\left(\hat{f}-\frac{1}{6}\mathcal{E}_3\right)=0\implies \hat{f}=\frac{1}{6}\mathcal{E}_3+g
\end{equation}
which determines $f$ up to an harmonic function $g$.  Putting everything together, we conclude that
\begin{equation}
    \delta K_{\text{SS}} = k_1 \frac{(U_3+\bar{U}_3)\mathcal{E}_3(i U_3)}{(T_1+\bar{T}_1)(T_2+\bar T_2)}
\end{equation}
provides the required matching to the string-derived one-loop vacuum energy.

\paragraph*{$\alpha'^3$-corrections}
To address the fate of these corrections,  it is  very useful to note that if a correction $\delta K$ is a real homogeneous function of $\phi^a$, $\delta K(\phi^a)$, with degree $k$ then we can apply Euler's theorem to evaluate $\Xi (\delta K)$.  Indeed, we then have
\begin{equation}\label{euler}
\begin{split}
    \Xi(\delta K)=-\left(2\,\phi^a\frac{\partial\delta K}{\partial\phi^a}+\phi^a\phi^b\frac{\partial^2\,\delta K}{\partial \phi^a\partial \phi^b}\right)&=-\left(2\,k+k(k-1)\right)\,\delta K\\\
      &=-k(k+1)\delta K\, ,
 \end{split}   
\end{equation}
from which we  infer that $\Xi(\delta K)$ vanishes whenever the homogeneous function $\delta K(\phi^a)$ has degree $k=0,-1$.
 It is readily checked that $\delta K_{\alpha'}(\phi^a)$ is indeed homogeneous of degree $k=-1$, hence
\begin{equation} \label{eq:noscalesimp}
\begin{split}
     \Xi (\delta K_{\alpha'})=0\,.
 \end{split}   
\end{equation}
 Because of this extended no-scale cancellation, there is no dangerous $\delta K_{\alpha'}\ab W_\tree\ab^2\subset \delta K_{\alpha'} |W|^2$ term arising at first order and $\delta K_{\alpha'}$ appears in the scalar potential only multiplying the non-perturbative superpotential.   Furthermore, the third and fourth term in (\ref{dV}) actually cancel each other: indeed, using (\ref{propktree}) and (\ref{euler}) it follows that
\begin{equation}
    K^{a\bar b}_{\tree}(\delta K_{\alpha'})_{\bar b a} K_{\text{tree}\,a}=-\phi^b(\delta K_{\alpha'})_{\bar b a}=-\frac{\partial}{\partial\Phi^a}\left(\phi^b\frac{\partial\delta K_{\alpha'}}{\partial \Phi^{\bar{b}}}\right)=\delta K_{\alpha'}\, .
\end{equation}
 
\paragraph*{The leading order correction to $V$}
 Finally, let us present the leading order correction to the scalar potential, from $\delta K$ and $W_{\text{np}}$.  Putting everything together, up to second order corrections the scalar potential reads
\begin{equation}\label{structurev1storder}
\begin{split}
V&=e^{K_\text{tree}}\left(\left(K_\text{tree}^{\alpha\bar{\beta}}(1+\delta K)-K_\tree^{\alpha \bar \gamma}(\delta K_{\alpha'})_{\bar \gamma \delta}K_\tree^{\delta\bar{\beta}}\right)D_\alpha^{(0)}W_{\np}D_{\bar{\beta}}^{(0)}\overline{W}_\np-\,\delta K_{\text{SS}}|W_\tree+W_{\text{np}}|^2\right)\\
&\quad+\mathcal{O}(\delta^2K)
\end{split}
\end{equation}
where we have used 
\begin{equation}
    \Xi(\delta K_{\text{SS}})=-\,\delta K_{\text{SS}}\,.
\end{equation}


 \bibliographystyle{JHEP}
 \bibliography{biblio.bib}

@article{dms-review,
    author = "Dudas, E. and Mourad, J. and Sagnotti, A.",
    title = "{Supersymmetry Breaking with Fields, Strings and Branes}",
    eprint = "2511.04367",
    archivePrefix = "arXiv",
    primaryClass = "hep-th",
    month = "11",
    year = "2025"
}

@article{Angelantonj:2003hr,
    author = "Angelantonj, Carlo and Antoniadis, Ignatios",
    title = "{Suppressing the cosmological constant in nonsupersymmetric type I strings}",
    eprint = "hep-th/0307254",
    archivePrefix = "arXiv",
    reportNumber = "CERN-TH-2003-165",
    doi = "10.1016/j.nuclphysb.2003.09.047",
    journal = "Nucl. Phys. B",
    volume = "676",
    pages = "129--148",
    year = "2004"
}

@article{Cascales:2003zp,
    author = "Cascales, Juan F. G. and Uranga, Angel M.",
    title = "{Chiral 4d string vacua with D branes and NSNS and RR fluxes}",
    eprint = "hep-th/0303024",
    archivePrefix = "arXiv",
    reportNumber = "FTUAM-03-03, IFT-UAM-CSIC-03-07",
    doi = "10.1088/1126-6708/2003/05/011",
    journal = "JHEP",
    volume = "05",
    pages = "011",
    year = "2003"
}

@article{Dienes:1994np,
    author = "Dienes, Keith R.",
    title = "{Modular invariance, finiteness, and misaligned supersymmetry: New constraints on the numbers of physical string states}",
    eprint = "hep-th/9402006",
    archivePrefix = "arXiv",
    reportNumber = "MCGILL-94-04",
    doi = "10.1016/0550-3213(94)90153-8",
    journal = "Nucl. Phys. B",
    volume = "429",
    pages = "533--588",
    year = "1994"
}

@article{Cribiori:2020sct,
    author = "Cribiori, Niccol\`o and Parameswaran, Susha and Tonioni, Flavio and Wrase, Timm",
    title = "{Misaligned Supersymmetry and Open Strings}",
    eprint = "2012.04677",
    archivePrefix = "arXiv",
    primaryClass = "hep-th",
    doi = "10.1007/JHEP04(2021)099",
    journal = "JHEP",
    volume = "04",
    pages = "099",
    year = "2021"
}

@inproceedings{Pradisi:Bfield,
    author = "Pradisi, Gianfranco",
    title = "{Magnetic fluxes, NS NS B field and shifts in four-dimensional orientifolds}",
    booktitle = "{2nd String Phenomenology 2003}",
    eprint = "hep-th/0310154",
    archivePrefix = "arXiv",
    reportNumber = "ROM2F-2003-26",
    pages = "304--314",
    month = "10",
    year = "2003"
}

@article{Kutasov:1990sv,
    author = "Kutasov, David and Seiberg, Nathan",
    title = "{Number of degrees of freedom, density of states and tachyons in string theory and CFT}",
    reportNumber = "PUPT-1221, RU-90-60",
    doi = "10.1016/0550-3213(91)90426-X",
    journal = "Nucl. Phys. B",
    volume = "358",
    pages = "600--618",
    year = "1991"
}

@article{Grimm:2004uq,
    author = "Grimm, Thomas W. and Louis, Jan",
    title = "{The Effective action of N = 1 Calabi-Yau orientifolds}",
    eprint = "hep-th/0403067",
    archivePrefix = "arXiv",
    reportNumber = "LPTENS-04-14",
    doi = "10.1016/j.nuclphysb.2004.08.005",
    journal = "Nucl. Phys. B",
    volume = "699",
    pages = "387--426",
    year = "2004"
}

@article{GVW,
    author = "Gukov, Sergei and Vafa, Cumrun and Witten, Edward",
    title = "{CFT's from Calabi-Yau four folds}",
    eprint = "hep-th/9906070",
    archivePrefix = "arXiv",
    reportNumber = "HUTP-99-A034, IASSNS-HEP-99-52, PUPT-1864",
    doi = "10.1016/S0550-3213(00)00373-4",
    journal = "Nucl. Phys. B",
    volume = "584",
    pages = "69--108",
    year = "2000",
    note = "[Erratum: Nucl.Phys.B 608, 477--478 (2001)]"
}

@article{nongeoref1,
    author = "de Carlos, Beatriz and Guarino, Adolfo and Moreno, Jesus M.",
    title = "{Flux moduli stabilisation, Supergravity algebras and no-go theorems}",
    eprint = "0907.5580",
    archivePrefix = "arXiv",
    primaryClass = "hep-th",
    reportNumber = "IFT-UAM-CSIC-09-36",
    doi = "10.1007/JHEP01(2010)012",
    journal = "JHEP",
    volume = "01",
    pages = "012",
    year = "2010"
}

@article{pr-nonlinear,
    author = "Pradisi, Gianfranco and Riccioni, Fabio",
    title = "{Geometric couplings and brane supersymmetry breaking}",
    eprint = "hep-th/0107090",
    archivePrefix = "arXiv",
    reportNumber = "ROM2F-01-24",
    doi = "10.1016/S0550-3213(01)00441-2",
    journal = "Nucl. Phys. B",
    volume = "615",
    pages = "33--60",
    year = "2001"
}

@article{nongeoref2,
    author = "de Carlos, Beatriz and Guarino, Adolfo and Moreno, Jesus M.",
    title = "{Complete classification of Minkowski vacua in generalised flux models}",
    eprint = "0911.2876",
    archivePrefix = "arXiv",
    primaryClass = "hep-th",
    reportNumber = "IFT-UAM-CSIC-09-51",
    doi = "10.1007/JHEP02(2010)076",
    journal = "JHEP",
    volume = "02",
    pages = "076",
    year = "2010"
}

@article{nongeoref3,
    author = "Dibitetto, Giuseppe and Linares, Roman and Roest, Diederik",
    title = "{Flux Compactifications, Gauge Algebras and De Sitter}",
    eprint = "1001.3982",
    archivePrefix = "arXiv",
    primaryClass = "hep-th",
    doi = "10.1016/j.physletb.2010.03.074",
    journal = "Phys. Lett. B",
    volume = "688",
    pages = "96--100",
    year = "2010"
}

@article{nongeoref4,
    author = {Bl{\r{a}}b{\"a}ck, Johan and Danielsson, Ulf and Dibitetto, Giuseppe},
    title = "{Fully stable dS vacua from generalised fluxes}",
    eprint = "1301.7073",
    archivePrefix = "arXiv",
    primaryClass = "hep-th",
    doi = "10.1007/JHEP08(2013)054",
    journal = "JHEP",
    volume = "08",
    pages = "054",
    year = "2013"
}

@article{nongeoref5,
    author = "Damian, Cesar and Diaz-Barron, Luis R. and Loaiza-Brito, Oscar and Sabido, Miguel",
    title = "{Slow-Roll Inflation in Non-geometric Flux Compactification}",
    eprint = "1302.0529",
    archivePrefix = "arXiv",
    primaryClass = "hep-th",
    doi = "10.1007/JHEP06(2013)109",
    journal = "JHEP",
    volume = "06",
    pages = "109",
    year = "2013"
}

@article{nongeoref6,
    author = "Damian, Cesar and Loaiza-Brito, Oscar",
    title = "{More stable de Sitter vacua from S-dual nongeometric fluxes}",
    eprint = "1304.0792",
    archivePrefix = "arXiv",
    primaryClass = "hep-th",
    doi = "10.1103/PhysRevD.88.046008",
    journal = "Phys. Rev. D",
    volume = "88",
    number = "4",
    pages = "046008",
    year = "2013"
}

@article{nongeoref7,
    author = "Cribiori, Niccol{\`o} and Kallosh, Renata and Linde, Andrei and Roupec, Christoph",
    title = "{de Sitter Minima from M theory and String theory}",
    eprint = "1912.02791",
    archivePrefix = "arXiv",
    primaryClass = "hep-th",
    doi = "10.1103/PhysRevD.101.046018",
    journal = "Phys. Rev. D",
    volume = "101",
    number = "4",
    pages = "046018",
    year = "2020"
}

@article{dudasshift,
    author = "Antoniadis, Ignatios and D'Appollonio, G. and Dudas, E. and Sagnotti, A.",
    title = "{Open descendants of Z(2) x Z(2) freely acting orbifolds}",
    eprint = "hep-th/9907184",
    archivePrefix = "arXiv",
    reportNumber = "CPTH-S725-0799, DFF-340-07-99, LPT-ORSAY-99-58, ROM2F-99-22",
    doi = "10.1016/S0550-3213(99)00616-1",
    journal = "Nucl. Phys. B",
    volume = "565",
    pages = "123--156",
    year = "2000"
}

@article{antoniadis2DD,
    author = "Anchordoqui, Luis and Antoniadis, Ignatios and Lust, Dieter",
    title = "{Two Micron-Size Dark Dimensions}",
    eprint = "2501.11690",
    archivePrefix = "arXiv",
    primaryClass = "hep-th",
    reportNumber = "MPP-2025-5, LMU-ASC 02/25",
    month = "1",
    year = "2025"
}

@article{nongeoflux,
    author = "Betzler, Philip and Plauschinn, Erik",
    title = "{Type IIB flux vacua and tadpole cancellation}",
    eprint = "1905.08823",
    archivePrefix = "arXiv",
    primaryClass = "hep-th",
    doi = "10.1002/prop.201900065",
    journal = "Fortsch. Phys.",
    volume = "67",
    number = "11",
    pages = "1900065",
    year = "2019"
}

@article{Kachru:2003aw,
    author = "Kachru, Shamit and Kallosh, Renata and Linde, Andrei D. and Trivedi, Sandip P.",
    title = "{De Sitter vacua in string theory}",
    eprint = "hep-th/0301240",
    archivePrefix = "arXiv",
    reportNumber = "SLAC-PUB-9630, SU-ITP-03-01, TIFR-TH-03-03",
    doi = "10.1103/PhysRevD.68.046005",
    journal = "Phys. Rev. D",
    volume = "68",
    pages = "046005",
    year = "2003"
}

@article{LVS,
    author = "Balasubramanian, Vijay and Berglund, Per and Conlon, Joseph P. and Quevedo, Fernando",
    title = "{Systematics of moduli stabilisation in Calabi-Yau flux compactifications}",
    eprint = "hep-th/0502058",
    archivePrefix = "arXiv",
    reportNumber = "DAMTP-2005-10, UNH-05-01, UPR-1109-T",
    doi = "10.1088/1126-6708/2005/03/007",
    journal = "JHEP",
    volume = "03",
    pages = "007",
    year = "2005"
}

@article{noscale5,
    author = "Kounnas, Costas and Partouche, Herve",
    title = "{Super no-scale models in string theory}",
    eprint = "1607.01767",
    archivePrefix = "arXiv",
    primaryClass = "hep-th",
    reportNumber = "LPTENS-16-04, CPHT-RR033.062016",
    doi = "10.1016/j.nuclphysb.2016.10.001",
    journal = "Nucl. Phys. B",
    volume = "913",
    pages = "593--626",
    year = "2016"
}

@article{Antoniadis:r4couplings,
    author = "Antoniadis, Ignatios and Ferrara, S. and Minasian, R. and Narain, K. S.",
    title = "{R**4 couplings in M and type II theories on Calabi-Yau spaces}",
    eprint = "hep-th/9707013",
    archivePrefix = "arXiv",
    reportNumber = "CERN-TH-97-094, CERN-TH-97-94, CPTH-S512-0697",
    doi = "10.1016/S0550-3213(97)00572-5",
    journal = "Nucl. Phys. B",
    volume = "507",
    pages = "571--588",
    year = "1997"
}

@article{Blumenhageninstantons,
    author = "Blumenhagen, Ralph and Cvetic, Mirjam and Kachru, Shamit and Weigand, Timo",
    title = "{D-Brane Instantons in Type II Orientifolds}",
    eprint = "0902.3251",
    archivePrefix = "arXiv",
    primaryClass = "hep-th",
    reportNumber = "MPP-2009-15, UPR-1205-T, SLAC-PUB-13531",
    doi = "10.1146/annurev.nucl.010909.083113",
    journal = "Ann. Rev. Nucl. Part. Sci.",
    volume = "59",
    pages = "269--296",
    year = "2009"
}

@article{AntoniadisnonCompactCY,
    author = "Antoniadis, Ignatios and Minasian, Ruben and Vanhove, Pierre",
    title = "{Noncompact Calabi-Yau manifolds and localized gravity}",
    eprint = "hep-th/0209030",
    archivePrefix = "arXiv",
    reportNumber = "CPHT-RR-067-0902, CERN-TH-2002-220, SACLAY-SPHT-T02-035, CPHT-RR-067.0902, SPHT-T02-035",
    doi = "10.1016/S0550-3213(02)00974-4",
    journal = "Nucl. Phys. B",
    volume = "648",
    pages = "69--93",
    year = "2003"
}

@article{Coudarchet:2021qwc,
    author = "Coudarchet, Thibaut and Dudas, Emilian and Partouche, Herv\'e",
    title = "{Geometry of orientifold vacua and supersymmetry breaking}",
    eprint = "2105.06913",
    archivePrefix = "arXiv",
    primaryClass = "hep-th",
    reportNumber = "CPHT-RR026.032021",
    doi = "10.1007/JHEP07(2021)104",
    journal = "JHEP",
    volume = "07",
    pages = "104",
    year = "2021"
}

@article{carlo1,
    author = "Angelantonj, Carlo and Cardella, Matteo",
    title = "{Vanishing perturbative vacuum energy in nonsupersymmetric orientifolds}",
    eprint = "hep-th/0403107",
    archivePrefix = "arXiv",
    reportNumber = "HU-EP-04-13, IFUM-786-FT",
    doi = "10.1016/j.physletb.2004.06.058",
    journal = "Phys. Lett. B",
    volume = "595",
    pages = "505--512",
    year = "2004"
}

@article{Abel:2017rch,
    author = "Abel, Steven and Stewart, Richard J.",
    title = "{Exponential suppression of the cosmological constant in nonsupersymmetric string vacua at two loops and beyond}",
    eprint = "1701.06629",
    archivePrefix = "arXiv",
    primaryClass = "hep-th",
    reportNumber = "IPPP-17-6, CERN-TH-2017-015",
    doi = "10.1103/PhysRevD.96.106013",
    journal = "Phys. Rev. D",
    volume = "96",
    number = "10",
    pages = "106013",
    year = "2017"
}

@article{Montero:dd,
    author = "Montero, Miguel and Vafa, Cumrun and Valenzuela, Irene",
    title = "{The dark dimension and the Swampland}",
    eprint = "2205.12293",
    archivePrefix = "arXiv",
    primaryClass = "hep-th",
    doi = "10.1007/JHEP02(2023)022",
    journal = "JHEP",
    volume = "02",
    pages = "022",
    year = "2023"
}

@article{add,
    author = "Arkani-Hamed, Nima and Dimopoulos, Savas and Dvali, G. R.",
    title = "{The Hierarchy problem and new dimensions at a millimeter}",
    eprint = "hep-ph/9803315",
    archivePrefix = "arXiv",
    reportNumber = "SLAC-PUB-7769, SU-ITP-98-13",
    doi = "10.1016/S0370-2693(98)00466-3",
    journal = "Phys. Lett. B",
    volume = "429",
    pages = "263--272",
    year = "1998"
}

@article{ddg,
    author = "Dienes, Keith R. and Dudas, Emilian and Gherghetta, Tony",
    title = "{Extra space-time dimensions and unification}",
    eprint = "hep-ph/9803466",
    archivePrefix = "arXiv",
    reportNumber = "CERN-TH-98-065",
    doi = "10.1016/S0370-2693(98)00977-0",
    journal = "Phys. Lett. B",
    volume = "436",
    pages = "55--65",
    year = "1998"
}

@article{aadd,
    author = "Antoniadis, Ignatios and Arkani-Hamed, Nima and Dimopoulos, Savas and Dvali, G. R.",
    title = "{New dimensions at a millimeter to a Fermi and superstrings at a TeV}",
    eprint = "hep-ph/9804398",
    archivePrefix = "arXiv",
    reportNumber = "SLAC-PUB-7801, SU-ITP-98-28, CPTH-S608-0498, IC-98-39",
    doi = "10.1016/S0370-2693(98)00860-0",
    journal = "Phys. Lett. B",
    volume = "436",
    pages = "257--263",
    year = "1998"
}

@article{Antoniadis:mediation,
    author = "Antoniadis, Ignatios and Taylor, Tomasz R.",
    title = "{Note on mediation of supersymmetry breaking from closed to open strings}",
    eprint = "hep-th/0509048",
    archivePrefix = "arXiv",
    reportNumber = "CERN-PH-TH-2005-155",
    doi = "10.1016/j.nuclphysb.2005.10.017",
    journal = "Nucl. Phys. B",
    volume = "731",
    pages = "164--170",
    year = "2005"
}

@article{Dixon:1990pc,
    author = "Dixon, Lance J. and Kaplunovsky, Vadim and Louis, Jan",
    title = "{Moduli dependence of string loop corrections to gauge coupling constants}",
    reportNumber = "SLAC-PUB-5138, UTTG-36-89",
    doi = "10.1016/0550-3213(91)90490-O",
    journal = "Nucl. Phys. B",
    volume = "355",
    pages = "649--688",
    year = "1991"
}

@article{tabletop1,
    author = "Adelberger, E. G. and Gundlach, J. H. and Heckel, B. R. and Hoedl, S. and Schlamminger, S.",
    title = "{Torsion balance experiments: A low-energy frontier of particle physics}",
    doi = "10.1016/j.ppnp.2008.08.002",
    journal = "Prog. Part. Nucl. Phys.",
    volume = "62",
    pages = "102--134",
    year = "2009"
}

@article{tabletop2,
    author = "Murata, Jiro and Tanaka, Saki",
    title = "{A review of short-range gravity experiments in the LHC era}",
    eprint = "1408.3588",
    archivePrefix = "arXiv",
    primaryClass = "hep-ex",
    doi = "10.1088/0264-9381/32/3/033001",
    journal = "Class. Quant. Grav.",
    volume = "32",
    number = "3",
    pages = "033001",
    year = "2015"
}

@article{Antoniadis:1998ep,
    author = "Antoniadis, Ignatios and D'Appollonio, G. and Dudas, E. and Sagnotti, A.",
    title = "{Partial breaking of supersymmetry, open strings and M theory}",
    eprint = "hep-th/9812118",
    archivePrefix = "arXiv",
    reportNumber = "CERN-TH-98-382, CPTH-S691-1198, LPTHE-ORSAY-98-70, ROM2F-98-41",
    doi = "10.1016/S0550-3213(99)00232-1",
    journal = "Nucl. Phys. B",
    volume = "553",
    pages = "133--154",
    year = "1999"
}

@article{Antoniadis:1998ki,
    author = "Antoniadis, Ignatios and Dudas, E. and Sagnotti, A.",
    title = "{Supersymmetry breaking, open strings and M theory}",
    eprint = "hep-th/9807011",
    archivePrefix = "arXiv",
    reportNumber = "CERN-TH-98-212, CPTH-S616-0698, LPTHE-ORSAY-98-42, ROM2F-98-21",
    doi = "10.1016/S0550-3213(98)00806-2",
    journal = "Nucl. Phys. B",
    volume = "544",
    pages = "469--502",
    year = "1999"
}

@article{Blum:1997cs,
    author = "Blum, Julie D. and Dienes, Keith R.",
    title = "{Duality without supersymmetry: The Case of the SO(16) x SO(16) string}",
    eprint = "hep-th/9707148",
    archivePrefix = "arXiv",
    reportNumber = "IASSNS-HEP-97-67",
    doi = "10.1016/S0370-2693(97)01172-6",
    journal = "Phys. Lett. B",
    volume = "414",
    pages = "260--268",
    year = "1997"
}

@article{Blum:1997gw,
    author = "Blum, Julie D. and Dienes, Keith R.",
    title = "{Strong / weak coupling duality relations for nonsupersymmetric string theories}",
    eprint = "hep-th/9707160",
    archivePrefix = "arXiv",
    reportNumber = "IASSNS-HEP-97-80",
    doi = "10.1016/S0550-3213(97)00803-1",
    journal = "Nucl. Phys. B",
    volume = "516",
    pages = "83--159",
    year = "1998"
}

@article{tabletop3,
    author = "Tan, Wen-Hai and Yang, Shan-Qing and Shao, Cheng-Gang and Li, Jia and Du, An-Bin and Zhan, Bi-Fu and Wang, Qing-Lan and Luo, Peng-Shun and Tu, Liang-Cheng and Luo, Jun",
    title = "{New Test of the Gravitational Inverse-Square Law at the Submillimeter Range with Dual Modulation and Compensation}",
    doi = "10.1103/PhysRevLett.116.131101",
    journal = "Phys. Rev. Lett.",
    volume = "116",
    number = "13",
    pages = "131101",
    year = "2016"
}

@article{tabletop4,
    author = "Lee, J. G. and Adelberger, E. G. and Cook, T. S. and Fleischer, S. M. and Heckel, B. R.",
    title = "{New Test of the Gravitational $1/r^2$ Law at Separations down to 52 $\mu$m}",
    eprint = "2002.11761",
    archivePrefix = "arXiv",
    primaryClass = "hep-ex",
    doi = "10.1103/PhysRevLett.124.101101",
    journal = "Phys. Rev. Lett.",
    volume = "124",
    number = "10",
    pages = "101101",
    year = "2020"
}

@article{tabletop5,
    author = "Adelberger, E. G. and Heckel, Blayne R. and Nelson, A. E.",
    title = "{Tests of the gravitational inverse square law}",
    eprint = "hep-ph/0307284",
    archivePrefix = "arXiv",
    doi = "10.1146/annurev.nucl.53.041002.110503",
    journal = "Ann. Rev. Nucl. Part. Sci.",
    volume = "53",
    pages = "77--121",
    year = "2003"
}

@article{Abel:2018zyt,
    author = "Abel, Steven and Dudas, Emilian and Lewis, Daniel and Partouche, Herv\'e",
    title = "{Stability and vacuum energy in open string models with broken supersymmetry}",
    eprint = "1812.09714",
    archivePrefix = "arXiv",
    primaryClass = "hep-th",
    reportNumber = "CPHT-RR117.122018, DCPT-18/37, IPPP/18/112",
    doi = "10.1007/JHEP10(2019)226",
    journal = "JHEP",
    volume = "10",
    pages = "226",
    year = "2019"
}

@article{Cicoli:2007xp,
    author = "Cicoli, Michele and Conlon, Joseph P. and Quevedo, Fernando",
    title = "{Systematics of String Loop Corrections in Type IIB Calabi-Yau Flux Compactifications}",
    eprint = "0708.1873",
    archivePrefix = "arXiv",
    primaryClass = "hep-th",
    reportNumber = "DAMTP-2007-75",
    doi = "10.1088/1126-6708/2008/01/052",
    journal = "JHEP",
    volume = "01",
    pages = "052",
    year = "2008"
}

@article{Angelantonj:2002ct,
    author = "Angelantonj, Carlo and Sagnotti, Augusto",
    title = "{Open strings}",
    eprint = "hep-th/0204089",
    archivePrefix = "arXiv",
    reportNumber = "CERN-TH-2002-025, ROM2F-2002-08, LPTENS-02-14, CPHT-RR-020-0202, CPHT-RR-020.0202",
    doi = "10.1016/S0370-1573(02)00273-9",
    journal = "Phys. Rept.",
    volume = "371",
    pages = "1--150",
    year = "2002",
    note = "[Erratum: Phys.Rept. 376, 407 (2003)]"
}

@article{Kiritsis:1997hf,
    author = "Kiritsis, E. and Obers, N. A.",
    title = "{Heterotic type I duality in D \ensuremath{<} 10-dimensions, threshold corrections and D instantons}",
    eprint = "hep-th/9709058",
    archivePrefix = "arXiv",
    reportNumber = "CERN-TH-97-233",
    doi = "10.1088/1126-6708/1997/10/004",
    journal = "JHEP",
    volume = "10",
    pages = "004",
    year = "1997"
}

@article{Trapletti:2002uk,
    author = "Trapletti, M.",
    title = "{On the unfolding of the fundamental region in integrals of modular invariant amplitudes}",
    eprint = "hep-th/0211281",
    archivePrefix = "arXiv",
    reportNumber = "SISSA-88-2002-EP",
    doi = "10.1088/1126-6708/2003/02/012",
    journal = "JHEP",
    volume = "02",
    pages = "012",
    year = "2003"
}

@article{Weinberg:1988cp,
    author = "Weinberg, Steven",
    editor = "Hsu, Jong-Ping and Fine, D.",
    title = "{The Cosmological Constant Problem}",
    reportNumber = "UTTG-12-88",
    doi = "10.1103/RevModPhys.61.1",
    journal = "Rev. Mod. Phys.",
    volume = "61",
    pages = "1--23",
    year = "1989"
}

@article{Villadoro:2005cu,
    author = "Villadoro, Giovanni and Zwirner, Fabio",
    title = "{N=1 effective potential from dual type-IIA D6/O6 orientifolds with general fluxes}",
    eprint = "hep-th/0503169",
    archivePrefix = "arXiv",
    reportNumber = "ROMA-1402-05",
    doi = "10.1088/1126-6708/2005/06/047",
    journal = "JHEP",
    volume = "06",
    pages = "047",
    year = "2005"
}

@article{quevedononreno,
    author = "Burgess, C. P. and Escoda, C. and Quevedo, F.",
    title = "{Nonrenormalization of flux superpotentials in string theory}",
    eprint = "hep-th/0510213",
    archivePrefix = "arXiv",
    reportNumber = "DAMTP-2005-82",
    doi = "10.1088/1126-6708/2006/06/044",
    journal = "JHEP",
    volume = "06",
    pages = "044",
    year = "2006"
}

@article{Dine:nonreno,
    author = "Dine, Michael and Seiberg, N.",
    title = "{Nonrenormalization Theorems in Superstring Theory}",
    reportNumber = "WIS/86/53-Ph",
    doi = "10.1103/PhysRevLett.57.2625",
    journal = "Phys. Rev. Lett.",
    volume = "57",
    pages = "2625",
    year = "1986"
}

@article{Derendinger:2004jn,
    author = "Derendinger, Jean-Pierre and Kounnas, Costas and Petropoulos, P. Marios and Zwirner, Fabio",
    title = "{Superpotentials in IIA compactifications with general fluxes}",
    eprint = "hep-th/0411276",
    archivePrefix = "arXiv",
    reportNumber = "NEIP-04-08, LPTENS-0440, CPTH-RR053-0904, CERN-PH-TH-2004-228, ROMA-1395-04",
    doi = "10.1016/j.nuclphysb.2005.02.038",
    journal = "Nucl. Phys. B",
    volume = "715",
    pages = "211--233",
    year = "2005"
}

@article{Sugimoto:1999tx,
    author = "Sugimoto, Shigeki",
    title = "{Anomaly cancellations in type I D-9 - anti-D-9 system and the USp(32) string theory}",
    eprint = "hep-th/9905159",
    archivePrefix = "arXiv",
    reportNumber = "YITP-99-25",
    doi = "10.1143/PTP.102.685",
    journal = "Prog. Theor. Phys.",
    volume = "102",
    pages = "685--699",
    year = "1999"
}

@article{Antoniadis:1999xk,
    author = "Antoniadis, Ignatios and Dudas, E. and Sagnotti, A.",
    title = "{Brane supersymmetry breaking}",
    eprint = "hep-th/9908023",
    archivePrefix = "arXiv",
    reportNumber = "CPHT-S727-0799, LPT-ORSAY-99-60, ROM2F-99-23",
    doi = "10.1016/S0370-2693(99)01023-0",
    journal = "Phys. Lett. B",
    volume = "464",
    pages = "38--45",
    year = "1999"
}

@article{Pradisi:1995pp,
    author = "Pradisi, G. and Sagnotti, A. and Stanev, Ya. S.",
    title = "{The Open descendants of nondiagonal SU(2) WZW models}",
    eprint = "hep-th/9506014",
    archivePrefix = "arXiv",
    reportNumber = "ROM2F-95-09, CPTH-RR-358-0695",
    doi = "10.1016/0370-2693(95)00840-H",
    journal = "Phys. Lett. B",
    volume = "356",
    pages = "230--238",
    year = "1995"
}

@article{Mourad:2017rrl,
    author = "Mourad, J. and Sagnotti, A.",
    title = "{An Update on Brane Supersymmetry Breaking}",
    eprint = "1711.11494",
    archivePrefix = "arXiv",
    primaryClass = "hep-th",
    month = "11",
    year = "2017"
}

@article{Hannestad:2003yd,
    author = "Hannestad, Steen and Raffelt, Georg G.",
    title = "{Supernova and neutron star limits on large extra dimensions reexamined}",
    eprint = "hep-ph/0304029",
    archivePrefix = "arXiv",
    reportNumber = "MPP-2003-19, MPP-2003-19",
    doi = "10.1103/PhysRevD.69.029901",
    journal = "Phys. Rev. D",
    volume = "67",
    pages = "125008",
    year = "2003",
    note = "[Erratum: Phys.Rev.D 69, 029901 (2004)]"
}

@article{Hardy:2025ajb,
    author = "Hardy, Edward and Sokolov, Anton and Stubbs, Henry",
    title = "{Stellar cooling limits on KK gravitons and dark dimensions}",
    eprint = "2510.18975",
    archivePrefix = "arXiv",
    primaryClass = "hep-ph",
    month = "10",
    year = "2025"
}

@article{Dudas:2000nv,
    author = "Dudas, E. and Mourad, J.",
    title = "{Consistent gravitino couplings in nonsupersymmetric strings}",
    eprint = "hep-th/0012071",
    archivePrefix = "arXiv",
    reportNumber = "LPT-ORSAY-00-128",
    doi = "10.1016/S0370-2693(01)00777-8",
    journal = "Phys. Lett. B",
    volume = "514",
    pages = "173--182",
    year = "2001"
}

@article{Witten:1997bs,
    author = "Witten, Edward",
    title = "{Toroidal compactification without vector structure}",
    eprint = "hep-th/9712028",
    archivePrefix = "arXiv",
    reportNumber = "IASSNS-HEP-97-129",
    doi = "10.1088/1126-6708/1998/02/006",
    journal = "JHEP",
    volume = "02",
    pages = "006",
    year = "1998"
}

@article{Parameswaran:2024mrc,
    author = "Parameswaran, Susha and Serra, Marco",
    title = "{On (A)dS solutions from Scherk-Schwarz orbifolds}",
    eprint = "2407.16781",
    archivePrefix = "arXiv",
    primaryClass = "hep-th",
    doi = "10.1007/JHEP10(2024)039",
    journal = "JHEP",
    volume = "10",
    pages = "039",
    year = "2024"
}

@article{Gallego:2011jm,
    author = "Gallego, Diego",
    title = "{On the Effective Description of Large Volume Compactifications}",
    eprint = "1103.5469",
    archivePrefix = "arXiv",
    primaryClass = "hep-th",
    doi = "10.1007/JHEP06(2011)087",
    journal = "JHEP",
    volume = "06",
    pages = "087",
    year = "2011"
}

@article{Ferrara:1988jx,
    author = "Ferrara, Sergio and Kounnas, Costas and Porrati, Massimo and Zwirner, Fabio",
    title = "{Superstrings with Spontaneously Broken Supersymmetry and their Effective Theories}",
    reportNumber = "UCB/PTH/88/19, LBL-25776, LPTENS-88/28, UCLA/88/TEP/27",
    doi = "10.1016/0550-3213(89)90048-5",
    journal = "Nucl. Phys. B",
    volume = "318",
    pages = "75--105",
    year = "1989"
}

@misc{Zwirner:2025ohv,
  author        = {Zwirner, Fabio},
  title         = {No-scale supergravity},
  eprint        = {2504.06190},
  archivePrefix = {arXiv},
  primaryClass  = {hep-th},
  year          = {2025}
}

@article{noscale-cremmer,
    author = "Cremmer, E. and Ferrara, S. and Kounnas, C. and Nanopoulos, Dimitri V.",
    title = "{Naturally Vanishing Cosmological Constant in N=1 Supergravity}",
    reportNumber = "CERN-TH-3667",
    doi = "10.1016/0370-2693(83)90106-5",
    journal = "Phys. Lett. B",
    volume = "133",
    pages = "61",
    year = "1983"
}

@article{noscale-ellis,
    author = "Ellis, John R. and Kounnas, C. and Nanopoulos, Dimitri V.",
    title = "{Phenomenological SU(1,1) Supergravity}",
    reportNumber = "CERN-TH-3768",
    doi = "10.1016/0550-3213(84)90054-3",
    journal = "Nucl. Phys. B",
    volume = "241",
    pages = "406--428",
    year = "1984"
}

@article{ValeixoBento:2025emh,
    author = "Valeixo Bento, Bruno and Chakraborty, Dibya and Parameswaran, Susha and Zavala, Ivonne",
    title = "{A guide to frames, 2{\ensuremath{\pi}}{\textquoteright}s, scales and corrections in string compactifications}",
    eprint = "2301.05178",
    archivePrefix = "arXiv",
    primaryClass = "hep-th",
    doi = "10.1142/S0218271825300034",
    journal = "Int. J. Mod. Phys. D",
    volume = "34",
    number = "10",
    pages = "2530003",
    year = "2025"
}

@article{carlo2,
    author = "Angelantonj, Carlo and Cardella, Matteo and Irges, Nikos",
    title = "{An Alternative for Moduli Stabilisation}",
    eprint = "hep-th/0608022",
    archivePrefix = "arXiv",
    reportNumber = "DFTT-16-2006, IFUM-869-FT",
    doi = "10.1016/j.physletb.2006.08.072",
    journal = "Phys. Lett. B",
    volume = "641",
    pages = "474--480",
    year = "2006"
}

@article{Frey:2002hf,
    author = "Frey, Andrew R. and Polchinski, Joseph",
    title = "{N=3 warped compactifications}",
    eprint = "hep-th/0201029",
    archivePrefix = "arXiv",
    reportNumber = "NSF-ITP-01-77",
    doi = "10.1103/PhysRevD.65.126009",
    journal = "Phys. Rev. D",
    volume = "65",
    pages = "126009",
    year = "2002"
}

@article{Aghababaie:2003wz,
    author = "Aghababaie, Y. and Burgess, C. P. and Parameswaran, S. L. and Quevedo, F.",
    title = "{Towards a naturally small cosmological constant from branes in 6-D supergravity}",
    eprint = "hep-th/0304256",
    archivePrefix = "arXiv",
    reportNumber = "MCGILL-03-08, DAMTP-2003-38",
    doi = "10.1016/j.nuclphysb.2003.12.015",
    journal = "Nucl. Phys. B",
    volume = "680",
    pages = "389--414",
    year = "2004"
}

@article{Bhattacharya:2024kxp,
    author = "Bhattacharya, Sukannya and Borghetto, Giulia and Malhotra, Ameek and Parameswaran, Susha and Tasinato, Gianmassimo and Zavala, Ivonne",
    title = "{Cosmological tests of quintessence in quantum gravity}",
    eprint = "2410.21243",
    archivePrefix = "arXiv",
    primaryClass = "astro-ph.CO",
    doi = "10.1088/1475-7516/2025/04/086",
    journal = "JCAP",
    volume = "04",
    pages = "086",
    year = "2025"
}

@article{Khoury:2025txd,
    author = "Khoury, Justin and Lin, Meng-Xiang and Trodden, Mark",
    title = "{Apparent w{\ensuremath{<}}-1 and a Lower S8 from Dark Axion and Dark Baryons Interactions}",
    eprint = "2503.16415",
    archivePrefix = "arXiv",
    primaryClass = "astro-ph.CO",
    doi = "10.1103/w4qb-plk8",
    journal = "Phys. Rev. Lett.",
    volume = "135",
    number = "18",
    pages = "181001",
    year = "2025"
}

@article{Burgess:2021qti,
    author = "Burgess, C. P. and Quevedo, F.",
    title = "{Axion homeopathy: screening dilaton interactions}",
    eprint = "2110.10352",
    archivePrefix = "arXiv",
    primaryClass = "hep-th",
    reportNumber = "CERN-TH-2021-176",
    doi = "10.1088/1475-7516/2022/04/007",
    journal = "JCAP",
    volume = "04",
    number = "04",
    pages = "007",
    year = "2022"
}

@article{Brax:2023qyp,
    author = "Brax, Philippe and Burgess, C. P. and Quevedo, F.",
    title = "{Axio-Chameleons: a novel string-friendly multi-field screening mechanism}",
    eprint = "2310.02092",
    archivePrefix = "arXiv",
    primaryClass = "hep-th",
    doi = "10.1088/1475-7516/2024/03/015",
    journal = "JCAP",
    volume = "03",
    pages = "015",
    year = "2024"
}

@article{Kachru:1998hd,
    author = "Kachru, Shamit and Kumar, Jason and Silverstein, Eva",
    title = "{Vacuum energy cancellation in a nonsupersymmetric string}",
    eprint = "hep-th/9807076",
    archivePrefix = "arXiv",
    reportNumber = "SLAC-PUB-7875, LBNL-41932, LBL-41932, SU-ITP-98-35, UCB-PTH-98-33",
    doi = "10.1103/PhysRevD.59.106004",
    journal = "Phys. Rev. D",
    volume = "59",
    pages = "106004",
    year = "1999"
}

@article{Harvey:1998rc,
    author = "Harvey, Jeffrey A.",
    title = "{String duality and nonsupersymmetric strings}",
    eprint = "hep-th/9807213",
    archivePrefix = "arXiv",
    reportNumber = "EFI-98-31",
    doi = "10.1103/PhysRevD.59.026002",
    journal = "Phys. Rev. D",
    volume = "59",
    pages = "026002",
    year = "1999"
}

@article{Blumenhagen:1998uf,
    author = "Blumenhagen, Ralph and Gorlich, Lars",
    title = "{Orientifolds of nonsupersymmetric asymmetric orbifolds}",
    eprint = "hep-th/9812158",
    archivePrefix = "arXiv",
    reportNumber = "HUB-EP-98-76",
    doi = "10.1016/S0550-3213(99)00241-2",
    journal = "Nucl. Phys. B",
    volume = "551",
    pages = "601--616",
    year = "1999"
}

@article{Angelantonj:1999gm,
    author = "Angelantonj, C. and Antoniadis, Ignatios and Forger, K.",
    title = "{Nonsupersymmetric type I strings with zero vacuum energy}",
    eprint = "hep-th/9904092",
    archivePrefix = "arXiv",
    reportNumber = "CPHT-S711-0299",
    doi = "10.1016/S0550-3213(99)00344-2",
    journal = "Nucl. Phys. B",
    volume = "555",
    pages = "116--134",
    year = "1999"
}

@article{Abel:2020ldo,
    author = "Abel, Steven and Coudarchet, Thibaut and Partouche, Herv{\'e}",
    title = "{On the stability of open-string orbifold models with broken supersymmetry}",
    eprint = "2003.02545",
    archivePrefix = "arXiv",
    primaryClass = "hep-th",
    reportNumber = "IPPP/20/6, CPHT-RR012.032020",
    doi = "10.1016/j.nuclphysb.2020.115100",
    journal = "Nucl. Phys. B",
    volume = "957",
    pages = "115100",
    year = "2020",
    note = "[Erratum: Nucl.Phys.B 1004, 116548 (2024)]"
}

@article{ParticleDataGroup:2024cfk,
    author = "Navas, S. and others",
    collaboration = "Particle Data Group",
    title = "{Review of particle physics}",
    doi = "10.1103/PhysRevD.110.030001",
    journal = "Phys. Rev. D",
    volume = "110",
    number = "3",
    pages = "030001",
    year = "2024"
}

@article{Burgess:2004ib,
    author = "Burgess, C. P.",
    editor = "Allen, R. E. and Nanopoulos, Dimitri V. and Pope, C. N.",
    title = "{Towards a natural theory of dark energy: Supersymmetric large extra dimensions}",
    eprint = "hep-th/0411140",
    archivePrefix = "arXiv",
    doi = "10.1063/1.1848343",
    journal = "AIP Conf. Proc.",
    volume = "743",
    number = "1",
    pages = "417--449",
    year = "2004"
}

@article{Burgess:2023pnk,
    author = "Burgess, C. P. and Quevedo, F.",
    title = "{Perils of towers in the swamp: dark dimensions and the robustness of EFTs}",
    eprint = "2304.03902",
    archivePrefix = "arXiv",
    primaryClass = "hep-th",
    doi = "10.1007/JHEP09(2023)159",
    journal = "JHEP",
    volume = "09",
    pages = "159",
    year = "2023"
}

@article{ValeixoBento:2025yhz,
    author = "Valeixo Bento, Bruno and Montero, Miguel",
    title = "{An M-theory dS maximum from Casimir energies on Riemann-flat manifolds}",
    eprint = "2507.02037",
    archivePrefix = "arXiv",
    primaryClass = "hep-th",
    reportNumber = "IFT-025-070",
    month = "7",
    year = "2025"
}

@article{CMS:2019gwf,
    author = "Sirunyan, Albert M and others",
    collaboration = "CMS",
    title = "{Search for high mass dijet resonances with a new background prediction method in proton-proton collisions at $\sqrt{s} =$ 13 TeV}",
    eprint = "1911.03947",
    archivePrefix = "arXiv",
    primaryClass = "hep-ex",
    reportNumber = "CMS-EXO-19-012, CERN-EP-2019-222",
    doi = "10.1007/JHEP05(2020)033",
    journal = "JHEP",
    volume = "05",
    pages = "033",
    year = "2020"
}

@article{ATLAS:2019fgd,
    author = "Aad, Georges and others",
    collaboration = "ATLAS",
    title = "{Search for new resonances in mass distributions of jet pairs using 139 fb$^{-1}$ of $pp$ collisions at $\sqrt{s}=13$ TeV with the ATLAS detector}",
    eprint = "1910.08447",
    archivePrefix = "arXiv",
    primaryClass = "hep-ex",
    reportNumber = "CERN-EP-2019-162",
    doi = "10.1007/JHEP03(2020)145",
    journal = "JHEP",
    volume = "03",
    pages = "145",
    year = "2020"
}

@article{GarciadelMoral:2017vnz,
    author = "Garcia del Moral, Maria P. and Parameswaran, Susha and Quiroz, Norma and Zavala, Ivonne",
    title = "{Anti-D3 branes and moduli in non-linear supergravity}",
    eprint = "1707.07059",
    archivePrefix = "arXiv",
    primaryClass = "hep-th",
    doi = "10.1007/JHEP10(2017)185",
    journal = "JHEP",
    volume = "10",
    pages = "185",
    year = "2017"
}

@article{Angelantonj:2023egh,
    author = "Angelantonj, Carlo and Florakis, Ioannis and Leone, Giorgio",
    title = "{Tachyons and misaligned supersymmetry in closed string vacua}",
    eprint = "2301.13702",
    archivePrefix = "arXiv",
    primaryClass = "hep-th",
    doi = "10.1007/JHEP06(2023)174",
    journal = "JHEP",
    volume = "06",
    pages = "174",
    year = "2023"
}

@article{Angelantonj:2010ic,
    author = "Angelantonj, Carlo and Cardella, Matteo and Elitzur, Shmuel and Rabinovici, Eliezer",
    title = "{Vacuum stability, string density of states and the Riemann zeta function}",
    eprint = "1012.5091",
    archivePrefix = "arXiv",
    primaryClass = "hep-th",
    reportNumber = "DFTT-26-2010",
    doi = "10.1007/JHEP02(2011)024",
    journal = "JHEP",
    volume = "02",
    pages = "024",
    year = "2011"
}

@article{Leone:2025mwo,
    author = "Leone, Giorgio and Raucci, Salvatore",
    title = "{Aspects of strings without spacetime supersymmetry}",
    eprint = "2509.24703",
    archivePrefix = "arXiv",
    primaryClass = "hep-th",
    reportNumber = "IFT-UAM/CSIC-25-100",
    month = "9",
    year = "2025"
}

@article{Buratti:2018onj,
    author = "Buratti, Ginevra and Garc{\'\i}a-Valdecasas, Eduardo and Uranga, Angel M.",
    title = "{Supersymmetry Breaking Warped Throats and the Weak Gravity Conjecture}",
    eprint = "1810.07673",
    archivePrefix = "arXiv",
    primaryClass = "hep-th",
    reportNumber = "IFT-UAM/CSIC-18-102",
    doi = "10.1007/JHEP04(2019)111",
    journal = "JHEP",
    volume = "04",
    pages = "111",
    year = "2019"
}

@article{Angelantonj:2007ts,
    author = "Angelantonj, Carlo and Dudas, Emilian",
    title = "{Metastable string vacua}",
    eprint = "0704.2553",
    archivePrefix = "arXiv",
    primaryClass = "hep-th",
    reportNumber = "CPHT-RR-017.0417, DFTT-2007-5, LPT-ORSAY-07-23",
    doi = "10.1016/j.physletb.2007.06.031",
    journal = "Phys. Lett. B",
    volume = "651",
    pages = "239--245",
    year = "2007"
}

@article{Fraiman:2025yrx,
    author = "Fraiman, Bernardo and Parra de Freitas, H{\'e}ctor",
    title = "{Symmetries and dualities in non-supersymmetric CHL strings}",
    eprint = "2511.01674",
    archivePrefix = "arXiv",
    primaryClass = "hep-th",
    reportNumber = "MPP-2025-79; IFT-UAM/CSIC-25-111",
    month = "11",
    year = "2025"
}

@article{DESI:2025zgx,
    author = "Abdul Karim, M. and others",
    collaboration = "DESI",
    title = "{DESI DR2 results. II. Measurements of baryon acoustic oscillations and cosmological constraints}",
    eprint = "2503.14738",
    archivePrefix = "arXiv",
    primaryClass = "astro-ph.CO",
    reportNumber = "FERMILAB-PUB-25-0169-PPD",
    doi = "10.1103/tr6y-kpc6",
    journal = "Phys. Rev. D",
    volume = "112",
    number = "8",
    pages = "083515",
    year = "2025"
}

@article{Burgess:2021obw,
    author = "Burgess, C. P. and Dineen, Danielle and Quevedo, F.",
    title = "{Yoga Dark Energy: natural relaxation and other dark implications of a supersymmetric gravity sector}",
    eprint = "2111.07286",
    archivePrefix = "arXiv",
    primaryClass = "hep-th",
    reportNumber = "CERN-TH-2021-192",
    doi = "10.1088/1475-7516/2022/03/064",
    journal = "JCAP",
    volume = "03",
    number = "03",
    pages = "064",
    year = "2022"
}


\end{document}